\documentclass[a4paper,11pt]{article}
\usepackage{jheppub} 
\usepackage{physics}
\usepackage{bbold}
\usepackage{subcaption}
\usepackage{xcolor}
\usepackage{array}
\usepackage{multirow}
\usepackage{setspace}
\usepackage{marginnote}
\usepackage{booktabs}
\usepackage{tikz}
\usepackage{arydshln}

\usetikzlibrary{calc}
\usetikzlibrary{decorations.markings,shapes.geometric}
\usepgflibrary{decorations.pathmorphing}

\tikzstyle{quark}=[thick,postaction={decorate,decoration={markings,mark=at position 0.5 with {\arrow{>}}}}]
\tikzstyle{photon}=[thick, decorate, decoration={snake, amplitude=0.5mm, segment length=1.5mm}]
\tikzstyle{vertex}=[thick, black,fill=white]

\DeclareMathOperator{\Pf}{Pf}

\newcommand{\rcstar}{$\textrm{RC}^\star$}
\newcommand{\cstar}{C^\star}
\newcommand{\Nf}{N_{\mathrm{f}}}

\arxivnumber{} 
\preprint{HU-EP-25/18-RTG}

\title{\boldmath Comparing QCD+QED via full simulation versus the RM123 method: \texorpdfstring{$U$}{U}-spin window contribution to \texorpdfstring{$a_\mu^\mathrm{HVP}$}{amuHVP}}

\newlength{\ftsize}

\collaboration{
\begin{center}
\includegraphics[height=\ftsize]{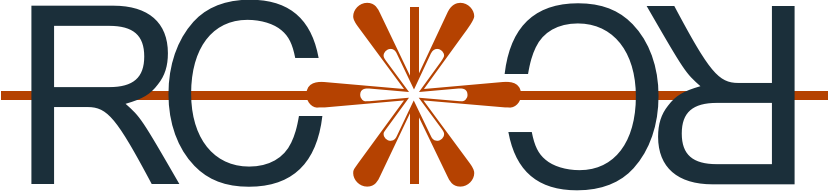}~\textnormal{collaboration}    
\end{center}
}

\author[1]{A.~Altherr,}
\author[2]{I.~Campos,}
\author[3,4]{A.~Cotellucci,}
\author[1]{R.~Gruber,}
\author[1]{T.~Harris,}
\author[1]{J.~Komijani,}
\author[5]{F.~Margari,}
\author[1]{M.~K.~Marinkovic,}
\author[1]{L.~Parato,}
\author[4,6]{A.~Patella,}
\author[2,4]{S.~Rosso,}
\author[5]{N.~Tantalo,}
\author[1]{P.~Tavella}

\affiliation[1]{Institut f\"ur Theoretische Physik, ETH Z\"urich, Wolfgang-Pauli-Str.~27, 8093 Z\"urich, Switzerland}
\affiliation[2]{Instituto de Física de Cantabria, CSIC, Avda.~de Los Castros s/n, 39005 Santander, Spain}
\affiliation[3]{Jülich Supercomputing Centre, Forschungszentrum Jülich, D-52428 Jülich, Germany}
\affiliation[4]{Humboldt Universität zu Berlin, Institut für Physik, Newtonstra{\ss}e 15, 12489 Berlin, Germany}
\affiliation[5]{Dipartimento di Fisica, Universit\`a di Roma ``Tor Vergata'' and INFN, Sezione di Roma ``Tor Vergata'', Via della Ricerca Scientifica 1, I-00133 Roma, Italy}
\affiliation[6]{DESY, Platanenallee 6, D-15738 Zeuthen, Germany}

\emailAdd{ptavella@phys.ethz.ch}

\abstract{
Electromagnetic corrections to hadronic vacuum polarization contribute
significantly to the uncertainty of the Standard Model prediction of the muon anomaly, which poses conceptual and numerical challenges for ab initio lattice determinations.
In this study, we compute the non-singlet contribution from intermediate Euclidean current
separations in quantum chromo- and electrodynamics (QCD+QED) using $\cstar$
boundary conditions in two ways: either non-perturbatively by sampling the
joint probability distribution directly or by perturbatively expanding from an
isospin-symmetric theory.
This allows us to compare the predictions and their uncertainties at a fixed
lattice spacing and volume, including fully the sea quarks effects in
both cases.
Treating carefully the uncertainty due to tuning to the same renormalized theory
with $\Nf=1+2+1$ quarks, albeit with unphysical masses, we find it advantageous
to simulate the full QCD+QED distribution given a fixed number of samples.
This study lays the ground-work for further applications of $\cstar$ boundary
conditions to study QCD+QED at the physical point, essential for the next
generation of precision tests of the Standard Model.
}

\begin{document}

\maketitle
\flushbottom

\section{Introduction}
\label{sec:intro}
The Standard Model (SM) prediction for the muon anomalous magnetic moment
$a_{\mu}=(g-2)_\mu/2$~\cite{Aoyama:2020ynm, Aliberti:2025beg} is coming into sharper focus since several lattice quantum chromodynamics (QCD) studies~\cite{Borsanyi_2021,blum2024long,
djukanovic2024hvp,bazavov2024long} have shown internal consistency and further highlighted the tension between lattice and data-driven dispersive evaluations indicated in ref.~\cite{Borsanyi_2021}.
The lattice QCD~\cite{Borsanyi_2021,blum2024long,
djukanovic2024hvp,bazavov2024long} results\footnote{For a recent updated SM prediction by the Muon $g-2$ Theory Initiative see~\cite{Aliberti:2025beg}.} including a recent high-precision hybrid calculation~\cite{boccaletti2024highprec} point towards compatibility with the experimental results from BNL and Fermilab~\cite{Muong-2:2006rrc,Muong-2:2021ojo,Muong-2:2023cdq,Muong-2:2025xyk}, and the no new physics scenario.
Further tensions in the $e^+e^-\rightarrow\mathrm{hadrons}$ cross sections have arisen in light of the CMD-3 data~\cite{Ignatov_2024}, which call for scrutiny of all assumptions underpinning the SM predictions from the data-driven determination of the HVP contribution~\cite{Aoyama:2020ynm,benton2023datadriven}.

While an impressive array of lattice computations have been able to pin down sub-contributions to the HVP~\cite{Lehner_2020,Aubin_2022,%
Ce:2022kxy,Alexandrou_2023,kuberski2024hvp,Wang_2023,blum2023update,%
bazavov2024hvp,lahert2024twopi, Bazavov_2023, ExtendedTwistedMass:2024nyi}, in particular from small and intermediate Euclidean separations, the so-called short-distance and window quantities, certain lacunae still exist.
In particular, the final result from Fermilab’s E989 experiment has now achieved a precision of 127 parts per billion (ppb)~\cite{Muong-2:2025xyk}, thereby reinforcing the need for a per mille theoretical determination of the HVP contribution that matches the experimental accuracy. Consequently, the inclusion of leading QED corrections in such ab initio computations remains essential.
On the lattice, this requires either computing the corrections to the
leading-order (in QED) result, defined usually in a scheme which naturally
incorporates isospin symmetry such that the corrections become so-called isospin-breaking effects, or else computing directly in QCD+QED without an intermediate determination in the isospin-symmetric theory.
In either case, the incorporation of QED is challenging on the lattice for a number of conceptual and technical reasons,  a contributing factor to the BMW computation remaining the only complete computation of the muon anomaly in QCD+QED that incorporates all dynamical isospin-breaking effects.

To compute hadronic quantities and their electromagnetic corrections in
numerical lattice simulations, it is convenient, although not necessary, to use
the same cutoff provided by the lattice size $L$ and lattice spacing $a$ for
both QCD and QED. Alternative infinite-volume QED approaches have also been explored~\cite{Feng:2018qpx,Parrino:2025afq}.
The implementation of QED in a finite volume is, however, not trivial due to constraints imposed by the long-range nature of electromagnetism, which forbid the existence of charged states.
This has been overcome in many ways~\cite{Uno_2008,Endres_2016,Blum_2017,Feng_2019,Kronfeld:1990qu,Kronfeld:1992ae,Polley:1993bn},
and in this work we make use of $\cstar$ boundary conditions, which permit a local and gauge-invariant formulation of finite-volume
QED~\cite{Lucini:2015hfa}.
These properties likely result in better finite-volume effects compared to other
formulations \cite{Martins:2022hqb} and good scaling behaviour towards the
continuum~\cite{Patella:2017fgk}, making this setup ideal in the pursuit of
per mille accuracy on the gold-plated hadronic quantities that form the input
for precision tests of the SM.

In this work, we consider an observable closely related to the muon's HVP
contribution in the intermediate window defined by the correlator of the
$U$-spin vector current
\begin{align}
    V_\mu(x) &= \tfrac{1}{2}\{\bar s(x)\gamma_\mu s(x) - \bar d(x)\gamma_\mu d(x)\},
    \label{eq:uspin_current}
\end{align} which is flavour non-singlet when $m_\mathrm{d}=m_\mathrm{s}$ even
when isospin-symmetry is explicitly broken, as in the setup we use in this
work.
Due to its flavour quantum numbers, its correlation function is represented by
quark-line connected diagrams which are significantly less complicated to
compute than the full electromagnetic current correlator, whose isospin-violating contributions have been recently considered in ref.~\cite{Erb:2025nxk}.
If the full $\mathrm{SU}(3)$ symmetry of the light quarks is manifest, the
contribution of this current correlator represents exactly 75\% of their
contribution to the hadronic vacuum polarization.
In addition to this simplification, the intermediate window we investigate
is dominated by current separations between
$0.4\mbox{--}1.0\,\mathrm{fm}$, meaning it is neither afflicted by either large
finite-volume effects nor discretization effects.
Therefore, this contribution to the muon anomaly, denoted $a_\mu^\mathrm{U,w}$,
represents a very good quantity to use as a probe observable to test our
approach to QCD+QED using $\cstar$ boundary conditions.

In the following, we compare and contrast two implementations of QCD+QED with
$\cstar$ boundary conditions, consistent at next-to-leading order in the
electromagnetic coupling, to test their efficiency and utility.
On one hand, we expand around an action defined by QCD in the isospin-symmetric
limit, \emph{\`a la} Rome-123
(RM123)~\cite{deDivitiis:2011eh,deDivitiis:2013xla}, while on the other hand we
simulate directly the joint QCD+QED distribution non-perturbatively.
In both cases, all effects from the sea quarks are completely included, making
the comparison unambiguous at leading order in the expansion parameters.
We make the comparison in two ways: (i) by fixing the bare parameters of the simulations in both implementations and comparing the resulting predictions for
both $a_\mu^\mathrm{U,w}$ as well as the hadronic observables used to define
the renormalized theory, and (ii) by fixing the line of constant physics using
those same hadronic observables and propagating their uncertainty to the
physical prediction of the muon anomaly.
Our final results incorporating the uncertainty due to fixing to the same lines of constant physics are
\begin{align*}
    a_\mu^\mathrm{U,w}\times 10^{11} &=
    \begin{cases}
        1094(21) & \qquad\textrm{RM123}\\
        1085(7)  & \qquad\textrm{non-perturbative}
    \end{cases}\quad.
\end{align*}
We find the uncertainty in the RM123 method completely dominated by the statistical uncertainty in the estimation of the isospin-breaking effects due to the sea quarks.
This uncertainty can only be reduced by sampling more gauge-field configurations. Moreover, the variance is expected to grow linearly with the volume in lattice units, up to logarithmic corrections in the lattice spacing, 
making the reduction of this uncertainty very costly toward the infinite-volume and continuum limits.
This work provides the first experience on the computation of physical
predictions with this setup and lays the crucial groundwork for future
computations and moving closer to the physical point. 

The rest of this paper is organized as follows.
In section~\ref{sec:sec2-lcp}, we describe the renormalization scheme used to define QCD+QED and isospin-symmetric QCD parameters.
Section~\ref{sec:sec3-latt} provides the details of the lattice setup and the discretization of the observable with $\cstar$ boundary conditions.
In section~\ref{sec:sec4-rm123}, we derive in detail the isospin-breaking corrections through the RM123 approach.
The numerical implementation is described in section~\ref{sec:numerics} while
the analysis and results are discussed in section~\ref{sec:sec6-res}, followed
by our concluding remarks.

\section{Parameterization of QCD+QED and isospin-symmetric QCD}
\label{sec:sec2-lcp}
The QCD+QED action with $\Nf=4$ quark flavors $f=\mathrm{u,d,s,c}$ contains six
bare parameters: four quark masses $m_f$ and the electromagnetic and strong
couplings $e_0^2$ and $g_0^2$.
The renormalized theory is defined by introducing constraints which fix the
bare masses along the lines of constant physics.
Physical predictions for QCD+QED are then unambiguous in the continuum limit
once the lattice scale has been set and ignoring the running of the
electromagnetic coupling, which goes beyond the target accuracy of
state-of-the-art computations.

In this work, we follow the hadronic renormalization scheme of
ref.~\cite{RCstar:2022yjz} to set the lattice scale and fix the quark masses. The lattice spacing is set using the gradient-flow scale $\sqrt{8t_0}$.
The bare masses of the up, down, strange, and charm quarks are tuned by fixing the following set of dimensionless hadronic quantities
\begin{equation}
\begin{aligned}
    \phi_0&=8t_0 (M_{K^{\pm}}^2-M_{\pi^{\pm}}^2),  \\
    \phi_1&=8t_0 (M_{K^{\pm}}^2+M_{\pi^{\pm}}^2+ M_{K^{0}}^2),  \\
    \phi_2&=8t_0 (M_{K^{0}}^2-M_{K^{\pm}}^2)/\alpha_\mathrm{R},  \\
    \phi_3&=\sqrt{8t_0}(M_{D_\mathrm{s}^{\pm}}+M_{D^{\pm}}+ M_{D^{0}}),
    \label{eq:phi-obs-def}
\end{aligned}
\end{equation}
where $\alpha_\mathrm{R}$ is the renormalized electromagnetic coupling computed at flow time $t_0$, and the masses denote those of the light
$\pi^{\pm}, K^{\pm,0}$ and charmed $D^{\pm,0}, D_\mathrm{s}^{\pm}$ pseudoscalar
mesons.
The $\phi_i$ quantities are
particularly sensitive to certain combinations of quark masses, as discussed
for example in refs.~\cite{Neufeld:1995mu,Bar:2013ora}.
Specifically, $\phi_0$, $\phi_1$, $\phi_2$, and $\phi_3$ probe $(m_\mathrm{s} - m_\mathrm{d})$, $ (m_\mathrm{u} + m_\mathrm{d} + m_\mathrm{s})$, $\alpha_R^{-1} (m_\mathrm{d} - m_\mathrm{u})$, and  $m_\mathrm{c}$, respectively. 
As we are only concerned with the theory accurate to next-to-leading order in the electromagnetic coupling, which is more than sufficient for the target
precision, we can safely set the bare parameter $e^2_0$ to its physical value $e^2=4\pi
\alpha=0.091701237$ in the Thomson limit.
In the following, we will use the notation $e^2$ or $\alpha$ to refer to the bare parameters for brevity.

The renormalization conditions we use to define our theory at unphysical quark
masses are given by matching the above quantities to the values
\begin{equation}
     \phi_0^\star = 0 , \qquad \phi_1^\star = 2.11 , \qquad \phi_2^\star = 2.36 ,  \qquad \phi_3^\star = 12.1 ,
      \label{eq:schemeunphys}
  \end{equation}
which do not match their physical values.
We set the gradient-flow scale to the central value of the $\Nf=3$ CLS
determination~\cite{Bruno:2016plf} 
\begin{equation}
    \sqrt{8t_0^\star} = 0.415 \; \text{fm}.
    \label{eq:scale}
\end{equation}
The specific choices for the $\phi_i$ and the scale correspond to a
pseudoscalar mass of approximately $M_{\pi^\pm}\approx M_{K^\pm}\approx400\,\mathrm{MeV}$ in
physical units.
In QCD+QED, where isospin symmetry is explicitly broken, the condition
$\phi_0=0$ corresponds to setting $m_\mathrm{d}=m_\mathrm{s}$ and there
remains a $\mathrm{SU}(2)$ flavour symmetry between the down and strange
quarks, which we refer to as $U$-spin symmetry.

We now turn our attention to QCD in the isospin-symmetric limit, which forms the
practical starting point of the perturbative approach to QCD+QED.
While the parameterization of QCD+QED is unambiguous at the level of accuracy that can be probed in Nature and the renormalization scheme can be chosen as
a matter of convenience, the same is not true for QCD at the target percent level of precision.
In ref.~\cite{FLAG:2024oxs}, a prescription
for lattice computations has been proposed to facilitate comparisons between high-precision lattice QCD computations.
In this work, because 
our renormalized theory is still far from the physical point, where such considerations are important, we impose for isospin-symmetric QCD the same conditions defined by the values in eq.~\eqref{eq:schemeunphys} and take the limit of $\alpha \to 0$ at a constant $\phi_2$, leading to $m_{\mathrm u} = m_{\mathrm d} = m_\mathrm{s}$. 
In the following, we denote the corresponding theory by isoQCD for brevity.
While the definition of a renormalized isoQCD theory is not strictly necessary to use as a starting point for QCD+QED computations as we perform in this study, it is required for computing renormalized results for isospin-breaking effects.

In this work, we will not examine the isospin-breaking effects directly but
compare results for a specific probe observable, $a^{\mathrm{U,w}}_{\mu}$,
described in detail in section~\ref{sec:Iwhvp}, computed in QCD+QED using either
the perturbative expansion around the isoQCD action or the non-perturbative
simulation of the joint distribution of QCD+QED.
Details about the implementation of the perturbative approach, originally
proposed by the RM123 Collaboration~\cite{deDivitiis:2011eh,
deDivitiis:2013xla}, are contained in section~\ref{sec:sec4-rm123}. Firstly, we will compare the results for the scale-setting quantities $t_0/a^2$ and $\phi_i$ and
for $a^{\mathrm{U,w}}_{\mu}$ at fixed bare parameters: in this case, both the
target observable and the scale-setting quantities will carry an uncertainty.
Secondly, we will compare results for $a^{\mathrm{U,w}}_{\mu}$ at fixed
renormalization scheme: this requires propagating the uncertainty on the
scale-setting quantities to our target observable.

\section{QCD+QED on the lattice with \texorpdfstring{$\cstar$}{C*} boundary conditions}
\label{sec:sec3-latt}
Implementing QED on a finite lattice presents additional challenges due to the
zero modes of the photon field, see  e.g. ref.~\cite{Patella:2017fgk} for a
discussion.
One approach for handling these zero modes is the application of ${\cstar}$ boundary conditions,
also known as $C$-periodic boundary conditions~\cite{Kronfeld:1990qu, Kronfeld:1992ae,Polley:1993bn,Lucini:2015hfa}.
This method provides a rigorous way to simulate QED on a lattice without compromising the theory's locality or requiring an additional regulator, such as a mass term for the photon field.  Although using ${\cstar}$ boundary conditions increases computational cost due to the need to effectively doubling the volume in the orbifold construction, studies suggest that finite volume effects are relatively suppressed in such simulations, compensating at least part of the additional cost~\cite{Martins:2022hqb}.
Here, we briefly outline our implementation of ${\cstar}$ boundary conditions; for a comprehensive review, we refer the reader to ref.~\cite{Lucini:2015hfa}.

On a lattice with a finite size $L$ in $\hat k$ direction,
$\cstar$ boundary conditions on fermionic and gauge fields are defined
using charge conjugation matrix $C$ as
\begin{equation}
\begin{aligned} \label{eq:orbi}
    \psi_f (x + L \hat{k}) &= \psi_f^{\mathcal{C}}(x) := C^{-1} \bar{\psi}_f^\top(x),
    \\
    \bar{\psi}_f (x + L \hat{k}) &= \bar{\psi}^{\mathcal{C}}(x) := - \psi_f^\top(x) C,
    \\
    U_{\mu}(x +  L \hat{k}) &= U_{\mu}^{\mathcal{C}}(x) := U_{\mu}^{\ast}(x),
    \\
    A_{\mu}(x +  L \hat{k}) &= A_{\mu}^{\mathcal{C}}(x) := -A_{\mu}(x).
\end{aligned}
\end{equation}
Here, $\psi_f$ and $\bar{\psi}_f$ are the fermionic fields of flavour $f$, $U_{\mu}(x) \in \,$$\mathrm{SU}(3)$ are the QCD lattice gauge fields, and $A_{\mu}(x)$ is the photon gauge field.
The symbols ${\mathcal{C}}$, $\top$, and $\ast$ denote charge conjugation, transposition, and complex conjugation, respectively.
Also note that in our convention the charge conjugation matrix $C$ obeys
$C \gamma_{\mu} C^{-1} = - \gamma_{\mu}^\top$
with the Euclidean gamma matrices $\gamma_{\mu}$.
By imposing the $\cstar$ boundary conditions in one or more directions,
the photon field $A_{\mu}(x)$ is antiperiodic in those directions,
and therefore the zero-modes are excluded by construction. In this work, we impose $\cstar$ boundary conditions in three spatial directions.  
The details of the computational implementation of $\cstar$ boundary conditions are discussed
in ref.~\cite{Campos:2019kgw}.

\subsection{Lattice action}\label{sec:lattact}
To simulate QCD+QED we discretize the Euclidean path integral of the theory,
\begin{equation}\label{eq:fullpathint}
    \mathcal{Z} =\int \mathcal{D}\psi  \mathcal{D}\bar{\psi} \mathcal{D} U  \mathcal{D} z \, 
    e^{-S_{\mathrm{F}}[U,z,\bar{\psi}, \psi]} \, 
    e^{-S_{\mathrm{g,SU(3)}}[U]} \,
    e^{-S_{\mathrm{g,U(1)}}[z]},
\end{equation}
on a $(T/a) \times (L/a)^3$ lattice using (anti-)periodic boundary conditions in the time direction and ${\cstar}$ boundary conditions in the spatial directions.
For the $\mathrm{SU}(3)$ gauge field we employ the L\"uscher-Weisz
discretization of the action~\cite{Weisz:1982zw,Luscher:1984xn}, while for the
$\mathrm{U}(1)$ gauge field we use the Wilson plaquette action in the compact
formulation
\begin{equation}\label{eq:U(1)act}
    S_{\text{g,U(1)}}(z) = \frac{1}{8\pi q_{\mathrm{el}}^2\alpha} \sum_{x} \sum_{\mu \neq \nu} [1-P_{\mu \nu}^{\mathrm{U(1)}}(x)],
\end{equation}
where $P^{\mathrm{U(1)}}_{\mu \nu}$ represents the plaquette built using the U(1) links $z_{\mu}(x)= e^{ie a q_{\mathrm{el}}A_{\mu}(x)}$. 
While the parameter $q_{\mathrm{el}}$ drops out of the action in the continuum limit,
at finite lattice spacing, the compact formulation implies that the electric charge is quantized: only states with a charge that is an integer multiple of the parameter $q_{\mathrm{el}}$ exist in the Hilbert space of the theory. In finite volume with ${\cstar}$ boundary conditions, we use $q_{\mathrm{el}}=1/6$ to construct gauge–invariant interpolating operators for charged hadrons, as explained in ref.~\cite{Lucini:2015hfa}.

The fermionic part of the QCD+QED action has to be modified because of the effect of the ${\cstar}$ boundary conditions. Indeed, the boundaries mix the $\psi$ and $\bar{\psi}$ degrees of freedom such that the Dirac operator does not act as a linear operator on the field $\psi$. The problem is overcome by defining an extended spinor that contains both the fermion field and its charge conjugate, i.e.
\begin{align}
    \chi(x) = \begin{pmatrix} \psi(x) \\\psi^{\mathcal{C}}(x) \end{pmatrix} ,
\end{align}
for which the boundary conditions in the spatial directions are set by
\begin{align}\label{eq:doubletbnd}
    \chi(x + L \hat{k}) &= K \chi (x), \quad K=
    \begin{pmatrix}
        0 & 1 \\
        1 & 0
    \end{pmatrix}.
\end{align}
In this new formulation, the measure of the path integral can be simply re-written as $\mathcal{D} \chi= \mathcal{D} \psi \mathcal{D} \bar{\psi}$, and the action reads
\begin{equation}\label{eq:fermsact}
    S_{\mathrm{F}} = -\sum_f a^4\sum_x \frac{1}{2} \chi_{f}^\top(x) KC D_f \chi_f(x).
\end{equation}
The massive $\mathrm O(a)$-improved Wilson-Dirac operator is defined by
\begin{equation}\label{eq:Diracop}
    D_f=D_{\mathrm{w},f}+\delta D_{\mathrm{sw},f} + m_f,
\end{equation}
where the Wilson-Dirac term is defined as
\begin{equation}\label{eq:Wilson-Diracop}
    D_{\mathrm{w},f}=\sum_{\mu=0}^{3} \frac{1}{2}\biggr[\gamma_{\mu}(\nabla^f_{\mu} + \nabla_{\mu}^{f  *}) - \nabla_{\mu}^{f *}\nabla^f_{\mu} \biggr],
\end{equation}
with the covariant forward finite-difference operator acting on the spinor as
\begin{equation}
    \nabla_{\mu}^f \chi_f(x) = a^{-1}\left[
    \begin{pmatrix}
        U_{\mu}(x) z^{\hat{q}_f}_{\mu}(x)& 0 \\
        0 & U^{*}_{\mu}(x)z^{-\hat{q}_f}_{\mu}(x) \\
    \end{pmatrix}
    \chi_f(x+a\hat{\mu}) - \chi_f(x)\right].
\end{equation}
The covariant derivative is not universal for all quarks due to the presence of the compact U(1) link $z_{\mu}^{\hat{q}_f}$, with $\hat{q}_f$ being the electric charge of the quark of flavour $f$ in units of the elementary charge $q_{\mathrm{el}}.$
The second term in equation \eqref{eq:Diracop} is the Sheikholeslami-Wohlert (SW) term
\begin{equation}\label{eq:Symterm}
     \delta D_{\mathrm{sw},f} = \frac{i}{4}  \sum_{\mu,\nu=0}^{3} \sigma_{\mu \nu} \left \{c_{\text{sw}}^{\text{SU(3)}}    
    \begin{pmatrix}
       \hat{\mathcal{G}}_{\mu \nu} & 0 \\
       0 & \hat{\mathcal{G}}_{\mu \nu}^{\ast}
    \end{pmatrix}
    + q_fc_{\text{sw}}^{\text{U(1)}}\ 
     \begin{pmatrix}
       \hat{\mathcal{F}}_{\mu \nu} & 0 \\
       0 & \hat{\mathcal{F}}_{\mu \nu}^\ast
    \end{pmatrix}\right\},
\end{equation}
which removes $\mathrm O(a)$ discretization effects from the action with
appropriately chosen coefficients.
$\hat{\mathcal{G}}_{\mu \nu}$ and $\hat{\mathcal{F}}_{\mu \nu}$ are the clover discretizations of the anti-hermitian $\mathrm{SU}(3)$ and $\mathrm U(1)$ tensors. 
The $\mathrm{SU}(3)$ tensor is defined as in ref.~\cite{Luscher:1996sc} while the $\mathrm U(1)$ tensor is constructed as
\begin{equation}\label{eq: explicit expression U(1) clover}
    \hat{\mathcal{F}}_{\mu \nu}(x) = \frac{i}{4 q_{\mathrm{el}}} \Im{z_{\mu \nu}(x) + z_{\mu \nu}(x-\hat{\mu})+ z_{\mu \nu}(x-\hat{\nu})+ z_{\mu \nu}(x-\hat{\mu}- \hat{\nu})}, 
\end{equation}
with $z_{\mu \nu}(x)= \exp{ieq_{\mathrm{el}}a[A_{\mu}(x)+A_{\nu}(x+\hat{\mu})-A_{\mu}(x+\hat{\nu}) - A_{\nu}(x)]}$.

The action of the Dirac operator on the doublet field $\chi_f$ may then be given through
\begin{align}\label{eq:Diracopexpl}
    D_f \chi_f(x) = &
    (m_f + 4) \chi_f(x)
    - \frac{1}{2} \sum_\mu (1-\gamma_\mu) \left[e^{i e q_f a A_\mu \tau_3} W_\mu\right](x) \chi_f(x+\hat{\mu})
    \nonumber \\ &
    - \frac{1}{2} \sum_\mu (1+\gamma_\mu) \left[e^{i e q_f a A_\mu \tau_3} W_\mu\right](x-\hat{\mu})^\dag  \chi_f(x-\hat{\mu}) +\delta D_{\mathrm{sw},f}\, \chi_f(x),
\end{align}
where $e^{ieq_fa A_{\mu}\tau_3}W_{\mu}$ is the $\mathrm{SU}(3) \times \mathrm{U}(1)$ field, with the following definitions of the $\mathrm{SU}(3)$ parallel transporter $W_{\mu}$ and the matrix $\tau_3$:
\begin{equation}\label{eq:defWmatrix}
     W_{\mu}(x)=\begin{pmatrix}
            U_{\mu}(x) & 0\\
            0 & U^{*}_{\mu}(x)
        \end{pmatrix},
        \qquad 
        \tau_3 = 
    \begin{pmatrix}
        1 & 0 \\
        0 & -1
    \end{pmatrix}.
\end{equation}
The same action is used for isoQCD simulations with the electromagnetic charge of the quarks in eq.~\eqref{eq:Diracopexpl} set to zero.

\subsection{Flavour non-singlet contribution in the intermediate window}\label{sec:Iwhvp}

In this work, we consider the $U$-spin current
\begin{equation}\label{eq:ucurr}
    V_{\mu}(x)=\frac{1}{2}\sum_{f=d,s} Q_f\bar\chi_f(x)\gamma_{\mu}\frac{\tau_3}{2} \chi_f(x),
\end{equation}
defined using the doublet notation introduced in the previous subsection, with $\bar{\chi}=-\chi^\top KC$, and the charge assignments $Q_\mathrm{s}=1=-Q_\mathrm{d}$,
and compute the Euclidean-time correlator in the time-momentum representation~\cite{Bernecker:2011gh}
\begin{equation}\label{eq:ucurrcorr}
    G^\mathrm{U}(t) =- \frac{1}{3} \sum_{k=1}^3\int\mathrm d^3x\, \ev{V_k(x)V_k(0)}.
\end{equation}
The intermediate window of this contribution to the muon's HVP is obtained via the integral
\begin{equation}\label{eq:amuhvp}
    a^{\mathrm{U,w}}_{\mu} =\biggr (\frac{\alpha}{\pi} \biggr )^2\int_0^\infty \mathrm dt \, G^\mathrm{U}(t) \tilde{K}(t; m_{\mu})w_{\mathrm{I}}(t).
\end{equation}
The kernel $\tilde K$ is computed following ref.~\cite{DellaMorte:2017dyu},
while the intermediate window $w_{\mathrm{I}}(t)$ is equal to the weight function 
\begin{align}\label{eq:IWdef}
    w(t;t_1,t_2,\Delta)&=\Theta(t,t_1,\Delta) - \Theta(t,t_2,\Delta),
\end{align}
with the choice $t_1=0.4 \,\mathrm{fm}, \,  t_2=1 \,\mathrm{fm},
\Delta=0.15 \,\mathrm{fm}$ \cite{Ce:2022kxy}, and the $\Theta$ function defined as
\begin{align}
    \Theta(t,t',\Delta)&=\frac{1}{2}\left(1+\tanh[(t-t')/\Delta]\right).
\end{align}
The intermediate window selects the contribution less susceptible to finite-volume and lattice discretization effects, and yet provides a substantial fraction of the total.

With this definition of the current, we examine a contribution which is proportional to the full light-quark contribution from the electromagnetic current when there is $\mathrm{SU}(3)$ symmetry. At the physical point, the light quarks contribute the most to the HVP, as indeed the charm and bottom quarks provide only 2\% of the total. With $\mathrm{SU}(3)$ symmetry, the contribution of the light quarks to the electromagnetic current is non-singlet thanks to the vanishing sum of their charges and thus the correlator is represented by a single Wick contraction (a quark-line connected diagram) once the quark fields have been integrated out.
When the $\mathrm{SU}(3)$ symmetry is broken, for example from isospin-breaking effects, quark-line disconnected diagrams, which are
computationally demanding to compute, no longer cancel.
In this investigation, even after including QED, we retain an $\mathrm{SU}(2)$
symmetry between the $d$ and $s$ quarks.
With this choice, the current introduced in eq.~\eqref{eq:ucurr} transforms in the non-singlet representation of
$\mathrm{SU(2)}$ when $m_\mathrm{d}=m_\mathrm{s}$, and we avoid computing the
associated disconnected diagrams.

Although the absence of the singlet part of the current prevents us from
computing corrections to the full electromagnetic current correlator, one could
still determine the quark-line connected part of it, which can be defined in a
partially-quenched theory.
However, we do not expect that the isospin-breaking effects to that quantity
to have any special significance over the corrections to the non-singlet
correlator, especially in the comparison between the two approaches.
Finally, we note that $U$-spin current is protected from mixing with singlet
operators, so there are no additive renormalizations required in our case,
unlike for the electromagnetic current, c.f. ref.~\cite{Collins_2006}.

\subsection{Lattice discretization of \texorpdfstring{$a_\mu^\mathrm{U,w}$}{amuwU}}\label{sec:defcorr}

On the lattice, we must choose a discretization of the operators and integral appearing in eq.~\eqref{eq:ucurrcorr}.
In this work, in addition to the local current defined in eq.~\eqref{eq:ucurr},
we make use of the point-split current 
\begin{align}\label{eq:pscurr}
    \tilde V_{\mu}(x)= 
    \frac{1}{2}\sum_{f=\mathrm{d,s}}Q_f \Big[&\bar{\chi}_f(x+\hat{\mu}) 
        \frac{(1+\gamma_{\mu})}{4}e^{-i e q_f a A_\mu (x)\tau_3} W^{\dagger}_{\mu}(x)\tau_3\chi_f(x) \nonumber \\
        &-\bar{\chi}_f(x)  \frac{(1-\gamma_{\mu})}{4} e^{i e q_f a A_\mu(x) \tau_3} W_{\mu}(x)\tau_3\chi_f(x+\hat{\mu})
    \Big].
\end{align}
While this current satisfies a Ward identity at finite lattice spacing,
and automatically has the correct normalization, the local current in eq.~\eqref{eq:ucurr} requires a finite multiplicative renormalization to match it.
Furthermore, both currents require $\mathrm O(a)$ improvement counterterms whose
coefficients are not yet known for $N_{\mathrm f}=4$ Wilson fermions and the
L\"uscher-Weisz gauge action.
Therefore, we use the unimproved currents.

We define the renormalized but unimproved local current via
\begin{align}\label{eq:rencurr}
    V_{\mu}^\mathrm{R}(x) &= Z_\mathrm{V}^\mathrm{m}(g_0^2,e^2,m_f) V_\mu(x),
\end{align}
where $Z_\mathrm{V}^\mathrm{m}(g_0^2,e^2,m_f)$ is the mass-dependent
non-singlet renormalization factor.
A suitable renormalization condition to determine $Z_\mathrm{V}^\mathrm{m}$ can
be constructed by imposing that the local and point-split discretizations agree at large Euclidean separations 
\begin{align}\label{eq:vectorrc}
    \lim_{t \to \infty} \frac{Z_\mathrm{V}^\mathrm{m}G_\mathrm{bare}^{\mathrm{U,l}}(t)}
    {G_\mathrm{bare}^{\mathrm{U,c}}(t)}
    = 1,
\end{align}
where the bare correlators of the local current is defined through
\begin{align}\label{eq:barellcorr}
    G^\mathrm{U,l}_\mathrm{bare}(t) &= -\frac{1}{3}a^3 \sum_k \sum_\mathbf{x}\ev{V_k(x)V_k(0)}
\end{align}
and the corresponding one with the point-split discretization at the sink
\begin{align}\label{eq:bareclcorr}
    G^\mathrm{U,c}_\mathrm{bare}(t) &= -\frac{1}{3}a^3 \sum_k \sum_\mathbf{x}\ev{\tilde V_k(x)V_k(0)}.
\end{align}
In practice, as our operators are not improved, this condition has the
unfortunate feature that it will be potentially subject to large cut-off
effects.
As we will use the same condition in both implementations and compare at finite
lattice spacing, this does not pose a particular problem for our study.

The renormalization thus defined, the two discretizations of the correlator we
employ are one using only the local currents
\begin{align}\label{eq:llcorr}
    G^\mathrm{U,l}(t) &= (Z_\mathrm{V}^\mathrm{m})^2G^\mathrm{U,l}_\mathrm{bare}(t)
\end{align}
and one using a local and a point-split current
\begin{align}\label{eq:clcorr}
    G^\mathrm{U,c}(t) &= Z_\mathrm{V}^\mathrm{m}G^\mathrm{U,c}_\mathrm{bare}(t).
\end{align}
In practice, the normalization condition ensures that they agree at long distances, but in this study we probe smaller Euclidean separations where they will differ due to cut-off effects.
Thus, we have two estimators for the lattice, finite-volume, $U$-spin observable
\begin{equation}\label{eq:latamu}
    a^{\mathrm{U,w}}_{\mu} =\biggr (\frac{\alpha}{\pi} \biggr )^2 a\sum_{t=0}^{T/2} \, G^{\mathrm{U},\ell}(t) \tilde{K}(t; m_{\mu})w_{\mathrm{I}}(t),\qquad \ell=\mathrm{l,c},
\end{equation}
for either the local or point-split discretizations of the sink current.

\section{Isospin-breaking effects \emph{à la} RM123}
\label{sec:sec4-rm123}
While in the non-perturbative QCD+QED approach the computation of the observables proceeds identically to isoQCD with no special treatment, the RM123 method requires the estimation of new classes of diagrams which parameterize the linearization of the observables in the bare parameters.
In this section, we derive the required diagrams that arise from expanding the lattice action and currents in the bare parameters connecting isoQCD and QCD+QED,
as described in section~\ref{sec:sec2-lcp}.
First, we introduce our notation for the leading corrections order in the expansion parameters.

By Taylor-expanding an observable $X$ at finite lattice spacing
in the changes of the bare parameters $\Delta\varepsilon = (e^2,\Delta
m_\mathrm{u},\Delta m_\mathrm{d},\Delta m_\mathrm{s},\Delta m_\mathrm{c})$ as
\begin{equation} \label{eq:genericObsExpansion}
    X(\Delta\varepsilon)
    = X(0)
    + e^2 \partial_{e^2}X(0)
    + \sum_{f=\mathrm{u,d,s,c}} \Delta m_f \partial_{m_f}X(0)
    + \mathrm O(e^4),
\end{equation}
where we denote the partial derivatives $\partial_{e^2}=\partial/\partial e^2$
and $\partial_{m_f}=\partial/\partial m_f$, we can estimate our observable via
$X(\Delta\varepsilon)\approx X^{(0)}+\delta X$, where
\begin{align}
    X^{(0)} =X(0),\qquad \delta X = e^2 \partial_{e^2}X(0)
    + \sum_{f=\mathrm{u,d,s,c}} \Delta m_f \partial_{m_f}X(0).
    \label{eq:firstsecond}
\end{align}
Both terms can be computed in the isoQCD theory defined by $\Delta\varepsilon=0$.
We note that given our definitions of isoQCD and QCD+QED we have $\Delta
m_f=\mathrm O(e^2)$, and so corrections to the above formula start at second
order in that parameter.
We recall that both theories are defined at the same bare coupling $g_0^2$, so
no change in this parameter is needed, but isoQCD and QCD+QED will have
different lattice spacings.
Furthermore, in the current definition, no change in the $\mathrm O(a)$
improvement coefficients are included.

For fixed line of constant physics, the change in the bare masses $\Delta m_f$
are computed by expanding the hadronic observables defined in
eq.~\eqref{eq:phi-obs-def} at first-order in $\Delta m_f$ and $e^2$ around the
isoQCD point and imposing the renormalization conditions as before.
This amounts to solving the system of equations
\begin{equation}\label{eq:linearizedMatchingConditions}
    \phi_i(0)
    + e^2 \partial_{e^2}\phi_i(0)
    + \sum_{f=\mathrm{u,d,s,c}} \Delta m_f \partial_{m_f}\phi_i(0)
    = {\phi}^\star_i,
\end{equation}
where the right-hand side ${\phi}^\star_i$ are the target values in the full
theory, given in eq.~\eqref{eq:schemeunphys}.
The derivatives $\partial_{e^2}\phi_i$ and $\partial_{m_f}\phi_i$ can be
related to the derivatives of the pseudoscalar meson masses that enter the
definitions of each $\phi_i$ and the derivatives of the scale setting
observable in lattice units $\hat{t}_0 = t_0 / a^2$.

When the same bare coupling is used in isoQCD and QCD+QED the lattice spacing $a$ also receives a correction to the value computed in the isoQCD ensemble
\begin{equation}\label{eq: corrections a}
    a^{(0)} = \sqrt{\frac{t_0^\star}{\hat t_0^{(0)}}}, \qquad \frac{\delta a}{a^{(0)}} = -\frac{1}{2} \frac{\delta \hat{t}_0}{\hat t_0^{(0)}}\,.
\end{equation}

In the following, we derive the form of the required corrections with $\cstar$
boundary conditions and $\mathrm O(a)$-improved Wilson fermions, where in
practice, we find it more convenient to expand the path integral after
integrating over the Grassmann fields.

\subsection{Derivation with \texorpdfstring{$\cstar$}{C*} boundary conditions}
\label{sec:sec4.1}
After integrating out the fermion fields, the expectation value of any
observable $\mathcal O[U,z]$ in QCD+QED with $\cstar$ boundary conditions may be written
\begin{equation}\label{eq:expvalueobs}
    \ev{\mathcal O[U,z]} = \mathcal Z^{-1}\int \mathcal D U \mathcal Dz
    \,\prod_f \Pf(CK D_{f}[U,z]) \, \mathcal O[U,z]
    \,\mathrm e^{-S_{\mathrm{g,SU(3)}}[U]-S_{\mathrm{g,U(1)}}[z]} ,
\end{equation}
where $\Pf(CKD_f)$ denotes the pfaffian of $CKD_f$, whose properties are discussed in ref.~\cite{RCstar:2022yjz}. 
Note that $\mathcal O[U,z]$ now depends explicitly on the $\mathrm{SU}(3)$ and $\mathrm U(1)$ gauge field variables, but not on the fermion fields and instead is in general a function of the inverse of the Dirac operator $D^{-1}_{f}$.

With the exception of the gradient flow scale $\hat t_0$, all hadronic observables we consider in this work are extracted from two-point functions of fermion bilinears which result in a single fermionic trace
\begin{equation}\label{eq:trace2pc}
    \mathcal O(x,y) = \Tr\{D^{-1}_f (y|x) \Gamma_{A} D^{-1}_g (x|y)\Gamma_{B}\}=
    \begin{tikzpicture}[baseline=-.1cm,scale=0.5]
            \coordinate (x) at (0,0);
            \draw[quark] (x) arc[start angle=-135, delta angle=90, radius=1.7] coordinate[at end] (y) node[black,below,pos=0.3] {$f$} ;
            \draw[quark] (y) arc[start angle=45, delta angle=90, radius=1.7] node[black,above,pos=0.3] {$g$} ;
            \draw[black, double] (x) arc[start angle=-135, delta angle=90, radius=1.7] coordinate[at end] (y) node[black,below,pos=0.3] {$f$} ;
            \draw[black, double] (y) arc[start angle=45, delta angle=90, radius=1.7] node[black,above,pos=0.3] {$g$} ;
            \draw[vertex] (x) circle (.1) node[left]{$x$};
            \draw[vertex] (y) circle (.1) node[right]{$y$};
        \end{tikzpicture}
\end{equation}
corresponding to quark-line connected diagram, where the trace is taken over the
color, Dirac and doublet spinor space.
We stress that, in the doublet formulation introduced in section \ref{sec:lattact}, the Dirac operator and its inverse appearing in eqs.~\eqref{eq:expvalueobs} and~\eqref{eq:trace2pc} are $24 \times 24$ matrices for fixed $x,y$.
The indices $f,g$ denote quark flavours, while $A,B=\mathrm{P,V,\tilde V}$ denote the pseudoscalar density, local vector and point-split vector current defined in eq.~\eqref{eq:pscurrent}.
The explicit form of the operators $\Gamma_\mathrm{P}, \Gamma_\mathrm{V}$ is 
\begin{align}
        \label{eq:GammasPV}
        \Gamma_\mathrm{P} = \tfrac{1}{2}\gamma_5 \mathbb{1}, \qquad
        \Gamma_\mathrm{V} = \tfrac{1}{2}\gamma_\mu \tau_3 \delta_{fg},
\end{align}
while the action of $\Gamma_\mathrm{\tilde{V}}$ on doublet spinors is
\begin{align}
    \label{eq:pscurrent}
    \eta^{\dagger}_f(x) \Gamma_\mathrm{\tilde{V}} \phi_g(x) &=  \eta^{\dagger}_f(x+\hat{\mu}) \tau_3 \frac{(1+\gamma_\mu)}{4}W^{\dagger}_{\mu}(x) e^{-ieq_fa A_{\mu}\tau_3} \phi_f(x)+ \nonumber\\
    & -\eta^{\dagger}_f(x) \tau_3 \frac{(1-\gamma_\mu)}{4}W_{\mu}(x) e^{ieq_fa A_{\mu}\tau_3} \phi_f(x+\hat{\mu}).
\end{align}
Given these definitions, it is clear that none of these vertices depends explicitly on the quark masses, and the only one depending on the electromagnetic coupling is $\Gamma_\mathrm{\tilde{V}}$.

In the following, we illustrate how to perturbatively expand the QCD+QED expectation value $\langle\mathcal O(x,y)\rangle$, which
requires expanding the inverse Dirac operator, the pfaffian and the vertex $\Gamma_\mathrm{\tilde V}$ around the isoQCD point.
We begin with the expansions of the inverse Dirac operator 
\begin{align} \label{eq:inv-D-exp}
    D_f^{-1}
    &= 
    (D_f^{(0)})^{-1} \left[ \mathbb{1} - \Delta D_f \, (D_f^{(0)})^{-1} + \left\{\Delta D_{f}  (D_f^{(0)})^{-1}\right\}^2   \right],
\end{align}
and the Pfaffian
\begin{align} 
    \label{eq:exp-pfaffian}
    \Pf(KCD_f)
    &= \Pf(KCD_f^{(0)})
    \left[ \mathbb{1} + \tfrac{1}{2} \Tr \left\lbrace(D_f^{(0)})^{-1} \, \Delta D_f\right\rbrace\right.
    \\ \nonumber
    & \qquad \left.+ \tfrac{1}{8}  \Tr \left\lbrace (D_f^{(0)})^{-1} \Delta D_{f} \right\rbrace^2 
    - \tfrac{1}{4} \Tr \left\lbrace (D_f^{(0)})^{-1} \Delta D_{f} (D_f^{(0)})^{-1}  \Delta D_{f} \right\rbrace\right]
    ,
\end{align}
which both can be expressed in terms of the Dirac operator at the isoQCD point $
D_f^{(0)}$, and its leading correction $\Delta D_{f}$.
Given the form of the Wilson-Dirac operator $D_f = D_{\mathrm{w},f}+\delta D_{\mathrm{sw},f} + m_f$  defined through eqs.~\eqref{eq:Diracop}, \eqref{eq:Wilson-Diracop}, and \eqref{eq:Symterm}, we expand the U(1) gauge links up to order $e^2$ and the quark masses to order $O(\Delta m_f)$, leading to the following three expressions:
\begin{align}
        &D_{\mathrm{w},f} = D^{(0)}_{\mathrm{w},f} + e q_f D^{(1)}_{\mathrm{w},f} + \frac{1}{2}e^2 q_f^2 D^{(2)}_{\mathrm{w},f} + O(e^3)
        , \label{eq:Dw-term-expan} \\
        &\delta D_{\mathrm{sw},f} = \delta D^{(0)}_{\mathrm{sw},f} + e q_f \, \delta D^{(1)}_{\mathrm{sw},f} + O(e^3)
        , \label{eq:Dsw-term-expan} \\
        &m_f = m^{(0)}_f + \Delta m_f + O((\Delta m)^2)
        , \label{eq:mass-term-expan}
\end{align}
where $D^{(0)}_{\mathrm{w},f}$ and $\delta D^{(0)}_{\mathrm{sw},f}$ are the terms at the isoQCD point, while $ q_f D^{(1)}_{\mathrm{w},f}, \, q_f \delta D^{(1)}_{\mathrm{sw},f}$ and $q_f^2 D^{(2)}_{\mathrm{w},f}$ denote the first and second derivatives with respect to $e$ at $e=0$.
Eq.~\eqref{eq:mass-term-expan} accounts for the shifts in the quark masses. 

We stress that the expansion of the SW term in eq.~\eqref{eq:Dsw-term-expan} relies solely on the expansion of the U(1) improvement term. 
In principle, the $\mathrm{SU}(3)$ improvement coefficient $c^{\text{SU(3)}}_{\text{sw}}$ could also be expanded around its isoQCD value.
However, since both isoQCD and QCD+QED simulations are performed with the same value for $c^{\text{SU(3)}}_{\text{sw}}$, in practice, we neglect QED corrections to this quantity in both approaches. Moreover, it can be shown that the expansion of eq.~\eqref{eq: explicit expression U(1) clover} contains only odd powers of $e$, which explains the absence of a term $ \propto e^2 q_f^2\delta D_{\mathrm{sw},f}^{(2)}$. 
Thus, $\delta D^{(0)}_{\mathrm{sw},f}$ in eq.~\eqref{eq:Dsw-term-expan} can be matched to the $\mathrm{SU}(3)$ improvement term in eq.~\eqref{eq:Symterm}, while $e q_f \, \delta D^{(1)}_{\mathrm{sw},f}$ is the first-order expansion of the U(1) improvement.

By using the definitions in eqs.~\eqref{eq:Dw-term-expan}-\eqref{eq:mass-term-expan}, we obtain $D_f^{(0)}$ and $\Delta D_f$ as
\begin{align}
    &D_f^{(0)}= D^{(0)}_{\mathrm{w},f}+  D^{(0)}_{\mathrm{sw},f} + m^{(0)}_f,
    \\
    \label{eq:DeltaDf}
    &\Delta D_f = \Delta m_f + e q_f D^{(1)}_{f} + \frac{1}{2}e^2 q_f^2 D^{(2)}_{\mathrm{w},f}  + O(e^3),
\end{align}
where we have collected the first order derivative $D^{(1)}_{f}  =  D^{(1)}_{\mathrm{w},f} + D^{(1)}_{\mathrm{sw},f}$. After factorizing out the photon fields, the three operators appearing in eq.~\eqref{eq:DeltaDf} can be represented diagrammatically as the following vertices:
\begin{align}
    \label{eq:massInsertionDiagram}
    \begin{tikzpicture}[baseline=-.1cm,scale=0.6]
        \draw[black,quark] (-1,0) -- (0,0) node[black,below,pos=0.5] {$f$};
        \draw[black,quark] (0,0) -- (1,0) node[black,below,pos=0.5] {$f$};
        \draw[fill=red, draw=black] (0,0.15) -- (-0.125,-0.075) -- (0.125,-0.075) -- cycle;
    \end{tikzpicture}
     & = \mathbb{1}\delta_{ff}
    \,,
    \\ \label{eq:chargeDerivativesDiagrams}
     \
    \begin{tikzpicture}[baseline=-.1cm,scale=0.6]
        \draw[black,quark] (-1,0) -- (0,0) node[black,below,pos=0.5] {$f$};
        \draw[black,quark] (0,0) -- (1,0) node[black,below,pos=0.5] {$f$};
        \draw[black,photon] (0,0) -- (0,1);
        \draw[fill=green, draw=black] (-0.1,-0.1) rectangle (0.1,0.1);
    \end{tikzpicture}
     & =\frac{\delta D_f^{(1)}}{\delta A_{\mu}}
    \,, \quad 
    \begin{tikzpicture}[baseline=-.1cm,scale=0.6]
        \draw[black,quark] (-1,0) -- (0,0) node[black,below,pos=0.5] {$f$};
        \draw[black,quark] (0,0) -- (1,0) node[black,below,pos=0.5] {$f$};
        \draw[black,photon] (0,0) -- (-.5,1);
        \draw[black,photon] (0,0) -- (0.5,1);
        \draw[fill=blue!60!cyan, draw=black, rotate=45] (-0.1,-0.1) rectangle (0.1,0.1);
    \end{tikzpicture} \
    = \frac{\delta D_{\mathrm{w},f}^{(2)}}{\delta A^2_{\mu}}\,,
\end{align}
where the identity in eq.~\eqref{eq:massInsertionDiagram} is an identity in Dirac, color and coordinate space.
The two operators in eq.~\eqref{eq:chargeDerivativesDiagrams} depend only on the $\mathrm{SU}(3)$ gauge field, and their action on spinor fields is readily obtained for the first derivative in $e$
\begin{align}\label{eq:Df-e-der}
    \eta^\dag(x) \frac{\delta D_f^{(1)}}{\delta A_{\mu}} \phi(x) &=                
    \frac{1}{2i} \eta^\dag(x) (1-\gamma_\mu) W_{\mu}(x) \tau_3 \phi(x+\hat{\mu})
    - \frac{1}{2i} \eta^\dag(x+\hat{\mu}) (1+\gamma_\mu) W_{\mu}(x)^\dag \tau_3 \phi(x)
    \nonumber                                                                        \\ 
    & \qquad
    - \frac{c_{\text{sw}}^{\text{U(1)}} }{8} \sum_\nu \sum_{ \substack{ \alpha=\pm 1 \\ \beta = 0,1} } \alpha \, \eta^\dag(x + \alpha \hat{\nu} + \beta \hat{\mu}) \sigma_{\mu\nu} \tau_3 \phi(x + \alpha \hat{\nu} + \beta \hat{\mu})
    ,
\end{align}
and for the second derivative in $e$
\begin{align}\label{eq:Df-e2-der}
    \eta^\dag(x) \frac{\delta D_{\mathrm{w},f}^{(2)}}{\delta A^2_{\mu}} \phi(x) &=
    \frac{1}{2} \eta^\dag(x) (1-\gamma_\mu) W_{\mu}(x) \phi(x+\hat{\mu})
    + \frac{1}{2} \eta^\dag(x+\hat{\mu})(1+\gamma_\mu) W_{\mu}(x)^\dag \phi(x)
    .
\end{align}
Although in our simulations we employ the leading order value in $e^2$ of the improvement coefficient $c^{\text{U(1)}}_{\text{sw}}=1$, we keep it generic in equation~\eqref{eq:Df-e-der} to highlight the part of the insertion that arises from improvement terms.

Finally, we expand $\Gamma_\mathrm{\tilde V}$ in powers of the electromagnetic coupling $e$:
\begin{equation}
    \label{eq:expansionVectorCurrent}
    \Gamma_\mathrm{\tilde V} = \Gamma_\mathrm{\tilde V}^{(0)} + e q_f \, \Gamma_\mathrm{\tilde V}^{(1)} + \frac{e^2 q_f^2}{2} \Gamma_\mathrm{\tilde V}^{(2)} + O(e^3),
\end{equation}
where $\Gamma_\mathrm{\tilde V}^{(0)}$ is the vertex defined in eq.~\eqref{eq:pscurrent} with $e$ set to 0, and $\Gamma_\mathrm{\tilde V}^{(1)}$ and $\Gamma_\mathrm{\tilde V}^{(2)}$ are the first and second derivatives of eq.~\eqref{eq:pscurrent} with respect to $eq_f$, evaluated at $e = 0$.
A diagrammatic expression for the vertices and their derivatives is:
\begin{align}
    \label{eq:VectorCurrentDiagrams}
    \Gamma_{A}^{(0)} =
    \begin{tikzpicture}[baseline=-.1cm,scale=0.6]
        \draw[black,quark] (0,0) -- (.7,0.7) node[above,pos=0.7] {};
        \draw[black,quark] (.7,-0.7) -- (0,0) node[below,pos=0.7] {};
        \fill[black] (0,0) circle (.1);
    \end{tikzpicture}
    \quad 
    A = \mathrm{P, V, \tilde V}\,, \qquad
    \frac{\delta \Gamma_\mathrm{\tilde V}^{(1)}}{\delta A_\mu} =
    \begin{tikzpicture}[baseline=-.1cm,scale=0.6]
        \draw[black,quark] (0,0) -- (.7,0.7) node[above,pos=0.5] {};
        \draw[black,quark] (.7,-0.7) -- (0,0) node[below,pos=0.5] {};
        \draw[photon] (-.7,0) -- (0,0);
        \node[regular polygon, regular polygon sides=5, draw=black, fill=orange, minimum size=5pt, inner sep=0pt, outer sep=0pt] at (0,0) {};
    \end{tikzpicture}
    \,, \qquad
    \frac{\delta \Gamma_\mathrm{\tilde V}^{(2)}}{\delta A_\mu^2} =
    \begin{tikzpicture}[baseline=-.1cm,scale=0.6]
        \draw[black,quark] (0,0) -- (.7,0.7) node[above,pos=0.5] {};
        \draw[black,quark] (.7,-0.7) -- (0,0) node[below,pos=0.5] {};
        \draw[photon] (-.7,0.7) -- (0,0);
        \draw[photon] (-.7,-0.7) -- (0,0);
        \node[star, star points=5, draw=black, fill=yellow, minimum size=5pt, inner sep=0pt, outer sep=0pt] at (0,0) {};
    \end{tikzpicture}
    \,.
\end{align}

{

\tikzstyle{quark}=[postaction={decorate,decoration={markings,mark=at position 0.5 with {\arrow{>}}}}]
\tikzstyle{photon}=[decorate, decoration={snake, amplitude=0.2mm, segment length=.7mm}]
\tikzstyle{vertex}=[draw=black,fill=white]
\tikzstyle{externalVertex}=[draw=black,fill=black]
\newcommand{\drawTriangle}[2]{%
    \pgfmathsetmacro{\h}{0.3}
    \draw[fill=red, draw=black]
    ($(#1)+(-0.5*\h,-0.289*\h)$) --
    ($(#1)+(0.5*\h,-0.289*\h)$) --
    ($(#1)+(0,0.577*\h)$) -- cycle;
}
\newcommand{\drawPhotonInsertion}[1]{%
    \draw[draw=black, fill=green] ($(#1)-(.1,.1)$) rectangle ($(#1)+(.1,.1)$);
}
\newcommand{\drawDoublePhotonInsertion}[1]{%
    \draw[draw=black, fill=blue!50!cyan, rotate=45] ($(#1)-(.1,.1)$) rectangle ($(#1)+(.1,.1)$);
}
\newcommand{\drawVertexPhotonInsertion}[1]{%
    \node[regular polygon, regular polygon sides=5, draw=black, fill=orange,
        minimum size=4pt, inner sep=0pt, outer sep=0pt] at (#1) {};%
}
\newcommand{\drawDoubleVertexPhotonInsertion}[1]{%
    \node[star, star points=5, draw=black, fill=yellow,
        minimum size=4pt, inner sep=0pt, outer sep=0pt] at (#1) {};%
}

\begin{table}[t]
    \renewcommand{\arraystretch}{1.5}\tiny
    \resizebox{\linewidth}{!}{%
    \begin{tabular}{lll|ll}
        \\ \hline \phantom{} & \phantom{} & \phantom{} & \phantom{} & \phantom{} \\[-1.5em]
        \begin{tikzpicture}[baseline,scale=0.5]
            \coordinate (x) at (0,0);
            \draw[quark] (x) arc[start angle=-135, delta angle=90, radius=1.7] coordinate[at end] (y) node[black,below,pos=0.7]{};
            \draw (y) arc[start angle=45, delta angle=90, radius=1.7]  coordinate[pos=0.5] (p) node[black,above,pos=0.7]{};
            \draw[externalVertex] (x) circle (.1) node[left]{};
            \draw[externalVertex] (y) circle (.1) node[left]{};
            \drawTriangle{p};
            \coordinate (oi) at ($(x) - (.1,-1)$);
            \draw[vertex,transparent] (oi) circle (.1) node[right] {};
            \coordinate (of) at ($(y) + (.1,1.2)$);
            \draw[vertex,transparent] (of) circle (.1) node[right] {};
        \end{tikzpicture}
            &
        \begin{tikzpicture}[baseline,scale=0.5]
            \coordinate (x) at (0,0);
            \draw[quark] (x) arc[start angle=-135, delta angle=90, radius=1.7] coordinate[at end] (y) node[black,below,pos=0.7] {};
            \draw[quark] (y) arc[start angle=45, delta angle=90, radius=1.7]  coordinate[pos=0.3] (p) node[black,above,pos=0.7] {};
            \draw[photon] (p) arc[start angle=255, delta angle=360, radius=.34];
            \draw[externalVertex] (x) circle (.1) node[left] {};
            \draw[externalVertex] (y) circle (.1) node[right] {};
            \drawDoublePhotonInsertion{p};
            \coordinate (oi) at ($(x) - (.1,-1)$);
            \draw[vertex,transparent] (oi) circle (.1) node[right] {};
            \coordinate (of) at ($(y) + (.1,1.2)$);
            \draw[vertex,transparent] (of) circle (.1) node[right] {};
        \end{tikzpicture}
          &
        \begin{tikzpicture}[baseline,scale=0.5]
            \coordinate (x) at (0,0);
            \draw[quark] (x) arc[start angle=-135, delta angle=90, radius=1.7] coordinate[at end] (y) coordinate[pos=0.3] (p1) node[black,below,pos=0.5] {};
            \draw[quark] (y) arc[start angle=45, delta angle=90, radius=1.7] coordinate[pos=0.3] (p2) node[black,above,pos=0.5] {};
            \draw[photon] (p1) -- (p2);
            \draw[externalVertex] (x) circle (.1) node[left] {};
            \draw[externalVertex] (y) circle (.1) node[right] {};
            \drawPhotonInsertion{p1};
            \drawPhotonInsertion{p2};
            \coordinate (oi) at ($(x) - (.1,-1)$);
            \draw[vertex,transparent] (oi) circle (.1) node[right] {};
            \coordinate (x) at (4,0);
            \draw[quark] (x) arc[start angle=-135, delta angle=90, radius=1.7] coordinate[at end] (y) node[black,below,pos=0.9] {};
            \draw[quark] (y) arc[start angle=45, delta angle=90, radius=1.7] coordinate[pos=0.3] (p1) coordinate[pos=0.7] (p2) node[black,above,pos=0.9] {};
            \draw[photon] (p1) .. controls ($(p1)!0.2!(p2) - (0,.5)$) and ($(p1)!0.8!(p2) - (0,.5)$) .. (p2);
            \draw[externalVertex] (x) circle (.1) node[left] {};
            \draw[externalVertex] (y) circle (.1) node[right] {};
            \drawPhotonInsertion{p1};
            \drawPhotonInsertion{p2};
            \coordinate (of) at ($(y) + (.1,1.2)$);
            \draw[vertex,transparent] (of) circle (.1) node[right] {};
        \end{tikzpicture}
          &
        \begin{tikzpicture}[baseline,scale=0.5]
            \coordinate (x) at (0,0);
            \draw[quark] (x) arc[start angle=-135, delta angle=90, radius=1.7] coordinate[at end] (y) node[black,below,pos=0.7] {};
            \draw[quark] (y) arc[start angle=45, delta angle=90, radius=1.7]  coordinate[pos=0.3] (p) node[black,above,pos=0.7] {};
            \draw[photon] (p) .. controls ($(p)!0.2!(x) - (0,.5)$) and ($(p)!0.8!(x)$) .. (x);
            \draw[externalVertex] (y) circle (.1) node[right] {};
            \drawPhotonInsertion{p};
            \drawVertexPhotonInsertion{x};
            \coordinate (oi) at ($(x) - (.1,-1)$);
            \draw[vertex,transparent] (oi) circle (.1) node[right] {};
            \coordinate (of) at ($(y) + (.1,1.2)$);
            \draw[vertex,transparent] (of) circle (.1) node[right] {};
        \end{tikzpicture}
          &
        \begin{tikzpicture}[baseline,scale=0.5]
            \coordinate (x) at (0,0);
            \draw[quark] (x) arc[start angle=-135, delta angle=90, radius=1.7] coordinate[at end] (y) node[black,below,pos=0.7] {};
            \draw[quark] (y) arc[start angle=45, delta angle=90, radius=1.7];
            \draw[externalVertex] (y) circle (.1) node[right] {};
            \draw[photon] (x) arc[start angle=0, delta angle=360, radius=.34];
            \drawDoubleVertexPhotonInsertion{x};
            \coordinate (oi) at ($(x) - (.1,-1)$);
            \draw[vertex,transparent] (oi) circle (.1) node[right] {};
            \coordinate (of) at ($(y) + (.1,1.2)$);
            \draw[vertex,transparent] (of) circle (.1) node[right] {};
        \end{tikzpicture}
        \\ \hline \phantom{} & \phantom{} & \phantom{} & \phantom{} & \phantom{} \\[-1.5em]
        \phantom{} &
        \phantom{} &
        \begin{tikzpicture}[baseline,scale=0.5]
            \coordinate (x) at (2,0);
            \draw[quark] (x) arc[start angle=-135, delta angle=90, radius=1.7] coordinate[at end] (y) node[black,below,pos=0.7] {};
            \draw[quark] (y) arc[start angle=45, delta angle=90, radius=1.7]  coordinate[pos=0.3] (p1);
            \coordinate (z) at ($(x)!0.5!(y) + (0,1)$);
            \draw[quark] (z) arc[start angle=0, delta angle=360, radius=.34];
            \draw[photon] (p1) .. controls ($(p1) + (0,.25)$) and ($(z) + (.25,0)$) .. (z);
            \drawPhotonInsertion{p1};
            \drawPhotonInsertion{z};
            \draw[externalVertex] (x) circle (.1) node[right] {};
            \draw[externalVertex] (y) circle (.1) node[right] {};
            \coordinate (oi) at ($(x) - (2.1,-1)$);
            \draw[vertex,transparent] (oi) circle (.1) node[right] {};
            \coordinate (of) at ($(y) + (2.1,1.5)$);
            \draw[vertex,transparent] (of) circle (.1) node[right] {};
        \end{tikzpicture}
          &
        \begin{tikzpicture}[baseline,scale=0.5]
            \coordinate (x) at (0,0);
            \draw[quark] (x) arc[start angle=-135, delta angle=90, radius=1.7] coordinate[at end] (y) node[black,below,pos=0.7] {};
            \draw[quark] (y) arc[start angle=45, delta angle=90, radius=1.7];
            \coordinate (z) at ($(x)!0.5!(y) + (0,1)$);
            \draw[quark] (z) arc[start angle=180, delta angle=360, radius=.34];
            \draw[photon] (x) .. controls ($(x) + (0,.5)$) and ($(z) - (.25,0)$) .. (z);
            \drawVertexPhotonInsertion{x};
            \drawPhotonInsertion{z};
            \draw[externalVertex] (y) circle (.1) node[right] {};
            \coordinate (oi) at ($(x) - (.1,-1)$);
            \draw[vertex,transparent] (oi) circle (.1) node[right] {};
            \coordinate (of) at ($(y) + (.1,1.5)$);
            \draw[vertex,transparent] (of) circle (.1) node[right] {};
        \end{tikzpicture}
          &
        \phantom{}
        \\ \phantom{} & \phantom{} &\phantom{} & \phantom{} & \phantom{} \\[-1.4em]
        \begin{tikzpicture}[baseline,scale=0.5]
            \coordinate (x) at (0,0);
            \draw[quark] (x) arc[start angle=-135, delta angle=90, radius=1.7] coordinate[at end] (y) node[black,below,pos=0.7] {};
            \draw[quark] (y) arc[start angle=45, delta angle=90, radius=1.7];
            \coordinate (z) at ($(x)!0.5!(y) + (0,.8)$);
            \draw[quark] (z) arc[start angle=270, delta angle=360, radius=.34];
            \drawTriangle{z};
            \draw[externalVertex] (x) circle (.1) node[right] {};
            \draw[externalVertex] (y) circle (.1) node[right] {};
            \coordinate (oi) at ($(x) - (.3,-1)$);
            \draw[vertex,transparent] (oi) circle (.1) node[right] {};
            \coordinate (of) at ($(y) + (.1,1.5)$);
            \draw[vertex,transparent] (of) circle (.1) node[right] {};
        \end{tikzpicture}
            &
        \begin{tikzpicture}[baseline,scale=0.5]
            \coordinate (x) at (0,0);
            \draw[quark] (x) arc[start angle=-135, delta angle=90, radius=1.7] coordinate[at end] (y) node[black,below,pos=0.7] {};
            \draw[quark] (y) arc[start angle=45, delta angle=90, radius=1.7];
            \coordinate (z) at ($(x)!0.5!(y) + (0,.8)$);
            \draw[quark] (z) arc[start angle=270, delta angle=360, radius=.34] coordinate[pos=0.25] (z1) coordinate[pos=0.75] (z2);
            \draw[photon] (z2) arc[start angle=0, delta angle=360, radius=.34];
            \draw[externalVertex] (x) circle (.1) node[right] {};
            \draw[externalVertex] (y) circle (.1) node[right] {};
            \drawDoublePhotonInsertion{z2};
            \coordinate (oi) at ($(x) - (.3,-1)$);
            \draw[vertex,transparent] (oi) circle (.1) node[right] {};
            \coordinate (of) at ($(y) + (0.1,1.5)$);
            \draw[vertex,transparent] (of) circle (.1) node[right] {};
        \end{tikzpicture}
          &
        \begin{tikzpicture}[baseline,scale=0.5]
            \coordinate (x) at (0,0);
            \draw[quark] (x) arc[start angle=-135, delta angle=90, radius=1.7] coordinate[at end] (y) node[black,below,pos=0.7] {};
            \draw[quark] (y) arc[start angle=45, delta angle=90, radius=1.7];
            \coordinate (z) at ($(x)!0.5!(y) + (0,.8)$);
            \draw[quark] (z) arc[start angle=270, delta angle=360, radius=.34] coordinate[pos=0.25] (z1) coordinate[pos=0.75] (z2);
            \draw[photon] (z1) -- (z2);
            \draw[externalVertex] (x) circle (.1) node[right] {};
            \draw[externalVertex] (y) circle (.1) node[right] {};
            \drawPhotonInsertion{z1};
            \drawPhotonInsertion{z2};
            \coordinate (oi) at ($(x) - (.1,-1)$);
            \draw[vertex,transparent] (oi) circle (.1) node[right] {};
            \coordinate (x) at (4,0);
            \draw[quark] (x) arc[start angle=-135, delta angle=90, radius=1.7] coordinate[at end] (y) node[black,below,pos=0.7] {};
            \draw[quark] (y) arc[start angle=45, delta angle=90, radius=1.7];
            \coordinate (z1) at ($(x)!0.8!(y) + (0,0.8)$);
            \coordinate (z2) at ($(x)!0.2!(y) + (0,0.8)$);
            \draw[quark] (z1) arc[start angle=270, delta angle=360, radius=.34] coordinate[pos=0.75] (p1);
            \draw[quark] (z2) arc[start angle=270, delta angle=360, radius=.34] coordinate[pos=0.25] (p2);
            \draw[photon] (p1) -- (p2);
            \draw[externalVertex] (x) circle (.1) node[right] {};
            \draw[externalVertex] (y) circle (.1) node[right] {};
            \drawPhotonInsertion{p1};
            \drawPhotonInsertion{p2};
            \coordinate (of) at ($(y) + (.1,1.5)$);
            \draw[vertex,transparent] (of) circle (.1) node[right] {};
        \end{tikzpicture}
          &
        \phantom{} &
        \phantom{} 
        \\ \hline
    \end{tabular}}
    \caption{Diagrammatic representations of the IB contributions to the valence-quark connected two-point functions. The diagrams in the top row are those included in the electro-quenched approximation, while the ones in the lower row represent the IB corrections from sea quarks. Vacuum terms from the expansion of the denominator in eq.~\eqref{eq:expvalueobs} must been subtracted from the four diagrams at the very bottom.
    The right column collects additional diagrams relevant when the point-split current is inserted at the sink. The meaning of the various vertices is explained in eqs.~\eqref{eq:massInsertionDiagram}, \eqref{eq:chargeDerivativesDiagrams}, and \eqref{eq:VectorCurrentDiagrams}: red triangle, green square and blue diamond for mass, single and double photon insertions from the expansion of $D_{f}$; orange pentagon and yellow star for single and double photon insertion in the sink, when the point-split current operator $\Gamma_{\tilde{\mathrm{V}}}$ is used.}
    \label{tab:IBdiagrams}
\end{table}
}

By inserting the expansions in eqs.~\eqref{eq:DeltaDf} and \eqref{eq:expansionVectorCurrent} into the expectation value \eqref{eq:expvalueobs} and into the trace \eqref{eq:trace2pc}, we obtain all the isospin-breaking corrections. Terms involving only a single photon vertex vanish when evaluated between vacuum states, and are therefore discarded. The remaining contributions yield the Feynman diagrams shown in table~\ref{tab:IBdiagrams}, and the expression for each diagram is detailed in appendix~\ref{sec:appendixa}. The symbols for vertices used in the table reflect the operator insertions defined in eqs.~\eqref{eq:massInsertionDiagram}, \eqref{eq:chargeDerivativesDiagrams}, and \eqref{eq:VectorCurrentDiagrams}.
Each diagram includes only the statistically connected contributions, with vacuum-disconnected terms from the Pfaffian expansion in the denominator of eq.~\eqref{eq:expvalueobs} being subtracted.

\section{Computational details}
\label{sec:numerics}
In this section, we describe the two ensembles used in this work and provide details about the numerical implementation of the RM123 approach.

\subsection{Ensembles}\label{sec:ensembles}

For this work, we perform measurements on two ensembles generated by the collaboration using the
\texttt{openQ*D} code~\cite{openQxD-csic}.
The parameters of the ensembles are summarized in table \ref{tab:action-param}.  
The two ensembles, here labeled \texttt{A400a00} and \texttt{A380a07}, correspond to \texttt{A400a00b324} and \texttt{A380a07b324+RW1} in ref.~\cite{RCstar:2022yjz}. For the latter, a non-perturbative reweighting in the bare mass has been
implemented to improve the consistency with the line of constant physics. The bare hopping parameters of \texttt{A380a07} shown in table \ref{tab:action-param} are the target quark hopping parameters obtained through the reweighting procedure. All quantities computed in this work on \texttt{A380a07} take into account this reweighting factor.

\begin{table}[t]
    \centering
    \resizebox{\linewidth}{!}{
        \renewcommand{\arraystretch}{1.2}
        \begin{tabular}{ccccccc}
            \toprule
            Ensemble         & $T/a\times (L/a)^3$  & $\beta=6/g_0^2$ & $\alpha$ & $\kappa_{\mathrm u}$ & $\kappa_{\mathrm d} = \kappa_{\mathrm s}$ & $\kappa_{\mathrm c}$ \\
            \midrule
            \texttt{A400a00} & 64 $\times$ 32$^3$ & 3.24    & 0        & 0.13440733           & 0.13440733                                & 0.12784              \\
            \texttt{A380a07} & 64 $\times$ 32$^3$ & 3.24    & 0.007299 & 0.13457969           & 0.13443525                                & 0.12806355           \\
            \bottomrule
        \end{tabular}}
        \caption{Parameters of the two ensembles used in this work. The \texttt{A400a00} ensemble has $\kappa_{\mathrm u} = \kappa_{\mathrm d} = \kappa_{\mathrm s}$, while the \texttt{A380a07} ensemble has $\kappa_{\mathrm u} > \kappa_{\mathrm d} = \kappa_{\mathrm s}$. The improvement coefficients used in the action are $c_{\text{sw}}^{\text{SU(3)}}=2.18859$ and $c_{\text{sw}}^{\text{U(1)}}=1$ for both cases. 
        }
    \label{tab:action-param}
\end{table}

Both ensembles have the same lattice volume and value of the strong coupling
constant $\beta=6/g_0^2$, while they differ for the electromagnetic coupling
constant $\alpha$ and the hopping parameters.
\texttt{A380a07} is an ensemble close to the physical value of $\alpha$, and with
an (unphysical) $\mathrm{SU}(2)$ symmetry in the down-strange quark sector.
On the other side, \texttt{A400a00} is an ensemble generated following the line of
constant physics with $\alpha_{\mathrm R} = 0 $ and degenerate masses for the
up and down quarks.
This leads to the $\mathrm{SU}(3)$-symmetric point
$\kappa_\mathrm{u}=\kappa_\mathrm{d}=\kappa_\mathrm{s}$.

We use the $\mathrm{SU}(3)$ improvement coefficient tuned in isosymmetric QCD
for both ensembles.
In this way, we do not remove all $\mathrm{O}(a)$ effects in QCD+QED in this
work.
Nevertheless, the definition of the action is identical in both non-perturbative
and perturbative implementations of QCD+QED.
The $\mathrm U(1)$ improvement coefficient provides tree-level improvement and
is considered in both non-perturbative and perturbative implementations.

\subsection{Computation of the RM123 diagrams}\label{sec: sec4.3}

In the perturbative RM123 approach, we use the non-compact formulation of the
$\mathrm U(1)$ gauge action
\begin{align} \label{eq:non-compact-U(1)act}
    S^{\mathrm{nc}}_{\text{g,U(1)}} = \frac{a^4}{4}\sum_{x} \sum_{\mu, \nu=0}^3 F^2_{\mu \nu}(x),
\end{align}
where the discretization of the field-strength tensor is chosen to be
$F_{\mu\nu}(x)=\partial_\mu A_\nu(x) - (\mu\leftrightarrow\nu)$, in terms of
the forward finite-difference operator $\partial_\mu f(x) =
a^{-1}\{f(x+a\hat\mu)-f(x)\}$.
This formulation requires us to fix the gauge, and on the lattice we adopt an
analogue of the Coulomb gauge fixing condition
\begin{equation}
    \sum_{k=1}^3\partial_k^* A_k(x) = 0,
\end{equation}
where $\partial_\mu^*$ is the backward finite-difference operator
$\partial^*_\mu f(x) = a^{-1}\{f(x)-f(x-a\hat\mu)\}$.
While all physical observables are independent of this choice, intermediate quantities may be gauge dependent.
In the non-compact formulation, the action is quadratic in the gauge potential,
and the photon field can be integrated out by hand.
Nevertheless, a stochastic representation is useful to estimate the integrals
over the vertices, where samples of the photon field distributed according to
the lattice action including the gauge-fixing term are generated by using the
momentum-space representation.
The coordinate-space fields can then be efficiently computed using the fast
Fourier transform.

In the rest of this section, we describe briefly the computation of the
diagrams required for the non-perturbative QCD+QED computation and the diagrams
in the R123 approach listed in table~\ref{tab:IBdiagrams}.
Using translation invariance, it is sufficient to fix one of the coordinates in
eq.~\eqref{eq:trace2pc}, which we choose to be $y$.
To reduce the variance we use an additional three translations of the
coordinate on every gauge field configuration for all diagrams, i.e. we use
$N_\mathrm{s}=4$ point sources for every diagram required for both the
non-perturbative and RM123 approaches.

In the perturbative RM123 approach, the diagrams with additional vertices
integrated over the space-time volume must be included as shown in
table~\ref{tab:IBdiagrams}.
Diagrams with a single fermion trace and just one insertion, i.e. the first two
diagrams in the first row, require just one additional inversion via the
sequential propagator method, and therefore, in addition to the last diagram of
the first row, can be computed exactly without further special treatment.
The remaining single fermion trace diagrams are computed with one sample of the
stochastic photon field at the vertex.
These diagrams constitute the contributions that remain in the so-called
electro-quenched approximation and require only the stochastic estimation of the
photon line to integrate exactly the additional vertices.

On the other hand, the diagrams in the second and third rows, which arise from the expansion of the Pfaffian, involve at least two fermion traces and are referred to as sea-valence (second row) and sea-sea (third row).
Thanks to the $\mathrm{SU(3)}$ symmetry of the isoQCD theory in our setup and
the vanishing sum of the light quarks' charges, $\sum_{f=\mathrm{u,d,s}}q_f=0$,
only the charm quark contributes to the additional traces in the sea-valence
diagrams and the final sea-sea diagram.
The additional fermion traces for all of diagrams involving the sea quarks are
estimated stochastically using pseudofermion fields.
The hopping parameter expansion was used for the charm-quark propagator where
the first few hopping terms have been estimated exactly using probing
vectors~\cite{Thron_1998,Gulpers:2013uca}.
One level of frequency-splitting was also applied for the light-quark
propagators~\cite{Giusti:2019kff}.
For the third diagram of the third row, only one of the two sea-quark fermion propagators was estimated using one level of frequency splitting, and only pseudofermion sources were used for the charm quark propagator.
A fixed number of $N_\mathrm{\eta}=160$ pseudofermion sources were used for all
estimators to reach the gauge noise where the variance is dominated by the
fluctuations of the gauge field.
The approach to the gauge variance for the corresponding contributions to our final observable is illustrated in figure~\ref{fig:varseasea}.

The photon propagators were also estimated differently for the sea-valence and
sea-sea diagrams.
For the sea-valence diagrams, the convolution of the photon propagator and the
additional fermion trace was computed using the fast Fourier transform and this
product was then inserted into the sequential propagator~\cite{Harris:2023zsl}.
In the sea-sea case, the stochastic photon field was used for the third and
fourth diagrams.
For the third diagram, an inversion is required for every photon field and every
pseudofermion field, so one photon field was used per pseudofermion field.
For the final diagram, the estimation of the photon propagator is independent
of the traces, and in this case we also choose $N_\mathrm{A}=160$ samples for the photon
field.
In the last diagram, clearly the photon propagator could have been estimated exactly by a convolution as in the sea-valence case.

In this work, the cost of the stochastic estimators was not optimized.
As will be illustrated later, we show, however, that the stochastic estimation
is sufficient to reach the gauge noise, and therefore, the dominant
fluctuations are driven by the fluctuations of the QCD gauge fields.
The gauge noise itself is expected to be large for the sea-sea effects, in
particular the variance will diverge with $(L/a)^4$.
For a detailed study and discussion, we refer to~\cite{Cotellucci_2025}.
Even disregarding the extra cost of the sea-sea diagrams, which are reused for every observable, the valence-valence and sea-valence diagrams require at least
5 additional sequential propagators for every isoQCD propagator, without
further differentiating the individual terms, greatly increasing the
computational cost of the measurement of the observables.

\section{Analysis and results}
\label{sec:sec6-res}
In this section, we present our analysis and results for the observables
$\phi_i$ defining the line of constant physics and $a^{\text{U,w}}_{\mu}$ in
QCD+QED using either the perturbative expansion around isoQCD or the fully
non-perturbative QCD+QED simulation.
This enables us to compare the two implementations that we perform at a
single lattice spacing and volume.
The physics parameters of the ensembles tuned to isoQCD and QCD+QED are provided in table~\ref{tab:physicsparms}, which of course may differ before matching.

\begin{table}[t]
    \centering
    \begin{tabular}{cccccc}
        \toprule
         Theory & Ensemble      & $a$ (fm) & $m_{\pi^\pm}$ (MeV) & $m_{K^\pm}$ (MeV) & $N_\mathrm{cfg}$ \\
        \midrule                                      
         isoQCD &\texttt{A400a00} & 0.05393(24)   &  398.5(4.7)                 &  398.5(4.7)               & 2000             \\
        QCD+QED & \texttt{A380a07}  &  0.05349(27)    &  398.8(3.7)                 & 403.1(3.8)               & 2000             \\
        \bottomrule
    \end{tabular}
    \caption{Physics parameters of the ensembles used in this work and presented in ref.~\cite{RCstar:2022yjz}. Note that there the two ensembles are referred to as \texttt{A400a00b324} and \texttt{A380a07b324+RW1}, with the latter including a mass-reweighting factor. }
    \label{tab:physicsparms}
\end{table}

As can be seen from table~\ref{tab:targetphi}, where we present the measured
$\phi_i$ as in ref.~\cite{RCstar:2022yjz}, even after a reweighting in the bare mass, there is still a
slight mistuning of the bare parameters compared with the target line of
constant physics for the \texttt{A380a07} ensemble.
Since in this work our goal is to compare the two implementations at fixed
lattice spacing, which will not depend on the precise definition of the line of
constant physics, we modify the renormalization condition, so the target matches
the central value of the measured values on the \texttt{A380a07} ensemble and no
further correction is required.

Explicitly expanding in the bare parameters to leading order around the
simulated parameters of the \texttt{A400a00} ensemble, then we have the conditions
\begin{equation}\label{eq:a380rcs}
    \begin{aligned}
        \phi_0 + e^2\partial_{e^2} \phi_0 + \sum_f \Delta m_f \, \partial_{m_f} \phi_0 - \Delta_L\phi_0 &= 0,\\
        \phi_1 + e^2\partial_{e^2} \phi_1 + \sum_f \Delta m_f \, \partial_{m_f} \phi_1 - \Delta_L\phi_1 &= 2.126, \\
        \phi_2 + e^2\partial_{e^2} \phi_2 + \sum_f \Delta m_f \, \partial_{m_f} \phi_2 - \Delta_L\phi_2 &= 2.13, \\
        \phi_3 + e^2\partial_{e^2} \phi_3 + \sum_f \Delta m_f \, \partial_{m_f} \phi_3 - \Delta_L\phi_3&= 12.122,\\
    \end{aligned}
\end{equation}
where the target values are slightly modified with respect to eq.~\eqref{eq:schemeunphys}. The last term on the left-hand side of each condition accounts for the subtraction of the universal QED finite-volume effects of the charged meson masses, which have already been subtracted for the $\phi_i$ quantities computed on \texttt{A380a07}. The finite-volume effects on the hadron masses in the case of $\cstar$ boundary conditions have been derived in ref.~\cite{Lucini:2015hfa}.
From the equations in \eqref{eq:a380rcs}, the shifts in the bare mass parameters can be determined.
We reiterate that no corrections have been considered to the $\mathrm O(a)$
improvement coefficients in either approach, so that the coefficients of the two ensembles are identical.

\begin{table}[t]
    \centering
    \begin{tabular}{cccc}
        \toprule
        $\phi_i$ & LCP  & \texttt{A400a00}  & \texttt{A380a07} \\
        \midrule
        $\phi_1$ & 2.11    & 2.107(50)             & 2.126(39)               \\
        $\phi_2$ & 2.36    & $-$                   & 2.13(17)                \\
        $\phi_3$ & 12.1    & 12.068(36)            & 12.122(47)              \\
        \hline
    \end{tabular}
    \caption{ $\phi_i$ measured on the \texttt{A400a00} and \texttt{A380a07} ensembles, together with the target values used to define
    the lines of constant physics~\cite{RCstar:2022yjz}. }
    \label{tab:targetphi}
\end{table}

When performing the comparison of the two methods at fixed line of constant
physics, in both the perturbative and non-perturbative approaches, we incorporate the uncertainty derived from fixing to the lines of constant physics.
In the perturbative approach, this is straightforwardly implemented by
propagating the uncertainty on the bare mass shifts obtained by solving the
system of eqs.~\eqref{eq:a380rcs}.
In the non-perturbative approach, however, we propagate the errors from the
determination of the $\phi_i$ and $\hat t_0$ to $a_{\mu}^{\mathrm{U,w}}$ assuming
Gaussian statistics
\begin{equation}\label{eq:errorprop}
    (\mathrm{d}{a^{\mathrm{U,w}}_\mu})^2=\sum_f  
    \left(\partial_{m_f} a_{\mu}^{\mathrm{U,w}} \right)^2(\mathrm{d}m_f)^2, \quad \mathrm{d}m_f(\vec{\phi})= \sum_i  (J^ {-1})_{fi} \, \mathrm{d}\phi_i,
\end{equation}
where $\mathrm{d}\phi_i$ denotes the statistical uncertainty on the measured
values in table \ref{tab:targetphi} and $J^{-1}$ is the inverse Jacobian of the change
of variables from the hadronic quantities to the bare quark masses.
In practice, the derivatives of $a^{\mathrm{U,w}}_\mu$ will be the ones computed
on the \texttt{A400a00} ensemble, and any associated error is higher order in the expansion in the bare parameters, so can be safely ignored.

An alternative way to compare the two strategies for making predictions in
QCD+QED is simply to fix the same bare parameters in the approaches and compare
the predictions for the $\phi_i$ and the scale $\hat t_0$ as well as
$a_\mu^\mathrm{U,w}$.
In this interpretation, the perturbative and non-perturbative approaches are
simply two algorithms for computing at fixed bare parameters, and such a
comparison uses exactly the same data as fixing to the lines of constant
physics, presented in a different manner.
In the former approach, all of the uncertainties are combined, whereas in the
latter, the uncertainties related to the tuning to the lines of constant physics are
presented separately.
Given that we find both presentations useful, we present both in the
following.
In particular, we show the results obtained for $a^{\mathrm{U,w}}_{\mu}$ in isoQCD and non-perturbative QCD+QED in section \ref{sec:5.1-nonpt}. In sections \ref{sec:5.2-t0der}
and \ref{sec:5.3-phider}, we present
the corrections to the scale-setting quantity and to the
hadronic observables defining the line of constant physics, then
the corrections to $a^{\mathrm{U,w}}_{\mu}$ are discussed in section~\ref{sec:5.4-amuder}.
In sections~\ref{sec:5.5-fixbare} and \ref{sec5.6-fixlcp} we compare the final results first
at fixed bare parameters, followed by fixed line of constant physics.

\subsection{Non-perturbative determination of \texorpdfstring{$a_\mu^\mathrm{U,w}$}{amuUw}}
\label{sec:5.1-nonpt}

As the analysis of the correlation functions in the non-perturbative QCD+QED
and isoQCD computations is identical, we present the two together.

\begin{figure}[t]
    \centering
    \includegraphics[width=1.\linewidth]{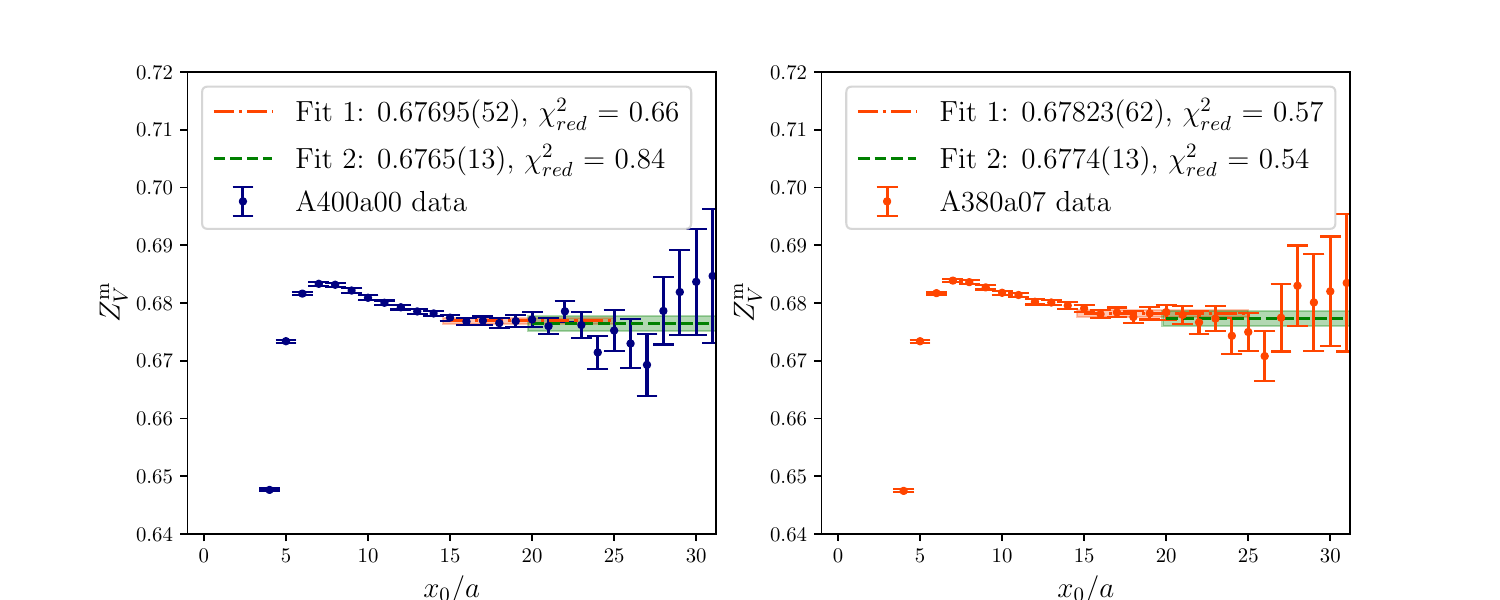}
    \caption{Renormalization constant of the local vector current computed on \texttt{A400a00} (left) and \texttt{A380a07} (right) according to the renormalization condition in eq. \eqref{eq:vectorrc}.}
    \label{fig:ZV33}
\end{figure}

\begin{figure}[t]
    \centering
    \includegraphics[width=0.9\linewidth]{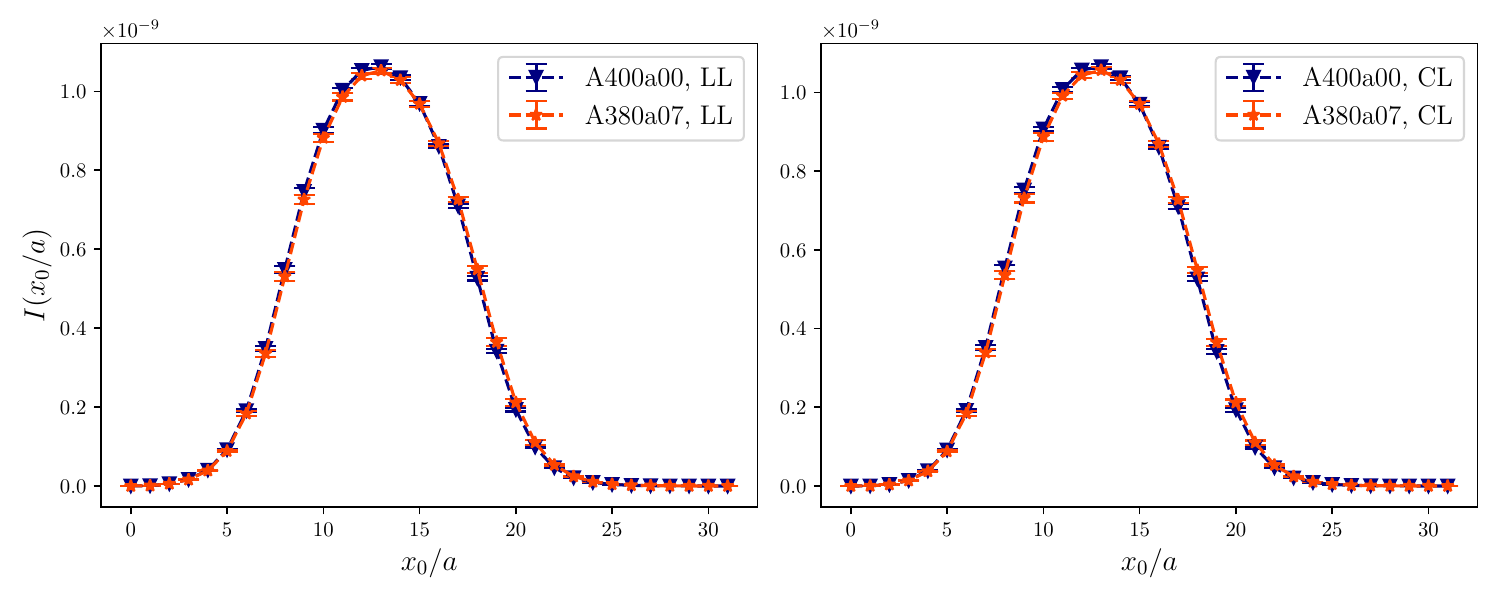}
    \caption{Integrand $I(\hat{t})=(\alpha/\pi)^2 \hat{G}^{\mathrm{U}, \ell}(\hat{t})\hat{\tilde{K}}(\hat{t}) w_{\mathrm{I}}(\hat{t})$ computed on the two ensembles \texttt{A400a00} and \texttt{A380a07} using the local-local (LL) and conserved-local (CL) discretization of the bare correlator.}
    \label{fig: plot b_mu LO}
\end{figure}

As described in section~\ref{sec:defcorr}, we have to determine the renormalization constant of the local $U$-spin current.
The renormalization factor is computed via the renormalization condition \eqref{eq:vectorrc}, which translates in the following relation between the bare correlators $G^{\mathrm{U},\ell}_\mathrm{bare}$
\begin{equation}\label{eq:vectorrcbare}
    Z_\mathrm{V}^{\mathrm{m}} = \lim_{t \to \infty} \frac{G^{\mathrm{U,c}}_{\mathrm{bare}}(t)}{G^{\mathrm{U,l}}_{\mathrm{bare}}(t)}.
\end{equation}
The renormalization constant is obtained by fitting the right-hand side at large $t$. The results of the fit are shown in figure \ref{fig:ZV33} for both ensembles and two different fit ranges.
We obtain the following result for the isoQCD theory 
\begin{align}
    Z^{\mathrm{m},(0)}_\mathrm{V} &=0.6767(10),
\end{align}
while using the same renormalization condition for the QCD+QED theory we have
\begin{align}
    Z^{\mathrm{m}}_\mathrm{V} &=0.6776(12).
\end{align}
The mean values and the errors are computed by combining the fit results for the different fit ranges, based on the associated AIC weights, as in ref.~\cite{Jay_2021}.
We stress that the isoQCD result represents only the leading-order piece of the renormalization factor. 

We compute $a_{\mu}^{\mathrm{U,w}}$ using the local-local and the point-split local discretization of the correlator defined in eq.~\eqref{eq:latamu}.
In figure~\ref{fig: plot b_mu LO}, we show the plots of the (renormalized)
integrands for the two ensembles and discretizations used in this work.
We do not need any extrapolation or model for the data at large distances since employing the intermediate window already removes this noisy region of the correlator. The results obtained by integrating in time are shown in table \ref{tab: amu res. LO}.
From comparing the two plots and the results in the table, we see that there are no visible differences between the local-local and the conserved-local estimators.
We also notice that the signals on the two ensembles agree with each other within the errors, although we have not yet considered the corrections for the
ensemble \texttt{A400a00}.
Thus, we expect that these effects will not significantly change the central value of $a_{\mu}^{\mathrm{U,w}}$, but they could still have an impact on the errors when computing them using the RM123 approach.

In the next sections, we present the corrections in the bare parameters that connect the \texttt{A400a00} to the \texttt{A380a07} ensemble, obtained by linearizing all of the observables and computing the derivatives with the method of insertions \emph{\`a la} RM123, as described in section \ref{sec:sec4-rm123}.
\begin{table}[t]
    \centering
    \begin{tabular}{ccc}
        \toprule
                      & \multicolumn{2}{c}{$ a_{\mu}^{\mathrm{U,w}}\times 10^{11}$}  \\
                      & $\ell=\mathrm l$ & $\mathrm c$ \\
         \midrule
         isoQCD       &       1083(6)     &     1086(5)                 \\
         non-perturbative QCD+QED &      1082(7)        & 1085(7)                       \\
         \midrule
    \end{tabular}
    \caption{Results obtained for $a_{\mu}^{\mathrm{U,w}} \times 10^{11}$, computed either in isoQCD or QCD+QED non-perturbatively. }
    \label{tab: amu res. LO}
\end{table}

\subsection{Corrections to the gradient-flow scale \texorpdfstring{$t_0$}{t0}} \label{sec:5.2-t0der}
In this work, we employ the gradient-flow scale $t_0$ for the scale setting.
Its value is defined implicitly through the condition
\begin{equation}\label{eq:t0def}
    t_0^2 \ev{E(t_0)} = 0.3, 
\end{equation}
where $E(t)$ is the action density at flow time $t$.
In the perturbative approach, the QCD+QED expectation value on the left-hand side of eq.~\eqref{eq:t0def} is expanded around the isoQCD point as explained in section~\ref{sec:sec4.1}, leading to scale corrections.
As $E(t)$ is an observable independent of the bare quark masses and the electromagnetic coupling, the corrections
to $t_0$ arise exclusively from the expansion of the pfaffian.

In table \ref{tab:t0der}, we show the results in lattice units for the scale at leading-order and its derivatives with respect to a bare parameter $\varepsilon_i=m_f,e^2$.
The errors are statistical and computed by using the $\Gamma$-method described in ref.~\cite{Wolff:2003sm} and exploiting the implementation of the \texttt{pyerrors} package~\cite{Joswig:2022qfe}. This applies to all statistical errors computed in our analysis.
\begin{table}[t]
    \centering
        \begin{tabular}{ccccc}
            \toprule
        $\hat{t}_0^{(0)}$ 
        & 
        $\partial_{am_{\mathrm{u}}} \hat{t}_0$ 
        & 
        ${\partial_{am_{\mathrm c}} \hat t_0}$ 
        & 
        $\partial_{e^2}  \hat{t}_0$ \\
                \midrule
                7.400(69)       & -76(24)                     & -26.5(8.1)                  & -6.1(1.9)     \\
            \bottomrule
        \end{tabular}
    \caption{Leading-order value of the dimensionless scale $\hat{t}_0 \equiv
        t_0/a^2$ and its derivatives $\partial_{\varepsilon_i} \hat t_0$
        evaluated at the isoQCD point for the ensemble \texttt{A400a00}. 
        Due to the $\mathrm{SU}(3)$ symmetry of the ensemble, the derivatives with respect to
    the light quark masses $am_{\mathrm u}$, $am_{\mathrm s}$ and $am_{\mathrm s}$ have the same value.}
    \label{tab:t0der}
\end{table}

\subsection{Corrections to the hadronic observables \texorpdfstring{$\phi_i$}{phii}}\label{sec:5.3-phider}
In the perturbative approach, we have to compute the derivatives of the hadronic observables defining the parametrization of QCD+QED. By employing the definitions in eq. \eqref{eq:phi-obs-def}, we obtain the explicit form of the derivatives with respect to the quark masses
\begin{equation}
    \begin{aligned}
        \partial_{m_f} \phi_0 &= 16 t_0  M_{\pi}  \left( \frac{\partial M_{K^{+}}}{\partial m_f} -  \frac{\partial M_{\pi^{+}}}{\partial m_f}  \right),\\
         \partial_{m_f} \phi_1 &= 16 t_0  M_{\pi}  \left( \frac{\partial M_{\pi^{+}}}{\partial m_f} +  \frac{\partial M_{K^{+}}}{\partial m_f} + \frac{\partial M_{K^{0}}}{\partial m_f} + \frac{3M_{\pi}}{2t_0} \frac{\partial t_0}{\partial m_f} \right),\\
          \partial_{m_f} \phi_2 &= \frac{16 t_0  M_{\pi}}{\alpha} \left( \frac{\partial M_{K^{0}}}{\partial m_f} - \frac{\partial M_{K^{+}}}{\partial m_f}   \right),\\
         \partial_{m_f} \phi_3 &= \sqrt{8 t_0}\left( \frac{\partial M_{D^{+}}}{\partial m_f} +  \frac{\partial M_{D^{+}_s}}{\partial m_f} + \frac{\partial M_{D^{0}}}{\partial m_f} + \frac{3 M_D}{2 t_0}  \frac{\partial t_0}{\partial m_f} \right) , \label{eq:phidersm}\\
    \end{aligned}
\end{equation}
and the derivatives with respect to $e^2$ 
\begin{equation}
    \begin{aligned}
        \partial_{e^2} \phi_0 &= 16 t_0  M_{\pi}  \left( \frac{\partial M_{K^{+}}}{\partial e^2} -  \frac{\partial M_{\pi^{+}}}{\partial e^2}  \right),\\
        \partial_{e^2} \phi_1 &=  16 t_0  M_{\pi}  \left( \frac{\partial M_{\pi^{+}}}{\partial e^2} + \frac{\partial M_{K^{+}}}{\partial e^2} + \frac{\partial M_{K^{0}}}{\partial e^2}  + \frac{3}{2} M_{\pi}\frac{\partial t_0}{\partial e^2} \right) ,\\
        \partial_{e^2} \phi_2 &= \frac{16 t_0  M_{\pi}}{\alpha} \left( \frac{\partial M_{K^{0}}}{\partial e^2} - \frac{\partial M_{K^{+}}}{\partial e^2}  \right),\\
        \partial_{e^2} \phi_3 &=  \sqrt{8 t_0} \left( \frac{\partial M_{D^{+}}}{\partial e^2} + \frac{\partial M_{D_s^{+}}}{\partial e^2} +  \frac{\partial M_{D^{0}}}{\partial e^2} + \frac{3 M_D}{2 t_0}  \frac{\partial t_0}{\partial e^2} \right ).\\
    \end{aligned} \label{eq:phiderse2}
\end{equation}
In the above equations, we exploit the $\mathrm{SU}(3)$ symmetry of the isoQCD ensemble to simplify the expressions, and denote $M_{\pi}=M_{\pi^+}=M_{K^{+}}=M_{K^0}$ and $M_{D}=M_{D^+}=M_{D^0}=M_{D_s^+}$ for the leading-order light and charmed meson masses. In addition, the contribution from the scale derivatives cancels out in $\phi_0$ and $\phi_2$ as they depend on the mass difference of mesons that are degenerate at the $\mathrm{SU(3)}$-symmetric point. 

To construct the quantities in eqs. \eqref{eq:phidersm} and \eqref{eq:phiderse2}, we compute the derivatives of the scale and meson masses and combine them. The computation of the scale derivatives was explained in the previous subsection, leading to the results of table \ref{tab:t0der}. Here we focus on the meson mass derivatives. We consider the flavor-charged pseudoscalar correlator projected to zero momentum in the doublet notation
\begin{equation}\label{eq:defpseudocorr}
    C(t) \equiv a^3\sum_{\vec{x}}\ev{O^{fg}(t,\vec{x})O^{fg, \dagger}(0)} = -a^3\sum_{\vec{x}}\ev{\Bar{\chi}^f(t,\vec{x}) \frac{\gamma_5}{2} \chi^g(t,\vec{x}) \Bar{\chi}^g(0) \frac{\gamma_5}{2} \chi^f(0)}.
\end{equation}
At large times, the correlator is dominated by the lowest-energy state in the spectrum. By taking into account the periodic boundary conditions in the temporal
direction of the lattice, it follows that, in the large time limit,
\begin{equation} \label{eq:fullcorr}
C(t) \to A (e^{-M (t-T/2)} +e^{M(t-T/2)}) , \end{equation}
where $A$ and $M$ are the amplitude of the correlator and the mass of the interpolated meson.
Assuming that the isospin-breaking corrections to the meson mass and the amplitude are small,
the correlator in equation \eqref{eq:fullcorr} expands as follows
\begin{equation}\label{eq:correlator-exp}
    C(t) \simeq C^{(0)}(t) + \delta C(t),
\end{equation}
with
\begin{align}\label{eq:LO-corr-exp}
    C^{(0)}(t) &= A^{(0)} \cosh{(M^{(0)} (t-T/2))}
\end{align}
and
\begin{align}\label{eq:corr-delta}
    \delta C(t) &= C^{(0)}(t) \left\{\frac{\delta A}{A^{(0)}}- \delta M \left(t-\frac{T}{2}\right) \tanh\left[M^{(0)}\left(\frac{T}{2} -t\right)\right]\right\}.
\end{align}
The effective derivatives with respect to a bare parameter $\varepsilon_i$ as functions of $t$ are derived from equation \eqref{eq:corr-delta} and take the form
\begin{equation}\label{eq:der-M-t}
\begin{aligned}
    \partial_{\varepsilon_i} M (t)& = \biggr[\frac{\partial_{\varepsilon_i} C(t)}{C^{(0)}(t)} - \frac{\partial_{\varepsilon_i} C(t+1)}{C^{(0)}(t+1)} \biggr] 
    \times \left[({T}/{2} - t) \tanh(M^{(0)}({T}/{2} -t)) \right.\\
    &\qquad\left.- ({T}/{2} - (t+1)) \tanh(M^{(0)}({T}/{2} -(t+1)))\right]^{-1}.
\end{aligned}
\end{equation}
The derivatives of the meson masses are computed by fitting the quantity on the RHS to a constant. The fits take as input the leading-order mass $M^{(0)}$ extracted from
\begin{equation}
    \frac{C^{(0)}(t)}{C^{(0)}(t+1)} = \frac{\cosh{(M^{(0)}(t-T/2))}}{\cosh{(M^{(0)}(t+1-T/2))}}.
\end{equation}
The results of the fits to eq.~\eqref{eq:der-M-t} are shown in the appendix~\ref{sec:appendix-mass}.

\begin{figure}[t] 
\begin{subfigure}{.332\linewidth}
  \includegraphics[width=\linewidth]{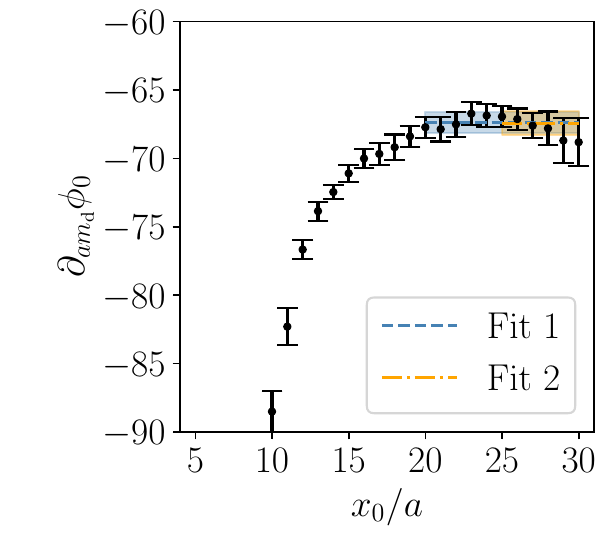}
  \label{fig: phi0 der. s}
\end{subfigure}
\begin{subfigure}{.332\linewidth}
  \includegraphics[width=\linewidth]{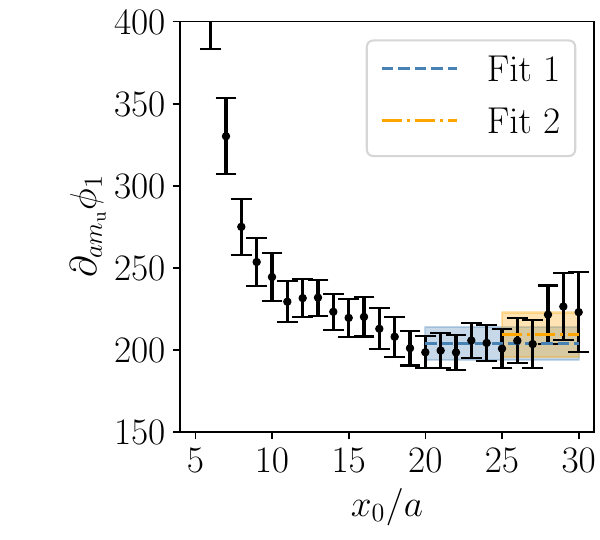}
 \label{fig: phi1 der. u}
\end{subfigure}%
\begin{subfigure}{.332\linewidth}
  \includegraphics[width=\linewidth]{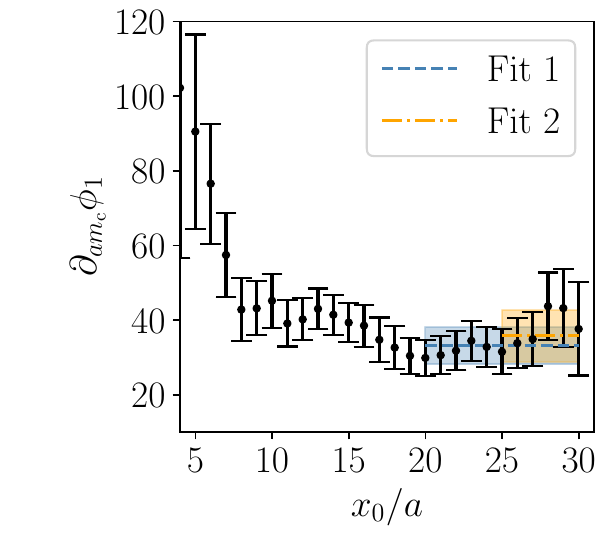}
 \label{fig: phi1 der. c}
\end{subfigure}
\begin{subfigure}{.332\linewidth}
  \includegraphics[width=\linewidth]{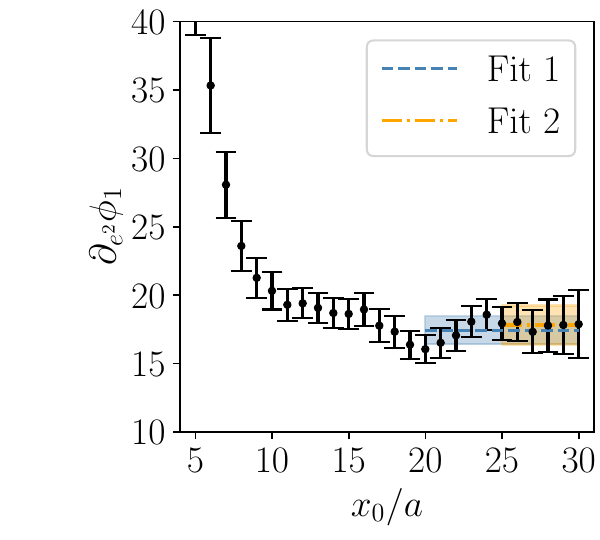}
  \label{fig: phi1 der. e}
\end{subfigure}
\begin{subfigure}{.332\linewidth}
  \includegraphics[width=\linewidth]{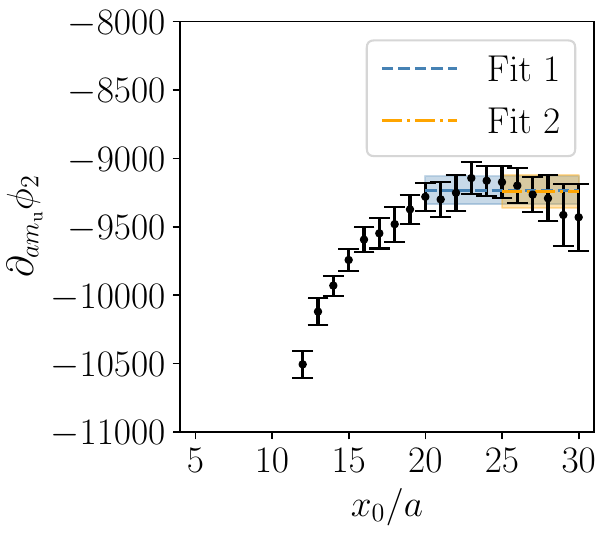}
 \label{fig: phi2 der. u}
\end{subfigure}%
\begin{subfigure}{.332\linewidth}
  \includegraphics[width=\linewidth]{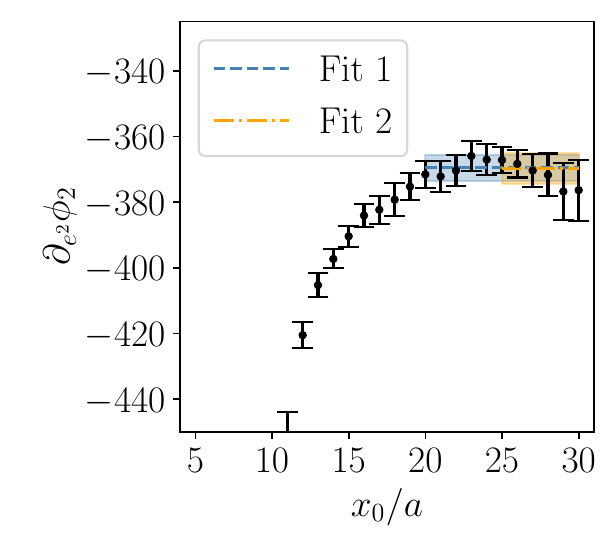}
 \label{fig: phi2 der. 4}
\end{subfigure}
\begin{subfigure}{.332\linewidth}
  \includegraphics[width=\linewidth]{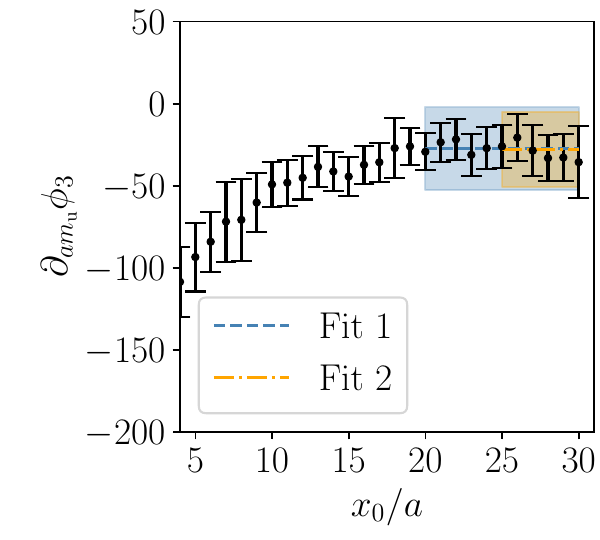}
  \label{fig: phi3 der. u}
\end{subfigure}
\begin{subfigure}{.332\linewidth}
  \includegraphics[width=\linewidth]{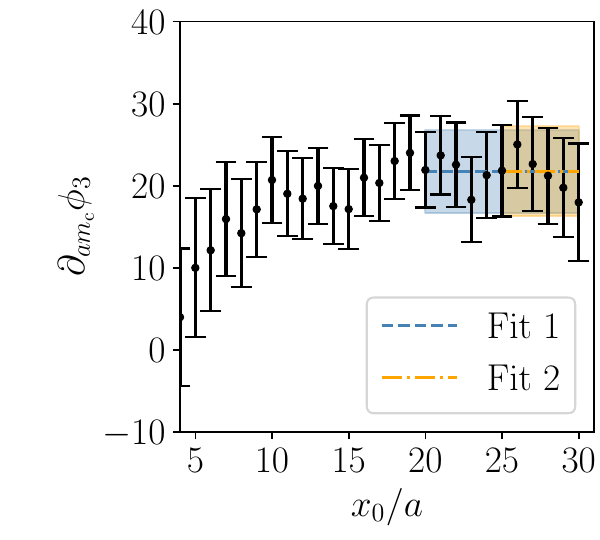}
 \label{fig: phi3 der. c}
\end{subfigure}%
\begin{subfigure}{.332\linewidth}
  \includegraphics[width=\linewidth]{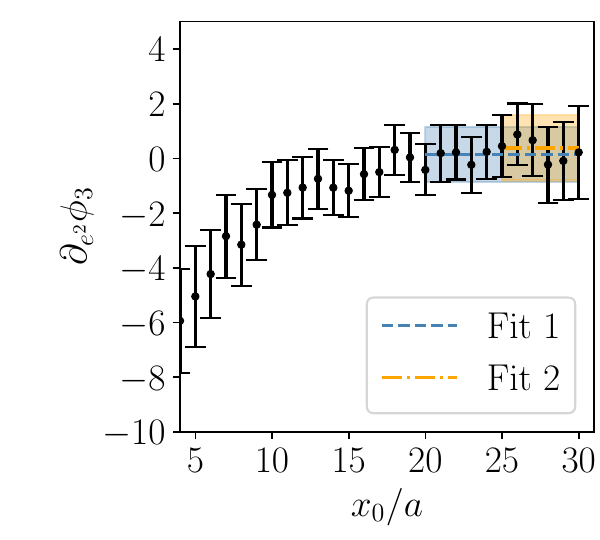}
 \label{fig: phi3 der. e}
\end{subfigure}
\caption{Fits of the $\phi_i$ derivatives with respect to the bare parameters $am_u,am_{d/s},am_c,e^2$, computed on \texttt{A400a00}. Note that all the other derivatives are identically zero.}
\label{fig:phiderfit}
\end{figure}

Here, we adopt a different strategy to compute the derivatives of $\phi_i$. In particular, we notice that we can replace all the meson mass derivatives in eqs.~\eqref{eq:phidersm}-\eqref{eq:phiderse2} with the corresponding time-dependent expressions in eq.~\eqref{eq:der-M-t}, obtaining the time-dependent quantities $\partial_{\varepsilon_i} \phi(t)$, which can be directly fitted to a constant at large $t$.
This strategy is preferred here because it enforces the cancellation of the sea-sea diagrams that contribute to $\phi_0,\phi_2$, due to the linearity of $\partial_{\varepsilon_i} M(t)$ in the correlator derivatives $\partial_{\varepsilon_i}C(t)$ and the $\mathrm{SU}(3)$ symmetry of the ensemble.
Figure~\ref{fig:phiderfit} shows the fits to the derivatives of the hadronic quantities and the results obtained by the fitting procedure are in table \ref{tab:derphis}. We use two fit ranges for each quantity and combine them based on the associated AIC weights~\cite{Jay_2021}.
\begin{table}[t]
    \centering
    \begin{tabular}{cccccc}
        \toprule
                        $\varepsilon$ & $am_\mathrm{u}$& $ am_\mathrm{d}$& $ am_\mathrm{s}$& $ am_\mathrm{c}$&$ e^2$ \\
                         \midrule                                
        $\partial_{\varepsilon} \phi_0$    & 0 & -67.4(9) & 67.4(9)  & 0 &  0     \\
        $\partial_{\varepsilon} \phi_1$    &204(11)& 204(11)& 204(11)& 33(6) & 17(1)    \\ $\partial_{\varepsilon} \phi_2$ & -9234(121) & 9234(121) & 0 & 0 & -370(5)    \\
        $\partial_{\varepsilon} \phi_3$    & -27(25)& -27(25)& -27(25)& 22(7)&0.2(1.5)    \\
        \bottomrule
    \end{tabular}
    \caption{Derivatives of the $\phi_i$ computed on the ensemble \texttt{A400a00}. The errors are obtained by summing in quadrature the statistical and systematic uncertainties. The former is computed using the $\Gamma$-method, while the latter is estimated by considering several fit-ranges for each meson mass's derivative. In all cases, we find that the statistical uncertainty is the dominant one.}
        \label{tab:derphis}
\end{table}

To compute the corrections to the hadronic quantities, together with the derivatives, we need to compute the finite-volume effects appearing in eq.~\eqref{eq:a380rcs}. The explicit form of these contributions on our $\mathrm{SU}(3)$-symmetric ensemble is given by
\begin{equation}
    \label{eq:fvphi}
    \begin{aligned}
    \Delta_L\phi_0 &=0, \\
     \Delta_L\phi_1 &=
    e^2\frac{8t_0 M_{\pi}}{\pi}\left(\frac{\zeta(1)}{2L}  + \frac{\zeta(2)}{\pi M_{\pi}L^2} \right), \\
     \Delta_L\phi_2 &= - e^2\frac{4t_0 M_{\pi}}{\alpha \pi} \left(\frac{\zeta(1)}{2L}  + \frac{\zeta(2)}{\pi M_{\pi}L^2} \right), \\
    \Delta_L\phi_3 &= e^2\frac{\sqrt{8t_0} }{2\pi}\left(\frac{\zeta(1)}{2L}  + \frac{\zeta(2)}{\pi M_{D}L^2} \right), 
    \end{aligned}
\end{equation}
and the results are shown in table~\ref{tab:fvphi}.
\begin{table}[t]
    \centering
    \begin{tabular}{ccc}
        \toprule
         $ \Delta_L\phi_1$& $ \Delta_L\phi_2$ & $ \Delta_L\phi_3$  \\
         \midrule
         -0.00651(4)& 0.446(3)& -0.00323(2) \\
         \bottomrule
    \end{tabular}
    \caption{Results for the universal QED finite-volume effects contribution to the $\phi_i$, computed on \texttt{A400a00}.}
    \label{tab:fvphi}
\end{table}

\subsection{Corrections to \texorpdfstring{$a_\mu^\mathrm{U,w}$}{amuUw}}
\label{sec:5.4-amuder}
The derivation of the isospin-breaking corrections to $a_\mu^\mathrm{U,w}$, namely 
\begin{equation}\label{eq:amucorr}
    \delta a_\mu^\mathrm{U,w} = \sum_i \Delta \varepsilon_i \partial_{\varepsilon_i} a_\mu^\mathrm{U,w},
\end{equation}
requires computing the derivatives of the renormalized $U$-spin correlator defined in eq.~\eqref{eq:latamu} and considering the effect of the scale corrections. 

In particular, we write the correction to the observable in the form
\begin{equation}\label{eq:deltaamudec}
    \delta a^{\mathrm{U,w}}_{\mu} = \delta_G a_\mu^\mathrm{U,w} +\delta_{Z_V} a_\mu^\mathrm{U,w} +\delta_a a_\mu^\mathrm{U,w},
\end{equation}
where the three contributions $\delta_G a_\mu^\mathrm{U,w}, \delta_{Z_V} a_\mu^\mathrm{U,w}$, and $\delta_a a_\mu^\mathrm{U,w}$ denote the corrections arising from the bare correlator, the renormalization constant and the scale.
We stress that this decomposition is unphysical as the individual contributions do not have a well-defined continuum limit. However, the separation highlights the role of the scale correction, which is neglected when the sea-quark effects are not considered. 

To give an explicit expression for these contributions, we introduce the following quantities in lattice units
\begin{equation}
    t = a \hat{t},  \quad 
    G^{\mathrm{U},\ell}(t)= \frac{\hat{G}^{\mathrm{U},\ell}\left(\hat{t}\right)}{a^3}, \quad   \tilde{K}(t; m_{\mu}) = \frac{\hat{\tilde{K}}\left(\hat{t}; am_{\mu}\right)}{a^2},
\end{equation}
and write the point-split local estimator of the observable as
\begin{align}
     a_\mu^\mathrm{U,w} &= \left(\frac{\alpha}{\pi} \right)^2 \sum^{\hat{T}/2}_{\hat{t}=0} \hat{\tilde{K}}\left(\hat{t}; am_{\mu}\right) w_{\mathrm{I}}\left(a\hat{t}\right) Z^{\mathrm{m}}_\mathrm{V}\hat{G}_{\mathrm{bare}}^{\mathrm{U},c}\left(\hat{t}\right) \label{eq:defamuluCL} 
\end{align}
and the local-local as
\begin{align}
     a_\mu^\mathrm{U,w} &= \left(\frac{\alpha}{\pi} \right)^2 \sum^{\hat{T}/2}_{\hat{t}=0} \hat{\tilde{K}}\left(\hat{t}; am_{\mu}\right) w_{\mathrm{I}}\left(a\hat{t}\right) (Z^{\mathrm{m}}_\mathrm{V})^2 \hat{G}_{\mathrm{bare}}^{\mathrm{U,l}}\left(\hat{t}\right),\label{eq:defamuluLL}
\end{align}
where the intermediate window is defined as in eq. \eqref{eq:IWdef}.

\begin{figure}[t]
    \centering
    \includegraphics[width=0.9\linewidth]{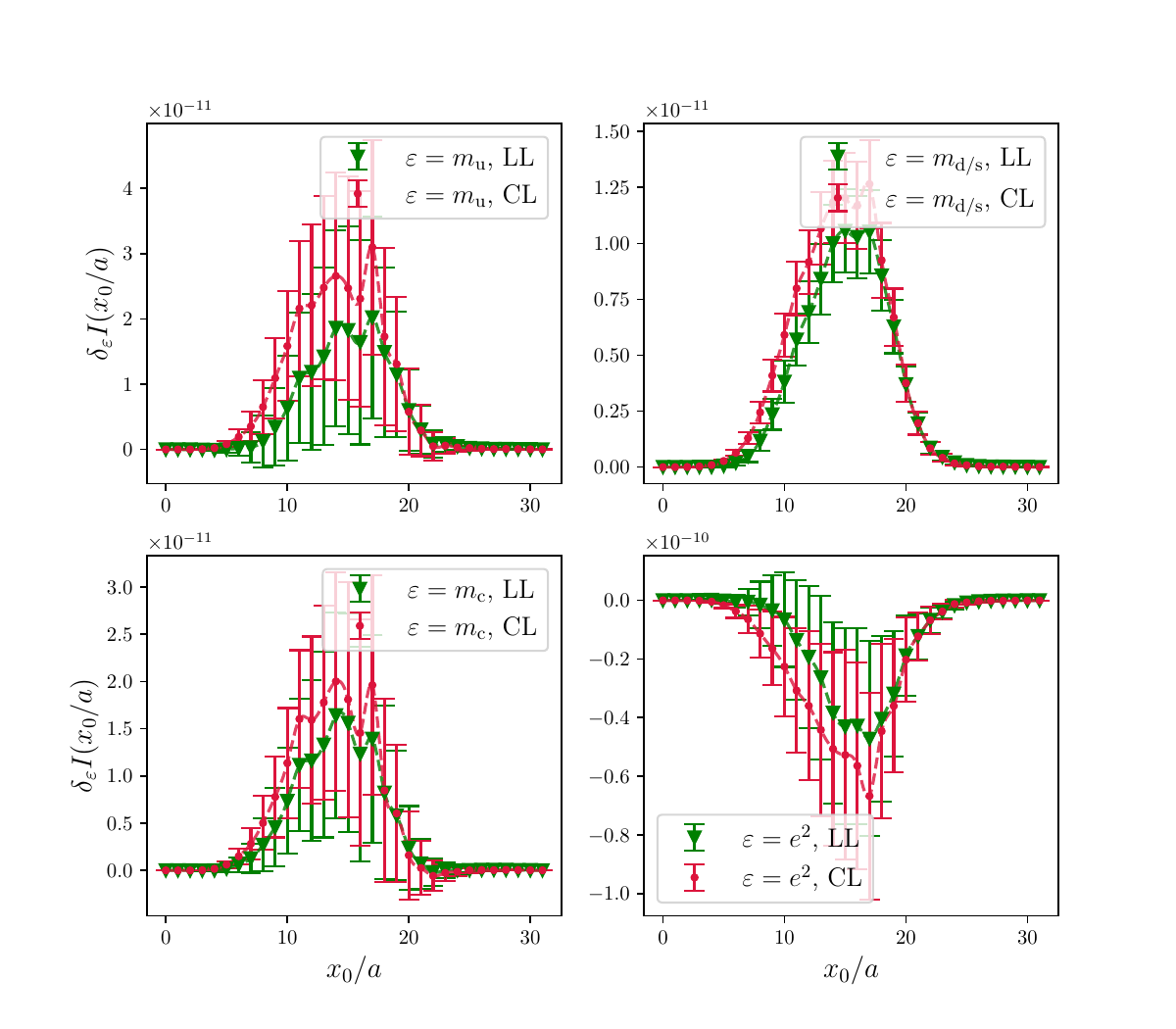}
    \caption{Corrections to the integrand $I(\hat{t})=(\alpha/\pi)^2 \hat{G}^{\mathrm{U}}(\hat{t})\hat{\tilde{K}}(\hat{t}) w_{\mathrm{I}}(\hat{t})$ due to the bare correlator correction $\delta \hat{G}_{\mathrm{bare}}^{\mathrm{U}}(\hat{t})$, computed using the local-local (green triangles) and conserved-local (red dots) discretization. The four plots show the contributions from the different bare parameters. To indicate the relative magnitude of the contributions, we have multiplied the correlator's derivatives by the exact quark mass shifts defined in \eqref{eq:shifts-fixbare}.}
    \label{fig: integrands corrections from G(t)}
\end{figure}

We define the contribution from the bare current correlator appearing in eq.~\eqref{eq:deltaamudec} as follows: for the point-split local estimator we have
\begin{align}
     \delta_G a_\mu^\mathrm{U,w} &\equiv \left(\frac{\alpha}{\pi} \right)^2 \sum^{\hat{T}/2}_{\hat{t}=0} \hat{\tilde{K}}\left(\hat{t}; a^{(0)}m_{\mu}\right)w_{\mathrm{I}}(a^{(0)}\hat{t}) Z^{\mathrm{m}}_\mathrm{V} \delta\hat{G}_{\mathrm{bare}}^{\mathrm{U,c}}\left(\hat{t}\right), \label{eq:GcorrectionsCL} 
\end{align}
and for the local-local one
\begin{align}
    \delta_G a_\mu^\mathrm{U,w} &\equiv \left(\frac{\alpha}{\pi} \right)^2 \sum^{\hat{T}/2}_{\hat{t}=0} \hat{\tilde{K}}\left(\hat{t}; a^{(0)}m_{\mu}\right) w_{\mathrm{I}}(a^{(0)} \hat{t}) (Z^{\mathrm{m}}_\mathrm{V})^2 \delta \hat{G}_{\mathrm{bare}}^{\mathrm{U,l}}\left(\hat{t}\right).
    \label{eq:GcorrectionsLL}
\end{align}
The derivatives of $\hat{G}_{\mathrm{bare}}^{\mathrm{U, \ell}}(\hat{t})$ with respect to the bare parameters are combinations of the diagrams in table \ref{tab:IBdiagrams}.
We obtain $\delta_G a^{\mathrm{U,w}}_{\mu}$ by integrating in time the signals shown in figure \ref{fig: integrands corrections from G(t)}. 
Each subplot represents the contribution from a different bare parameter,
computed with the two discretizations of the correlator.
We observe that, in general, the signal corresponding to the point-split local discretization
is larger in magnitude than the local-local one.
This difference is compensated by the different signs of the $m_f$- and
$e^2$-insertion diagrams and by the effect of the renormalization constant's
correction in \eqref{eq:ZcorrectionsCL}-\eqref{eq:ZcorrectionsLL}.

Secondly, the contribution to the observable from the renormalization constant is given by
\begin{align}
     \delta_{Z_V} a_\mu^\mathrm{U,w} &\equiv \left(\frac{\alpha}{\pi} \right)^2 \sum^{\hat{T}/2}_{\hat{t}=0} \hat{\tilde{K}}\left(\hat{t}; a^{(0)}m_{\mu}\right)w_{\mathrm{I}}(a^{(0)} \hat{t}) \delta Z^\mathrm{m}_V \hat{G}_{\mathrm{bare}}^{\mathrm{U,c}}\left(\hat{t}\right) \label{eq:ZcorrectionsCL}
\end{align}
for the point-split local discretization, and
\begin{align}
     \delta_{Z_V} a_\mu^\mathrm{U,w} &\equiv 2\left(\frac{\alpha}{\pi} \right)^2 \sum^{\hat{T}/2}_{\hat{t}=0} \hat{\tilde{K}}\left(\hat{t}; a^{(0)}m_{\mu}\right) w_{\mathrm{I}}(a^{(0)}\hat{t}) Z^\mathrm{m}_\mathrm{V} \delta Z^\mathrm{m}_V  \hat{G}_{\mathrm{bare}}^{\mathrm{U,l}}\left(\hat{t}\right),\label{eq:ZcorrectionsLL}
\end{align}
for the local-local one. 
We obtain the derivative of the renormalization constant with respect to a bare parameter by differentiating equation~\eqref{eq:vectorrcbare}, which leads to
\begin{equation}
    \partial_{\varepsilon_i}  Z^{\mathrm{m}}_\mathrm{V} = \lim_{t \to \infty} 
    \left[
    \frac{\partial G^{\mathrm{U,c}}_{\mathrm{bare}}}{\partial \varepsilon_i}(t) - G^{\mathrm{U,c}}_{\mathrm{bare}}(t) 
    \left( G^{\mathrm{U,l}}_{\mathrm{bare}}(t) \right)^{-1} 
    \frac{\partial G^{\mathrm{U,l}}_{\mathrm{bare}}}{\partial \varepsilon_i}(t)
    \right] \left( G^{\mathrm{U,l}}_{\mathrm{bare}}(t) \right)^{-1}. \label{eq:Zvder}
\end{equation}
As the renormalization condition involves the correlators, it is computed using
the same graphs that contribute to the corrections of the bare
correlators.
We perform again a constant fit of the right-hand side of equation \eqref{eq:Zvder}, choosing appropriate fit ranges.
In table~\ref{tab: der. ren. const}, we show the derivatives of 
$Z^\mathrm{m}_\mathrm{V}$, obtained through the fit procedure.
\begin{table}[t]
    \centering
    \begin{tabular}{ccccc}
        \toprule
         $\partial_{am_{\mathrm u}}Z^{\mathrm{m}}_{\mathrm V}$ &
         $\partial_{am_{\mathrm d}}Z^{\mathrm{m}}_{\mathrm V}$ &
         $\partial_{am_{\mathrm s}}Z^{\mathrm{m}}_{\mathrm V}$ &
         $\partial_{am_{\mathrm c}}Z^{\mathrm{m}}_{\mathrm V}$ &
         $\partial_{e^2} Z^{\mathrm{m}}_V$ \\
         \midrule
  -1.30(26)& -1.75(27) & -1.75(27) & -0.39(10) & -0.11(2)  \\
        \bottomrule
    \end{tabular}
    \caption{Derivatives of the renormalization constant $Z^{\mathrm{m}}_\mathrm{V}$ with respect to the bare parameters computed on the ensemble \texttt{A400a00}.}
    \label{tab: der. ren. const}
\end{table}

\begin{figure}[t]
    \centering
    \includegraphics[width=0.8\linewidth]{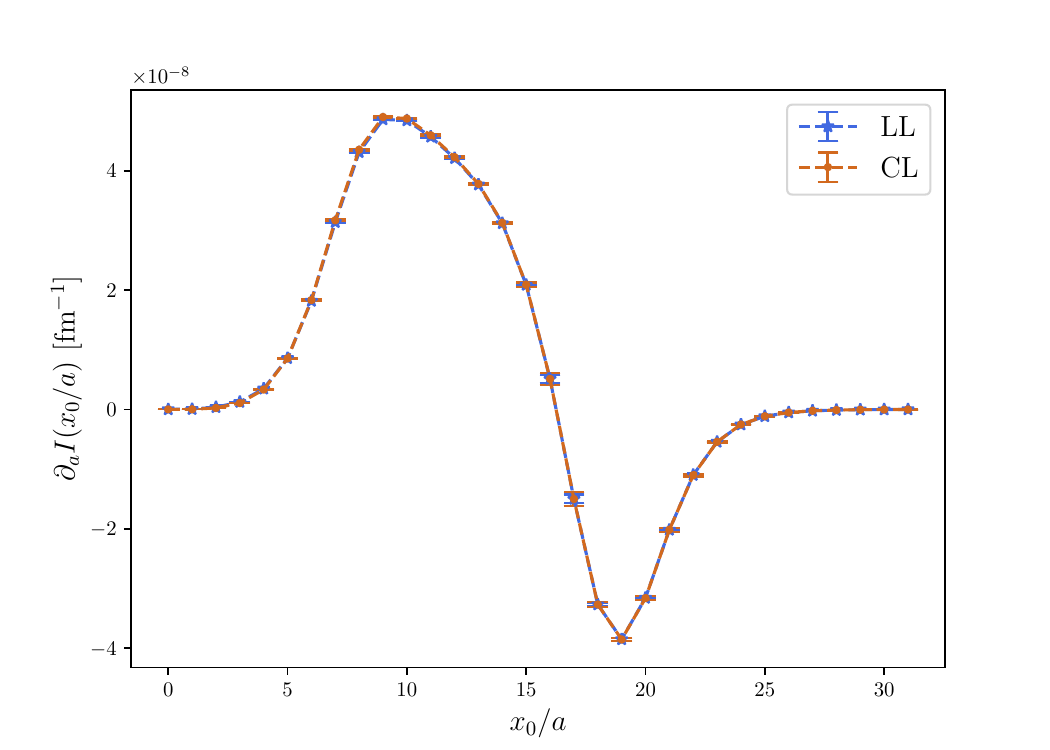}
    \caption{Derivative of the integrand $I(\hat{t})=(\alpha/\pi)^2 \hat{G}^{\mathrm{U}}(\hat{t})\hat{\tilde{K}}(\hat{t}) w_{\mathrm{I}}(\hat{t})$ with respect to the lattice spacing $a$, computed using the local-local (blue triangles) and conserved-local (orange dots) discretization of the correlator.}
    \label{fig:t0der}
\end{figure}

Finally, the last contribution in eq.~\eqref{eq:deltaamudec} arises from the correction to the scale and reads
\begin{align}
    \quad \delta_a a_\mu^\mathrm{U,w} &= \left(\frac{\alpha}{\pi} \right)^2 \sum^{\hat{T}/2}_{\hat{t}=0} \delta \left[\hat{\tilde{K}}\left(\hat{t}; am_{\mu}\right) w_{\mathrm{I}}(a\hat{t}) \right]\hat{G}^{\mathrm{U,\ell}}\left(\hat{t}\right), \label{eq:acorr} 
\end{align}
with
\begin{align}
     \delta \left[\hat{\tilde{K}}\left(\hat{t}; am_{\mu}\right) w_{\mathrm{I}}(a\hat{t}) \right] &= 
     \delta a \times \frac{\partial}{\partial a}
     \left[\hat{\tilde{K}}\left(\hat{t}; am_{\mu}\right) w_{\mathrm{I}}(a \hat{t}) \right].
\end{align}
The correction to the lattice spacing $\delta a$ is defined in eq.~\eqref{eq: corrections a}, while the derivative of $\hat{\tilde{K}}(\hat{t};
am_{\mu}) w_{\mathrm{I}}(a\hat{t})$ is computed as finite difference at each timeslice.
Although our observable is dimensionless, its definition depends on an external
scale via the muon mass $m_\mu$ and therefore on our scale setting.
The signal corresponding to the derivative of the integrand is shown in figure \ref{fig:t0der}. 
The integrand has the same behavior for the two discretizations since they
only differ in discretization effects, which are reduced in the
intermediate window.
Also, the relative errors are the same as for the leading-order correlators. 
The main contribution to the error on $\delta_a a^{\mathrm{U,w}}_{\mu}$ will come
instead from the uncertainty on  $\delta a$.
We stress again that the corrections to the scale arise from the expansion of the pfaffian, and therefore are due only to sea-sea effects given our definition of the lattice scale.

\subsection{Comparison at fixed bare parameters}\label{sec:5.5-fixbare}

Now we turn to the first comparison between the two implementations of QCD+QED, obtained by fixing the bare parameters and comparing the results for the
observables $\phi_i$ defining the renormalization scheme, the hadronic scale
$\hat t_0$ and the observable $a_\mu^\mathrm{U,w}$.
The bare parameter shifts between the \texttt{A400a00} and \texttt{A380a07} ensembles
can be worked out from table~\ref{tab:action-param} and are
explicitly
\begin{equation}\label{eq:shifts-fixbare}
    a\Delta m_{\mathrm u} = -0.00476435, \quad
    a\Delta m_{\mathrm {d,s}} = -0.00077259,\quad
    a\Delta m_{\mathrm c} = -0.00682735.
\end{equation}
In the perturbative approach, we use these values for the shifts and obtain predictions to compare with the corresponding quantities measured in the non-perturbative QCD+QED approach.

We show the results of this first comparison in table~\ref{tab:final-fixbare}. The values in the first row are computed using the perturbative approach in the electro-quenched setup. In this case, the scale $t_0$ does not receive corrections, and the other observables are computed by neglecting the sea-quark effects. The isoQCD+RM123 results in the second row include the effect of the sea quarks and are obtained by adding the corrections, computed as described in the previous subsections, to the isoQCD values of tables \ref{tab:targetphi} and \ref{tab: amu res. LO}.
The non-perturbative results are instead obtained from direct calculation
on the \texttt{A380a07} ensemble.
The errors are given by the quadrature of the statistical uncertainties,
computed using the $\Gamma$-method~\cite{Wolff:2003sm,Joswig:2022qfe}, and the systematic errors, estimated by
varying the fit range in all fit procedures involved in the computation and
then computing their standard deviation.

We observe a good agreement between the two approaches for the scale parameter $\hat t_0$ and $\phi_i$ quantities.
Specifically, the results for $\hat t_0$ computed through the perturbative approach perfectly agree with the non-perturbative value, and the errors are of similar size. In addition, by comparing the precision of the full and electro-quenched results, we see that the impact of the sea-quark effects on $\hat t_0$ and $\phi$ observables is relatively small, except for $\phi_1$, where including the sea-quark effects results in almost three times larger error. 

\begin{figure}[t]
    \centering
    \includegraphics[width=0.8\linewidth]{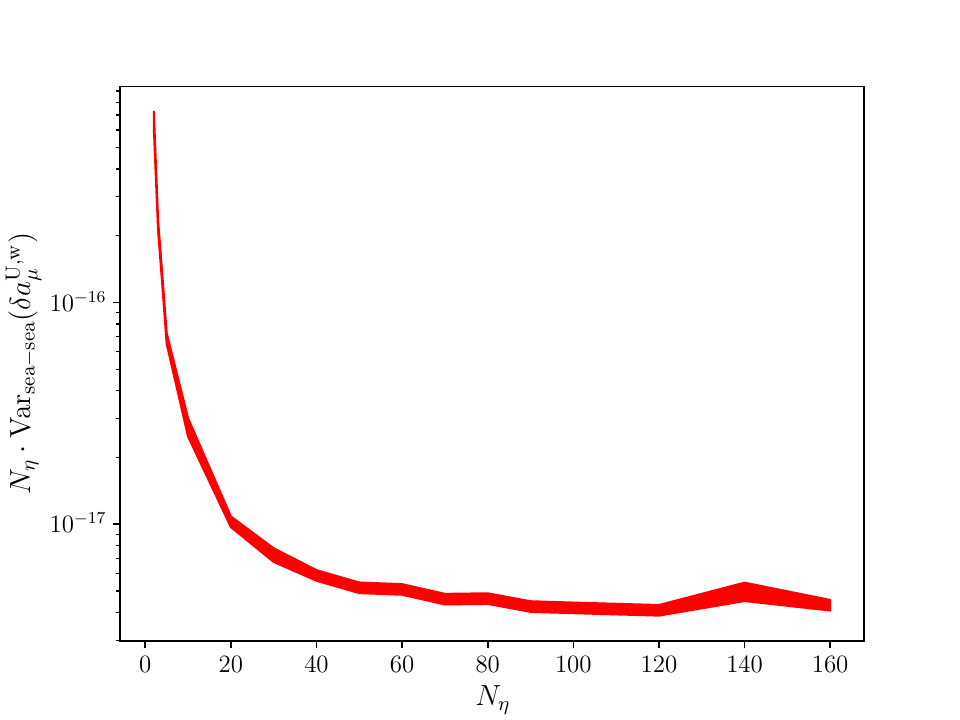}
    \caption{Variance of the sea-sea contribution to $\delta a_{\mu}^{\mathrm{U,w}}$ as a function of the number of pseudofermions sources for 2000 configurations.}
    \label{fig:varseasea}
\end{figure}

The electro-quenched results for $a^{\mathrm{U,w}}_{\mu}$ have the same precision as the observable computed at the isoQCD point.
The non-perturbative results for $a^{\mathrm{U,w}}_{\mu}$ also show sub-percent precision, confirming that the non-perturbative QCD+QED simulations at the physical value of $\alpha$ yield precision comparable to that of the isoQCD computations, as previously observed~\cite{RCstar:2022yjz,Altherr:2022Gr}. 
In contrast, by using the perturbative approach including also the sea-quark effects, we obtain results in agreement with the
non-perturbative computation but with 2.5-2.6 larger relative errors.
In particular, the final uncertainty on the perturbative result amounts to 1.6\% of the central value and is
dominated by the isospin-breaking correction term.
A closer examination shows us that the sea-quark effects are unequivocally the dominant source of errors for the correction, as it is clear from the comparison with the electro-quenched results. We stress that the uncertainty due to sea-quark effects can only be reduced by sampling more gauge-field configurations, as the number of stochastic sources used for estimating the sea-sea diagrams is sufficient to reach the gauge noise in this setup.
This is displayed in figure~\ref{fig:varseasea}, where we plot the variance of $a_{\mu}^{\mathrm{U,w}}$ due to sea-sea contributions and rescaled by the number of pseudofermion sources $N_\eta$ as a function of $N_\eta$ for a fixed number of configurations. We observe that the variance saturates for $N_\eta \gtrsim 100$. In this work, we employ $N_\eta=160$. 

\begin{table}[t]
    \centering
    \scalebox{0.9}{
    \begin{tabular}{ccccccc}
        \toprule
        & & & & & \multicolumn{2}{c}{$a_{\mu}^{\mathrm{U,w}}\times 10^{-11}$} \\
                         & $\hat t_0$      & $\phi_1$  & $\phi_2$ & $\phi_3$   & $\ell=\mathrm{l}$ &  $\mathrm c$ \\
                         \midrule
        isoQCD+RM123$|_{\mathrm{eq}}$ & 7.400(69) &  2.257(34)   &   2.20(14) &  12.100(44)  & 1078(5) & 1080(5)  \\
        isoQCD+RM123     & 7.502(81)  &  2.198(92)     & 2.53(14)  & 12.151(66) &  1090(18) & 1092(18)                        \\
        non-perturbative QCD+QED & 7.523(94)  & 2.128(34) &     2.37(12) &  12.103(47) & 1082(7) & 1085(7)\\
        \bottomrule
    \end{tabular}}
    \caption{Results obtained for the hadronic quantities $\hat t_0,\phi_1,\phi_2,\phi_3$ and $a_{\mu}^\mathrm{U,w}$ computed either non-perturbatively or by exploiting the RM123 method with the local current $\ell=\mathrm l$ or conserved current $\ell=\mathrm c$ at the sink. For the latter, we show results both in the electro-quenched setup and including the contributions from sea quarks. In both cases, we incorporate the corrections at fixed bare parameters, using the shifts in eq.~\eqref{eq:shifts-fixbare}.}
    \label{tab:final-fixbare}
\end{table}

We also want to highlight that the $\phi$ observables in isoQCD and non-perturbative QCD+QED, and their derivatives used for the RM123 approach, have been computed in this work using the same setup as our main observable $a^{\mathrm{U,w}}_{\mu}$, i.e., by employing 4 quark point sources per configuration. 

For completeness, in table \ref{tab:amu-corr-fixbare}, we provide the results for the individual corrections $\delta_G a_{\mu}^{\mathrm{U,w}}$, $\delta_{Z_V}
a_{\mu}^{\mathrm{U,w}}$ and $\delta_a a_{\mu}^{\mathrm{U,w}}$, as introduced in equation~\eqref{eq:deltaamudec}, together with
the total correction.
The two columns correspond to the two discretizations employed for the vector correlator.
We observe that the three contributions are of the same order of magnitude and partially cancel out.
At the current precision, the correction to
$a_{\mu}^{\mathrm{U,w}}$ amounts to ($0.6 \pm 1.7$)\% of the leading-order value and so is consistent with zero. 
\begin{table}[t]
    \centering
    \begin{tabular}{ccc}
        \toprule
        units of $10^{-11}$                                  & $\ell=\mathrm l$  & $\mathrm c$\\
        \midrule
         $\delta_G a_{\mu}^{\mathrm{U,w}}$    &    11(16)         & 12(17)     \\
         $\delta_{Z} a_{\mu}^{\mathrm{U,w}}$ &  3.8(3.4)         & 1.9(1.7)   \\
        $\delta_a a_{\mu}^{\mathrm{U,w}}$      &  -8.4(2.3)        & -8.4(2.4)  \\
         \midrule
         $\delta a_\mu^\mathrm{U,w}$           &     7(18)         &  6(18)    \\
        \bottomrule
    \end{tabular}
    \caption{Results obtained for the  corrections to $a_{\mu}^{\mathrm{U,w}} \times 10^{11}$ computed  by exploiting the RM123 method. We incorporate the isospin-breaking corrections at fixed bare parameters. The three corrections come from three sources: derivatives of the bare correlator, derivatives of the renormalization constant, and derivatives of the scale.}
    \label{tab:amu-corr-fixbare}
\end{table}

\subsection{Comparison at fixed line of constant physics}\label{sec5.6-fixlcp}

In the previous section, we compared the quantities computed in QCD+QED with
fixed bare parameters.
Typically, in a lattice simulation, to approach the continuum limit we prefer
to fix the lines of constant physics defined by the renormalization scheme and
propagate the associated uncertainty to the physical prediction, to avoid a
joint extrapolation to the physical point.
Here, we do such an exercise, which combines the resulting uncertainty onto our
physical prediction.

In the RM123 approach, this is simple to implement, given that we can fix to any scheme a posteriori.
The quark mass shifts are obtained by solving the system in eq.~\eqref{eq:a380rcs} using the target values and the derivatives of the $\phi_i$ in tables \ref{tab:targetphi} and \ref{tab:derphis}.
We derive the following quark mass shifts
\begin{equation}\label{eq: computed quark mass shifts}
    \begin{aligned}
        a\Delta m_{\mathrm u} &= -0.00477(17)_{\mathrm{stat}}(4)_{\mathrm{sys}},\\
        a\Delta m_\mathrm{d,s} &= -0.00082(17)_{\mathrm{stat}}(4)_{\mathrm{sys}},\\
        a\Delta m_\mathrm{c} &= -0.0083(28)_{\mathrm{stat}}(5)_{\mathrm{sys}},
    \end{aligned}
\end{equation}
which, as expected, agree with those shown in
eq.~\eqref{eq:shifts-fixbare} within the quoted uncertainty. 
The results for the individual contributions $\delta_G a_{\mu}^{\mathrm{U,w}}$,
$\delta_{Z_V} a_{\mu}^{\mathrm{U,w}}$ and $\delta_a a_{\mu}^{\mathrm{U,w}}$,
together with the total correction, are shown in
table~\ref{tab:amu-corr-fixlcp}.

For the non-perturbative prediction we consider the uncertainties on the $\phi_i$
observable and propagate them to $a_{\mu}^{\mathrm{U,w}}$ via eq.~\eqref{eq:errorprop}.
To this aim, we reuse the derivative of the $\phi_i$ and the derivatives of
$a^{\mathrm{U,w}}_{\mu}$ computed at the isoQCD point (see tables~\ref{tab:derphis} and~\ref{tab: der. a^{U,w}_mu}). 
This approximation is valid at first order in the bare parameters.
For both discretizations of the sink operator, we obtain the same result
\begin{equation}
    \mathrm{d}a^{\mathrm{U,w}}_{\mu}= 2(3)\times 10^{-11}.
\end{equation}
Since this value is consistent with zero, we take the $1\sigma$ error $3 \times 10^{-11}$ as an estimate of the uncertainty due to the inexact tuning to the line of constant physics. 
This uncertainty is summed in quadrature with the errors on $a_{\mu}^{\mathrm{U,w}}$ in the last row of table~\ref{tab:final-fixbare}.
The effect turns out to be negligible.
\begin{table}[t]
    \centering
    \begin{tabular}{ccc}
        \toprule
         units of $10^{-11}$           &  $\ell=\mathrm l$ & $\mathrm c$  \\
                    \midrule
         $\delta_G a_{\mu}^{\mathrm{U,w}}$ & 15(18) & 17(19)  \\
         $\delta_{Z} a_{\mu}^{\mathrm{U,w}}$ & 6(5) & 3(2) \\
         $\delta_a a_{\mu}^{\mathrm{U,w}}$ & -12(6)   & -12(6)   \\
         \midrule
         $\delta a_\mu^\mathrm{U,w}$ & 9(19) & 8(19)  \\
         \bottomrule
    \end{tabular}
    \caption{Results obtained for the corrections to $a_{\mu}^{\mathrm{U,w}} \times 10^{11}$ computed by exploiting the RM123 method. We incorporate the isospin-breaking corrections at fixed line of constant physics. The three contributions arise from derivatives of the bare correlator, derivatives of the renormalization constant, and derivatives of the scale.}
    \label{tab:amu-corr-fixlcp}
\end{table}

\begin{table}[t]
        \centering
        \begin{tabular}{cccccc}
        \toprule
        $\ell$ &
        $\partial_{am_\mathrm{u}} a^{\mathrm{U,w}}_{\mu}$ &
        $\partial_{am_\mathrm{d}} a^{\mathrm{U,w}}_{\mu}$ &
        $\partial_{am_\mathrm{s}} a^{\mathrm{U,w}}_{\mu}$ &
        $\partial_{am_\mathrm{c}} a^{\mathrm{U,w}}_{\mu}$ &
        $\partial_{e^2} a^{\mathrm{U,w}}_{\mu}$ \\
        \midrule
        $\mathrm l$ &  -12(31) & -113(32) & -113(32) & -9(15)  & -2.5(3.4) \\
        $\mathrm c$ & -11(31) & -112(32) & -112(32) & -9(15)  & -2.5(3.4) \\
        \bottomrule
        \end{tabular}
        \caption{Derivatives of  $a^{\mathrm{U,w}}_{\mu} \times 10^{9}$ computed on the ensemble \texttt{A400a00}.}
        \label{tab: der. a^{U,w}_mu}
    \end{table}

\begin{table}[t]
    \centering
    \begin{tabular}{ccc}
    \toprule
         $a_{\mu}^{\mathrm{U,w}}\times 10^{11}$                         &  $\ell=\mathrm{l}$ & $\mathrm c$ \\
                  \midrule
        isoQCD+RM123$|_{\mathrm{eq}}$ & 1084(5) & 1087(5)  \\
         isoQCD+RM123             &  1093(20) &  1094(21) \\
        non-perturbative QCD+QED & 1082(8) &  1085(7) \\
         \bottomrule
    \end{tabular}
    \caption{Final results obtained for $a_{\mu}^{\mathrm{U,w}} \times 10^{11}$, computed either in the full theory or by exploiting the RM123 method. For the latter, we show results both in the electro-quenched setup and including the contributions from sea quarks. In both cases, we incorporate the corrections at fixed line of constant physics, using the shifts derived by solving the system in eq.~\eqref{eq:a380rcs}}
    \label{tab: final results strategy 2}
\end{table}

In table~\ref{tab: final results strategy 2}, we show the final results for
$a_{\mu}^{\mathrm{U,w}}$, including the uncertainties from the tuning
propagated as described above.
By comparing with the results in table~\ref{tab:final-fixbare}, we
observe that the relative error on the isoQCD+RM123 results goes from 1.6\% to
1.9\%, while the relative error on the QCD+QED results remain stable to 0.6--0.7\%.
Thus, we find a reduction in the total uncertainty of 2.5--3 when using the
non-perturbative simulation compared to the RM123 method including carefully the
tuning to the same line of constant physics.
We stress that this factor considers only the different precision obtained on
the final observable using the same statistics, i.e., number of quark sources
per configuration and number of configurations, while the differing
computational costs and costs associated with tuning are not taken into account.

\section{Conclusions}
\label{sec:sec7-concl}
In this work, we computed the window contribution for a flavour non-singlet
current to the muon magnetic anomaly, $a_{\mu}^{\mathrm{U,w}}$, with
$N_\mathrm{f}=1+2+1$ quarks using two implementations of QCD+QED.
The lattice simulations were based on two ensembles generated by the \rcstar{}
collaboration: one QCD+QED ensemble and one isoQCD ensemble used as the starting
point for the perturbative RM123 approach.
Both employ ${\cstar}$ boundary conditions in the spatial directions that allow the photon field to be included in a local, gauge-invariant formulation, which also preserves lattice translational symmetry.
The two ensembles share the same fermion discretization and lattice volume and have similar bare parameters. Therefore, it is possible to compare the results obtained in the non-perturbative QCD+QED setup to the ones obtained with the
RM123 approach by matching either the bare parameters or the renormalization conditions defining a fixed line of constant physics.

In the perturbative approach, we considered all effects at order $\mathrm
O(\Delta m_{f})$ and $\mathrm O(e^2)$, including the complete contributions from
sea quarks, which represent the most numerically challenging components of the
RM123 method.
Given the renormalization schemes we used for defining isoQCD and QCD+QED, we find the expected 1\% correction to the isoQCD result.
Here we focus on the prediction of the full result in QCD+QED and the comparison
between the two implementations as the definition of the isospin-breaking
corrections is delicate at finite lattice spacing, and our simulated pion
mass of around $400\,\mathrm{MeV}$ is far from the physical point needed for phenomenological predictions.

In the first instance, we match the bare parameters of the QCD+QED simulation
and compare the result for $a_\mu^\mathrm{U,w}$ along with the hadronic
quantities defining the renormalized theory $\phi_i$ and the lattice scale $\hat
t_0$.
On one hand, we find good compatibility between the full QCD+QED results from the RM123 method $a_\mu^\mathrm{U,w}=1092(18)\times10^{-11}$ and the
non-perturbative simulation $a_\mu^\mathrm{U,w}=1085(7)\times10^{-11}$, in this
case for the point-split discretization of the sink current.
On the other hand, we see that the RM123 method has a 2.5 times larger
uncertainty, which can be understood to originate from the sea-quark diagrams, given that the electro-quenched result $a_\mu^\mathrm{U,w}|_\mathrm{eq}=1080(5)\times10^{-11}$, where they are neglected, has a similar
uncertainty to the non-perturbative result.
The hadronic observables $\phi_i$ and the lattice scale $\hat t_0$ that define our line of constant physics all exhibit good consistency between both approaches.
The observable $\phi_1$, related to the squared (hyper-)charged pseudoscalar
meson masses, has a 4\% uncertainty in the RM123 approach in contrast to a 1.6\%
relative precision in the non-perturbative simulation, which is again due to the sea-quark effects by comparing to the electro-quenched result.

In order to determine the significance of the determination of the lines of
constant physics, in the second comparison we impose the same renormalization
conditions described in the text in both setups and propagate the uncertainty
from the hadronic quantities $\phi_i$ and $\hat t_0$ to our prediction for
$a_\mu^\mathrm{U,w}$.
After this exercise, we see a mild increase in the uncertainty in the RM123 method prediction $a_\mu^\mathrm{U,w}=1094(21)\times 10^{-11}$, but no change in the relative uncertainty in the non-perturbative approach, where we obtain
$a_\mu^\mathrm{U,w}=1085(7)\times 10^{-11}$.
Therefore, we find a final result with 1.9\% uncertainty in the RM123 method and
0.6\% precision with non-perturbative QCD+QED. 
To perform this exercise, we use the same derivatives for the non-perturbative
QCD+QED ensemble as those worked out in the isoQCD ensemble for convenience, but in principle they could be estimated in many other ways, and likely they do not need to be precise in any case.

We would like to emphasize that we have so far compared the total uncertainty in both methods with a fixed number of gauge field configurations, but have made no attempt to quantify the true cost of both approaches, which is subtle and likely not universal.
However, the cost of the generation of the gauge field configurations has been investigated in ref.~\cite{RCstar:2022yjz}, which estimated that the generation of the non-perturbative QCD+QED simulations costs a factor 2.5 times more than
isoQCD, given the orbifold construction used here.
This does not account for the fact that the tuning of the non-perturbative simulation with more parameters is an onerous task and, in practice, even here we needed to include a small reweighting in the bare masses as in a realistic situation.
Instead, in the RM123 method, the cost of computing the required extra diagrams is not negligible:
the valence diagrams need to be computed for every observable, and while the sea-sea may be reused, typically the cost to reach the gauge noise is extensive, given the approximations required to perform the volume averages.
While some suggestions have been put forward to reduce the
cost~\cite{Harris:2023zsl}, it is expected that the gauge variance will be large
in large volumes~\cite{Harris:2023zsl,Cotellucci_2025}.
In particular, in our setup with $\mathrm{SU}(3)$ flavour symmetry in the isoQCD
setup, some classes of diagrams do not contribute at all in the RM123 method.
Therefore, we expect the RM123 method to become challenging close to the
physical point and in large volumes, even though the variances of the diagrams presented here should be largely insensitive to the quark mass.

As a final remark, in the non-perturbative QCD+QED case, we applied bare-mass reweighting to ensemble \texttt{A380a07}, reproducing a typical step when tuning an ensemble to specific lines of constant physics. This reweighting has hardly any impact on the statistical uncertainty of the observables we compute. In contrast, the precision lost from applying the RM123 method including all sea effects to the isoQCD ensemble is significantly larger, although this may depend on the scheme used to define the isoQCD point. To ensure the generality of our results, it would be valuable to explore how different definitions of isoQCD affect our conclusions and whether certain trajectories in the parameters space are more favorable for the RM123 method. We leave this exploration for future work.

This work represents one of the first experiences in making predictions in QCD+QED with
$\cstar$ boundary conditions.
The main conclusion that can be drawn is the apparent advantage of the
non-perturbative simulation method over the RM123 approach, as may be expected on theoretical grounds.
The difficulty in the non-perturbative tuning of the simulation parameters with more bare parameters is an issue that still needs to be addressed.
The \rcstar{} collaboration is generating new ensembles with smaller lattice spacings, larger volumes, and smaller quark masses to approach the physical point and have an impact on state-of-the-art and phenomenogically-relevant computations for precision physics of the Standard Model.
In particular, for the muon anomaly, the extension to compute the full
electromagnetic current correlator including singlet contributions is
essential, as well as extending the work to the long-distance window, where QED effects remain very challenging~\cite{Aliberti:2025beg}.

\acknowledgments

{We are grateful to the members of the RC* collaboration and the Muon $g-2$ Theory Initiative for insightful discussions. We acknowledge access to Piz Daint and Alps at the Swiss National Supercomputing Centre (CSCS) (projects eth8, c21, and s1196) and to Lise and Emmy at NHR@ZIB and NHR@Göttingen as part of the NHR infrastructure (projects bep00085, bep00102 and bep00116). The support by SNSF via Project No.~200021\_200866 is gratefully acknowledged. This work was supported by the Platform for Advanced Scientific Computing (PASC) project ``Efficient QCD+QED Simulations with openQ*D software''. This research was supported in part by grant NSF PHY-2309135 to the Kavli Institute for Theoretical Physics (KITP). JK, MKM and ICP acknowledge the support received from the Horizon Europe project interTwin, funded by the European Union Grant Agreement Number \href{https://cordis.europa.eu/project/id/101058386}{101058386}. F.M.~and N.T. are supported by the Italian Ministry of University and Research (MUR) and the European Union (EU) – Next Generation EU, Mission 4, Component 1, PRIN 2022, CUP F53D23001480006 and CUP D53D23002830006. A.C.’s research is funded by the Deutsche Forschungsgemeinschaft (DFG, German Research Foundation) - Projektnummer 417533893/GRK2575 “Rethinking Quantum Field Theory”.} 

\appendix
\section{Isospin-breaking corrections}
\label{sec:appendixa}
{
\tikzstyle{quark}=[postaction={decorate,decoration={markings,mark=at position 0.5 with {\arrow{>}}}}]
\tikzstyle{photon}=[decorate, decoration={snake, amplitude=0.2mm, segment length=.7mm}]
\tikzstyle{vertex}=[draw=black,fill=white]
\tikzstyle{externalVertex}=[draw=black,fill=black]
\newcommand{\drawTriangle}[2]{%
    \pgfmathsetmacro{\h}{0.3}
    \draw[fill=red, draw=black]
    ($(#1)+(-0.5*\h,-0.289*\h)$) --
    ($(#1)+(0.5*\h,-0.289*\h)$) --
    ($(#1)+(0,0.577*\h)$) -- cycle;
}
\newcommand{\drawPhotonInsertion}[1]{%
    \draw[draw=black, fill=green] ($(#1)-(.1,.1)$) rectangle ($(#1)+(.1,.1)$);
}
\newcommand{\drawDoublePhotonInsertion}[1]{%
    \draw[draw=black, fill=blue!50!cyan, rotate=45] ($(#1)-(.1,.1)$) rectangle ($(#1)+(.1,.1)$);
}
\newcommand{\drawVertexPhotonInsertion}[1]{%
    \node[regular polygon, regular polygon sides=5, draw=black, fill=orange,
        minimum size=4pt, inner sep=0pt, outer sep=0pt] at (#1) {};%
}
\newcommand{\drawDoubleVertexPhotonInsertion}[1]{%
    \node[star, star points=5, draw=black, fill=yellow,
        minimum size=4pt, inner sep=0pt, outer sep=0pt] at (#1) {};%
}

In this appendix, we provide the expressions for the isospin-breaking diagrams represented in table \ref{tab:IBdiagrams}.
We recall that our goal is to compute the QCD+QED expectation value
\begin{equation}\label{eq:fullexptrace}
    \ev{\mathcal{O}(x,y)} = \ev{\Tr[D^{-1}_f (y|x) \Gamma_{A} D^{-1}_g (x|y)\Gamma_{B}]}
\end{equation}
using the perturbative approach, at leading order in $e^2$ and $\Delta m_f$.
To simplify the notation, and given that $e^2 \sim \Delta m_f$, we adopt the convention that $O((\Delta m_f)^k) = O(e^{2k})$; thus, terms of $O(e^3)$ also account for higher-order mass corrections. 
We recall that the specific interpolators $\Gamma_{A,B}$ used in this work are the pseudoscalar ($\Gamma_{\mathrm{P}}$), and the local and point-split vector currents ($\Gamma_{\mathrm{V}}$ and $\Gamma_\mathrm{\tilde V}$) defined in eqs.~\eqref{eq:GammasPV} and~\eqref{eq:pscurrent}.

By using the expansions in eqs.~\eqref{eq:inv-D-exp} and \eqref{eq:expansionVectorCurrent}, we can write the observable as
\begin{align} \label{eq-app-a-exp-obs}
    \mathcal{O}=\mathcal{O}_0 + \mathcal{O}_1 +\mathcal{O}_2 + O(e^3). 
\end{align}
The operator $\mathcal{O}_0$ represents the isoQCD observable
\begin{align}
\mathcal{O}_{0}(x,y) =
\Tr[(D^{(0)}_f)^{-1} (y|x) \Gamma^{(0)}_{A} (D^{(0)}_g)^{-1} (x|y)\Gamma^{(0)}_{B}],
\end{align}
while $\mathcal{O}_k$ denotes the correction at order $e^k$. $\Gamma^{(0)}_{A,B}$ coincide with $\Gamma_{A,B}$ for pseudoscalar and local vector interpolators.
We also define the reweighting factor
\begin{equation}
\label{eq-app-a-r}
\mathcal{R} = \frac{ \prod_f \Pf(CK D_{f}) }{ \prod_f \Pf (CK D^{(0)}_{f}) },
\end{equation}
where the nominator is expanded according to eq.~\eqref{eq:exp-pfaffian}, leading to 
\begin{equation}
    \mathcal{R}= 1 + \mathcal{R}_1+ \mathcal{R}_2+O(e^3),
\end{equation}
where $\mathcal{R}_k$ is the correction at order $e^k$.

Given eqs.~\eqref{eq-app-a-exp-obs}-\eqref{eq-app-a-r}, we write the perturbative expansion for~\eqref{eq:fullexptrace} as
\begin{align}
    \ev{\mathcal{O}} = \ev{\mathcal{O}_0}_0 + \ev{\mathcal{O}_2}_{0+\gamma} +\ev{\mathcal{O}_1 \mathcal{R}_1}_{0+\gamma} +\ev{\mathcal{O}_0 \mathcal{R}_2}_{0+\gamma,c} +O(e^4),
\end{align}
where $\ev{}_0, \ev{}_{\gamma}$ denote the expectation value in isoQCD and over the free photon field distribution. The subscript $c$ refers to the connected expectation value, where the vacuum-disconnected term has been subtracted.

In writing the equation above, we have noticed that $\mathcal{O}_0$ does not depend on the photon field, while $\mathcal{R}_1$ and $\mathcal{O}_1$ are linear in the photon field, and therefore, the expectation values $\ev{ \mathcal{O}_0 \mathcal{R}_1 }_{0+\gamma}$ and $\ev{ \mathcal{O}_1 }_{0+\gamma}$ vanish. The correction term $\ev{ \mathcal{O}_2 }_{0+\gamma}$ arises purely from the expansion of the observable and is dubbed \textit{valence-valence} contribution. We refer to the correction term $\ev{ \mathcal{O}_1 \mathcal{R}_1 }_{0+\gamma}$ as \textit{sea-valence} contribution, since it involves a photon propagator connecting a quark line in the observable to the quark line arising from the pfaffian. Finally, we refer to the correction term $\ev{ \mathcal{O}_0 \mathcal{R}_2 }_{0+\gamma}$ as \textit{sea-sea} contribution. The latter involves either a mass term or a photon propagator connecting quark lines from the fermionic pfaffian.

\subsection{Valence-valence diagrams}
The first set of diagrams comes from the expansion of the inverse Dirac operators (quark propagators) 
 or the $\Gamma$ in the trace~\eqref{eq:fullexptrace}. We first consider the cases 
 where $\Gamma_A,\Gamma_B =\Gamma_{\mathrm{P}},\Gamma_{\mathrm{P}}$ or $\Gamma_A,\Gamma_B =\Gamma_{\mathrm{V}},\Gamma_{\mathrm{V}}$, 
 i.e., none of the $\Gamma$ depends on $e$. Diagrammatically, we have the following expression: 
\begin{align}
\langle \mathcal{O}_2(x,y) \rangle_\gamma
= \ 
&- \Delta m_f \ 
\begin{tikzpicture}[baseline=-.1cm,scale=0.5]\footnotesize
            \coordinate (x) at (0,0);
            \draw[quark] (x) arc[start angle=-135, delta angle=90, radius=1.7] coordinate[at end] (y) node[black,below,pos=0.7]{$g$} ;
            \draw (y) arc[start angle=45, delta angle=90, radius=1.7]  coordinate[pos=0.5] (p) node[black,above,pos=0.7]{$f$};
            \draw[externalVertex] (x) circle (.1) node[left]{$x$};
            \draw[externalVertex] (y) circle (.1) node[right]{$y$};
            \drawTriangle{p};
        \end{tikzpicture}
\ - e^2 q_f^2\ 
\begin{tikzpicture}[baseline=-.1cm,scale=0.5]\footnotesize
            \coordinate (x) at (0,0);
            \draw[quark] (x) arc[start angle=-135, delta angle=90, radius=1.7] coordinate[at end] (y) node[black,below,pos=0.7] {$g$};
            \draw[quark] (y) arc[start angle=45, delta angle=90, radius=1.7]  coordinate[pos=0.3] (p) node[black,above,pos=0.7] {$f$};
            \draw[photon] (p) arc[start angle=255, delta angle=360, radius=.34];
            \draw[externalVertex] (x) circle (.1) node[left] {$x$};
            \draw[externalVertex] (y) circle (.1) node[right] {$y$};
            \drawDoublePhotonInsertion{p};
        \end{tikzpicture}
\ + e^2 q_f^2 \ 
\begin{tikzpicture}[baseline=-.1cm,scale=0.5]\footnotesize
\coordinate (x) at (0,0);
            \draw[quark] (x) arc[start angle=-135, delta angle=90, radius=1.7] coordinate[at end] (y) coordinate[pos=0.3] node[black,below,pos=0.7] {$g$} ;
            \draw[quark] (y) arc[start angle=45, delta angle=90, radius=1.7] coordinate[pos=0.3] (p1) node[black,below,pos=0.9] {} coordinate[pos=0.7] (p2) node[black,above,pos=0.5] {$f$};
            \draw[photon] (p1) .. controls ($(p1)!0.2!(p2) - (0,.5)$) and ($(p1)!0.8!(p2) - (0,.5)$) .. (p2);
            \draw[externalVertex] (x) circle (.1) node[left] {$x$}; 
            \draw[externalVertex] (y) circle (.1) node[right] {$y$};
            \drawPhotonInsertion{p1};
            \drawPhotonInsertion{p2};
\end{tikzpicture} 
\\ &  \nonumber
\ +
\  \{x \leftrightarrow y, \,f \leftrightarrow g\}
\ + e^2 q_f q_g\ 
\begin{tikzpicture}[baseline=-.1cm,scale=0.5]\footnotesize
            \coordinate (x) at (0,0);
            \draw[quark] (x) arc[start angle=-135, delta angle=90, radius=1.7] coordinate[at end] (y) coordinate[pos=0.3] (p1) node[black,below,pos=0.7] {$g$};
            \draw[quark] (y) arc[start angle=45, delta angle=90, radius=1.7] coordinate[pos=0.3] (p2) node[black,above,pos=0.7] {$f$};
            \draw[photon] (p1) -- (p2);
            \draw[externalVertex] (x) circle (.1) node[left] {$x$};
            \draw[externalVertex] (y) circle (.1) node[right] {$y$};
            \drawPhotonInsertion{p1};
            \drawPhotonInsertion{p2};
        \end{tikzpicture},
\end{align}
which applies for any $f\neq g$ for the pseudoscalar correlator or any $f=g$ for the local-local vector correlator.
We give explicit expressions for the diagrams:
\begin{align}
\begin{tikzpicture}[baseline=-.1cm,scale=0.5]\footnotesize
            \coordinate (x) at (0,0);
            \draw[quark] (x) arc[start angle=-135, delta angle=90, radius=1.7] coordinate[at end] (y) node[black,below,pos=0.7]{$g$} ;
            \draw (y) arc[start angle=45, delta angle=90, radius=1.7]  coordinate[pos=0.5] (p) node[black,above,pos=0.7]{$f$};
            \draw[externalVertex] (x) circle (.1) node[left]{$x$};
            \draw[externalVertex] (y) circle (.1) node[right]{$y$};
            \drawTriangle{p};
        \end{tikzpicture}
\ = & \,
\Tr \left\{ \left[ (D^{(0)}_{f})^{-1} (D^{(0)}_{f})^{-1} \right](y|x) \Gamma_A (D_{g}^{(0)})^{-1}(x|y) \Gamma_B \right\}\,,
\\ 
\begin{tikzpicture}[baseline=-.1cm,scale=0.5]\footnotesize
            \coordinate (x) at (0,0);
            \draw[quark] (x) arc[start angle=-135, delta angle=90, radius=1.7] coordinate[at end] (y) node[black,below,pos=0.7] {$g$};
            \draw[quark] (y) arc[start angle=45, delta angle=90, radius=1.7]  coordinate[pos=0.3] (p) node[black,above,pos=0.7] {$f$};
            \draw[photon] (p) arc[start angle=255, delta angle=360, radius=.34];
            \draw[externalVertex] (x) circle (.1) node[left] {$x$};
            \draw[externalVertex] (y) circle (.1) node[right] {$y$};
            \drawDoublePhotonInsertion{p};
        \end{tikzpicture}
\ = & \, 
\sum_{w\mu} \Lambda_{\mu\mu}(0) \Tr \bigg\lbrace \left[ (D^{(0)}_{f})^{-1} \tfrac{\delta D_f^{(2)}}{\delta A^2_{\mu}}(w) (D^{(0)}_{f})^{-1} \right](y|x) 
\nonumber \\ & \hspace{54pt} \times
\Gamma_A (D_{g}^{(0)})^{-1}(x|y) \Gamma_B \bigg\rbrace\,,
\\ 
\begin{tikzpicture}[baseline=-.1cm,scale=0.5]\footnotesize
\coordinate (x) at (0,0);
            \draw[quark] (x) arc[start angle=-135, delta angle=90, radius=1.7] coordinate[at end] (y) coordinate[pos=0.3] node[black,below,pos=0.7] {$g$} ;
            \draw[quark] (y) arc[start angle=45, delta angle=90, radius=1.7] coordinate[pos=0.3] (p1) node[black,below,pos=0.9] {} coordinate[pos=0.7] (p2) node[black,above,pos=0.5] {$f$};
            \draw[photon] (p1) .. controls ($(p1)!0.2!(p2) - (0,.5)$) and ($(p1)!0.8!(p2) - (0,.5)$) .. (p2);
            \draw[externalVertex] (x) circle (.1) node[left] {$x$}; 
            \draw[externalVertex] (y) circle (.1) node[right] {$y$};
            \drawPhotonInsertion{p1};
            \drawPhotonInsertion{p2};
\end{tikzpicture}
\ = & \, 
\sum_{wz\mu\nu} \Lambda_{\mu\nu}(w-z) \Tr \bigg\lbrace \left[ (D^{(0)}_{f})^{-1} \tfrac{\delta D_f^{(1)}}{\delta A_{\mu}}(w) (D^{(0)}_{f})^{-1} \tfrac{\delta D_f^{(1)}}{\delta A_{\nu}}(z) (D^{(0)}_{f})^{-1} \right](y|x) 
\nonumber \\ &  \hspace{79pt} \times
\Gamma_A (D_{g}^{(0)})^{-1}(x|y) \Gamma_B \bigg\rbrace\,,
\\ 
\begin{tikzpicture}[baseline=-.1cm,scale=0.5]\footnotesize
            \coordinate (x) at (0,0);
            \draw[quark] (x) arc[start angle=-135, delta angle=90, radius=1.7] coordinate[at end] (y) coordinate[pos=0.3] (p1) node[black,below,pos=0.7] {$g$};
            \draw[quark] (y) arc[start angle=45, delta angle=90, radius=1.7] coordinate[pos=0.3] (p2) node[black,above,pos=0.7] {$f$};
            \draw[photon] (p1) -- (p2);
            \draw[externalVertex] (x) circle (.1) node[left] {$x$};
            \draw[externalVertex] (y) circle (.1) node[right] {$y$};
            \drawPhotonInsertion{p1};
            \drawPhotonInsertion{p2};
        \end{tikzpicture}
\ = & \,
\sum_{wz\mu\nu} \Lambda_{\mu\nu}(w-z) \Tr \bigg\lbrace \left[ (D^{(0)}_{f})^{-1} \tfrac{\delta D_f^{(1)}}{\delta A_{\mu}}(w) (D^{(0)}_{f})^{-1} \right](y|x) 
\nonumber \\ &  \hspace{79pt} \times
\Gamma_A \left[ (D_{g}^{(0)})^{-1} \tfrac{\delta D_f^{(1)}}{\delta A_{\nu}}(z) (D_{g}^{(0)})^{-1} \right](x|y) \Gamma_B \bigg\rbrace\,,
\end{align}
where $\Lambda_{\mu \nu}(x-y) = \ev{A_{\mu}(x) A_{\nu}(y)}_{\gamma}$
The remaining diagrams are obtained from the first three by exchanging $x \leftrightarrow y$ and $f \leftrightarrow g$.

For $f=g$ and $A,B ={\tilde{\mathrm{V}}},{\mathrm V}$, there are three additional diagrams contributing to $\ev{\mathcal{O}_2}_\gamma$ with coefficient $-e^2q_f^2$, which are the following two
\begin{align}
    \begin{tikzpicture}[baseline=-.1cm,scale=0.5]\footnotesize
            \coordinate (x) at (0,0);
            \draw[quark] (x) node[left] {$x$} arc[start angle=-135, delta angle=90, radius=1.7] coordinate[at end] (y) node[black,below,pos=0.7] {$g$};
            \draw[quark] (y) node[right] {$y$} arc[start angle=45, delta angle=90, radius=1.7] node[black,above,pos=0.7] {$f$};
            \draw[externalVertex] (y) circle (.1) node[right] {};
            \draw[photon] (x) arc[start angle=0, delta angle=360, radius=.43];
            \drawDoubleVertexPhotonInsertion{x};
    \end{tikzpicture}
\ = &  \, 
 \Lambda_{\mu\mu}(0) \Tr \left\{ (D^{(0)}_{f})^{-1}(y|x) \Gamma_{\tilde{\mathrm{ V}},\mu}^{(2)} (D_{g}^{(0)})^{-1}(x|y) \Gamma_{\mathrm{V}, \nu} \right\}
\\ 
\begin{tikzpicture}[baseline=-.1cm,scale=0.5]\footnotesize
            \coordinate (x) at (0,0);
            \draw[quark] (x) node[left] {$x$} arc[start angle=-135, delta angle=90, radius=1.7] coordinate[at end] (y) node[black,below,pos=0.7] {$g$};
            \draw[quark] (y) node[right] {$y$} arc[start angle=45, delta angle=90, radius=1.7]  coordinate[pos=0.3] (p) node[black,above,pos=0.7] {$f$};
            \draw[photon] (p) .. controls ($(p)!0.2!(x) - (0,.5)$) and ($(p)!0.8!(x)$) .. (x);
            \draw[externalVertex] (y) circle (.1) node[right] {};
            \drawPhotonInsertion{p};
            \drawVertexPhotonInsertion{x};
        \end{tikzpicture}
\ = & \, 
\sum_{z \rho} \Lambda_{\mu\rho}(x-z) \Tr \bigg\lbrace \left[ (D^{(0)}_{f})^{-1} \tfrac{\delta D_f^{(1)}}{\delta A_{\mu}}(z) (D^{(0)}_{f})^{-1} \right](y|x) 
\nonumber \\ &  \hspace{72pt} \times
\Gamma^{(1)}_{\tilde{\mathrm{V}},\mu}  (D_{g}^{(0)})^{-1}(x|y) \Gamma_{\mathrm{V}, \nu} \bigg\rbrace
\,, \label{eq-app-a-sinkprop}
\end{align}
and the last one is obtained from \eqref{eq-app-a-sinkprop} with $x \leftrightarrow y$ and $f \leftrightarrow g$.

\subsection{Sea-valence diagrams}
The sea-valence contributions are obtained by combining the $O(e)$ contributions from the observable and the pfaffian 
\begin{align}
\label{eq:evO1R1}
\ev{ \mathcal{O}_1(x,y) \mathcal{R}_1 }_\gamma
= & \ 
- e^2 q_f \sum_h q_h \ 
        \begin{tikzpicture}[baseline=-.1cm,scale=0.5]\footnotesize
            \coordinate (x) at (2,0);
            \draw[quark] (x) node[left] {$x$} arc[start angle=-135, delta angle=90, radius=1.7] coordinate[at end] (y) node[black,below,pos=0.7] {$g$};
            \draw[quark] (y) node[right] {$y$} arc[start angle=45, delta angle=90, radius=1.7] node[black,above,pos=0.7] {$f$} coordinate[pos=0.3] (p1);
            \coordinate (z) at ($(y) + (0,1)$);
            \draw[quark] (z) arc[start angle=180, delta angle=360, radius=.4] node[black,above,pos=0.75] {$h$};
            \draw[photon] (p1) .. controls ($(p1) + (0,.25)$) and ($(z) - (.25,0)$) .. (z);
            \drawPhotonInsertion{p1};
            \drawPhotonInsertion{z};
            \draw[externalVertex] (x) circle (.1) node[right] {};
            \draw[externalVertex] (y) circle (.1) node[right] {};
        \end{tikzpicture}
- e^2 q_g \sum_h q_h 
        \begin{tikzpicture}[baseline=-.1cm,scale=0.5]\footnotesize
            \coordinate (x) at (2,0);
            \draw[quark] (x) node[left] {$x$} arc[start angle=-135, delta angle=90, radius=1.7] coordinate[at end] (y) node[black,below,pos=0.7] {$g$} coordinate[pos=0.3] (p1);
            \draw[quark] (y) node[right] {$y$} arc[start angle=45, delta angle=90, radius=1.7] node[black,above,pos=0.7] {$f$};
            \coordinate (z) at ($(x) - (0,1)$);
            \draw[quark] (z) arc[start angle=0, delta angle=360, radius=.4] node[black,below,pos=0.75] {$h$};
            \draw[photon] (p1) .. controls ($(p1) - (0,.25)$) and ($(z) + (.25,0)$) .. (z);
            \drawPhotonInsertion{p1};
            \drawPhotonInsertion{z};
            \draw[externalVertex] (x) circle (.1) node[right] {};
            \draw[externalVertex] (y) circle (.1) node[right] {};
        \end{tikzpicture}
\ ,
\end{align}

The first diagram has the following explicit expression
\begin{align}
\begin{tikzpicture}[baseline=-.1cm,scale=0.5]\footnotesize
            \coordinate (x) at (2,0);
            \draw[quark] (x) node[left] {$x$} arc[start angle=-135, delta angle=90, radius=1.7] coordinate[at end] (y) node[black,below,pos=0.7] {$g$};
            \draw[quark] (y) node[right] {$y$} arc[start angle=45, delta angle=90, radius=1.7] node[black,above,pos=0.7] {$f$} coordinate[pos=0.3] (p1);
            \coordinate (z) at ($(y) + (0,1)$);
            \draw[quark] (z) arc[start angle=180, delta angle=360, radius=.4] node[black,above,pos=0.75] {$h$};
            \draw[photon] (p1) .. controls ($(p1) + (0,.25)$) and ($(z) - (.25,0)$) .. (z);
            \drawPhotonInsertion{p1};
            \drawPhotonInsertion{z};
            \draw[externalVertex] (x) circle (.1) node[right] {};
            \draw[externalVertex] (y) circle (.1) node[right] {};
        \end{tikzpicture}
=
\frac{1}{2} \sum_{wz\mu\nu} &
\Tr \left\{ \left[ (D^{(0)}_{f})^{-1}  \tfrac{\delta D_f^{(1)}}{\delta A_{\mu}}(w)(D^{(0)}_{f})^{-1} \right](y|x) \Gamma_A (D_{g}^{(0)})^{-1}(x|y) \Gamma_B \right\}
\nonumber \\[-2mm] & \times
\Lambda_{\mu\nu}(w-z) 
\Tr \left\{ (D^{(0)}_{f})^{-1} \tfrac{\delta D_f^{(1)}}{\delta A_{\nu}}(z) \right\}
,
\end{align}
while the second diagram is obtained from the first one by exchanging $x \leftrightarrow y$ and $f \leftrightarrow g$.

When ${A},B = {\tilde{\mathrm{V}}},\mathrm{V}$, an additional diagram should be added to eq.~\eqref{eq:evO1R1}, leading to
\begin{align}
\ev{ \mathcal{O}_1(x,y) \mathcal{R}_1 }_\gamma = 
- 2 e^2 q_f \sum_h q_h \ 
        \Re\bigg[\begin{tikzpicture}[baseline=-.1cm,scale=0.5]\footnotesize
            \coordinate (x) at (2,0);
            \draw[quark] (x) node[left] {$x$} arc[start angle=-135, delta angle=90, radius=1.7] coordinate[at end] (y) node[black,below,pos=0.7] {$f$};
            \draw[quark] (y) node[right] {$y$} arc[start angle=45, delta angle=90, radius=1.7] node[black,above,pos=0.7] {$f$} coordinate[pos=0.3] (p1);
            \coordinate (z) at ($(y) + (0,1)$);
            \draw[quark] (z) node[above] {} arc[start angle=180, delta angle=360, radius=.35] node[black,above,pos=0.6] {$h$};
            \draw[photon] (p1) .. controls ($(p1) + (0,.25)$) and ($(z) - (.25,0)$) .. (z);
            \drawPhotonInsertion{p1};
            \drawPhotonInsertion{z};
            \draw[externalVertex] (x) circle (.1) node[right] {};
            \draw[externalVertex] (y) circle (.1) node[right] {};
        \end{tikzpicture}
        \ \bigg]
- e^2 q_f \sum_h q_h \
        \begin{tikzpicture}[baseline=-.1cm,scale=0.5]\footnotesize
            \coordinate (x) at (0,0);
            \draw[quark] (x) node[left] {$x$} arc[start angle=-135, delta angle=90, radius=1.7] node[black,below,pos=0.2] {$f$}  coordinate[at end] (y) node[black,below,pos=0.7] {};
            \draw[quark] (y) node[right] {$y$} arc[start angle=45, delta angle=90, radius=1.7] node[black,above,pos=0.2] {$f$};
            \coordinate (z) at ($(x)!0.2!(y) + (0,1)$);
            \draw[quark] (z) node[above] {} arc[start angle=180, delta angle=360, radius=.35] node[black,above,pos=0.6] {$h$};
            \draw[photon] (x) .. controls ($(x) + (0,.5)$) and ($(z) - (.25,0)$) .. (z);
            \drawVertexPhotonInsertion{x};
            \drawPhotonInsertion{z};
            \draw[externalVertex] (y) circle (.1) node[right] {};
        \end{tikzpicture}
\end{align}
where 
\begin{equation}
\begin{split}
    \begin{tikzpicture}[baseline=-.1cm,scale=0.5]\footnotesize
            \coordinate (x) at (0,0);
            \draw[quark] (x) node[left] {$x$} arc[start angle=-135, delta angle=90, radius=1.7] node[black,below,pos=0.2] {$f$}  coordinate[at end] (y) node[black,below,pos=0.7] {};
            \draw[quark] (y) node[right] {$y$} arc[start angle=45, delta angle=90, radius=1.7] node[black,above,pos=0.2] {$f$};
            \coordinate (z) at ($(x)!0.2!(y) + (0,1)$);
            \draw[quark] (z) node[above] {} arc[start angle=180, delta angle=360, radius=.35] node[black,above,pos=0.6] {$h$};
            \draw[photon] (x) .. controls ($(x) + (0,.5)$) and ($(z) - (.25,0)$) .. (z);
            \drawVertexPhotonInsertion{x};
            \drawPhotonInsertion{z};
            \draw[externalVertex] (y) circle (.1) node[right] {};
        \end{tikzpicture}
        = 
        \frac{1}{2} \sum_{z\rho} 
        & \Tr \left\{ (D_{f}^{(0)})^{-1}(y|x)\Gamma_{\tilde{\mathrm{V}},\,\mu}^{(1)} (D_{f}^{(0)})^{-1}(x|y) \Gamma_{\mathrm{V},\nu} \right\}
        \\[-2mm] & \times
        \Lambda_{\mu\rho}(x-z) \Tr \left\{ (D^{(0)}_{f})^{-1} \tfrac{\delta D_f^{(1)}}{\delta A_{\rho}}(z) \right\}.
\end{split}
\end{equation}

\subsection{Sea-sea diagrams}
The sea-sea contributions $\ev{\mathcal{O}_0 \mathcal{R}_2}_{0+\gamma,c}$ can also be written as $\ev{\mathcal{O}_0 \ev{\mathcal{R}_2}_{\gamma}}_{0,c}$ as $\mathcal{O}_0$ does not depend on the photon field.
The contribution of the reweighing factor turns out to be:
\begin{align}
\ev{ \mathcal{R}_2 }_{\gamma} = &
\sum_f \Delta m_f \ 
\begin{tikzpicture}[baseline=-.1cm,scale=0.6]\footnotesize
    \coordinate (z) at (0,0);
    \draw[quark] (z) arc[start angle=180, delta angle=360, radius=.5] node[right,pos=0.5] {$f$};
    \drawTriangle{z};
\end{tikzpicture}
\ + e^2 \sum_f q_f^2 \ 
\begin{tikzpicture}[baseline=-.1cm,scale=0.6]\footnotesize
    \coordinate (z) at (0,0);
    \draw[quark] (z) arc[start angle=180, delta angle=360, radius=.5] node[right,pos=0.5] {$f$};
    \draw[photon] (z) arc[start angle=0, delta angle=360, radius=.35];
    \drawDoublePhotonInsertion{z};
\end{tikzpicture}
\nonumber \\ &
+ e^2 \sum_f q_f^2 \ 
\begin{tikzpicture}[baseline=-.1cm,scale=0.6]\footnotesize
    \coordinate (z1) at (0,0);
    \draw[quark] (z1) arc[start angle=180, delta angle=180, radius=.5]  node[below,pos=0.25] {$f$} coordinate[at end] (z2);
    \draw[quark] (z2) arc[start angle=0, delta angle=180, radius=.5] node[above,pos=0.25] {$f$};
    \draw[photon] (z1) -- (z2);
    \drawPhotonInsertion{z1};
    \drawPhotonInsertion{z2};
\end{tikzpicture}
\ + e^2 \sum_{fg} q_f q_g \ 
\begin{tikzpicture}[baseline=-.1cm,scale=0.6]\footnotesize
    \coordinate (z1) at (-.5,0);
    \coordinate (z2) at (.5,0);
    \draw[quark] (z1) node[left] {} arc[start angle=0, delta angle=360, radius=.5] node[left,pos=0.5] {$f$};
    \draw[quark] (z2) node[right] {} arc[start angle=180, delta angle=360, radius=.5] node[right,pos=0.5] {$g$};
    \draw[photon] (z1) -- (z2);
    \drawPhotonInsertion{z1};
    \drawPhotonInsertion{z2};
\end{tikzpicture}
\ ,
\end{align}
where the diagrams are given by the following explicit expressions
\begin{align}
\begin{tikzpicture}[baseline=-.1cm,scale=0.6]\footnotesize
    \coordinate (z) at (0,0);
    \draw[quark] (z) node[left] {} arc[start angle=180, delta angle=360, radius=.5] node[right,pos=0.5] {$f$};
    \drawTriangle{z};
\end{tikzpicture}
\ = &
\frac{1}{2} \sum_z \Tr[(D^{(0)}_{f})^{-1}(z|z)]
\ , \\
\begin{tikzpicture}[baseline=-.1cm,scale=0.6]\footnotesize
    \coordinate (z) at (0,0);
    \draw[quark] (z) node[right] {} arc[start angle=180, delta angle=360, radius=.5] node[right,pos=0.5] {$f$};
    \draw[photon] (z) arc[start angle=0, delta angle=360, radius=.35];
    \drawDoublePhotonInsertion{z};
\end{tikzpicture}
\ = &
\frac{1}{2} \sum_{z\mu} \Lambda_{\mu\mu}(0) \Tr [ (D^{(0)}_{f})^{-1}(z|z) \tfrac{\delta D_f^{(2)}}{\delta A^2_{\mu}}(z) ]
\ , \\
\begin{tikzpicture}[baseline=-.1cm,scale=0.6]\footnotesize
    \coordinate (z1) at (0,0);
    \draw[quark] (z1) node[left] {}  arc[start angle=180, delta angle=180, radius=.5]  node[below,pos=0.25] {$f$} coordinate[at end] (z2);
    \draw[quark] (z2) node[right] {} arc[start angle=0, delta angle=180, radius=.5] node[above,pos=0.25] {$f$};
    \draw[photon] (z1) -- (z2);
    \drawPhotonInsertion{z1};
    \drawPhotonInsertion{z2};
\end{tikzpicture}
\ = &
- \frac{1}{4} \sum_{zw\mu\nu} \Lambda_{\mu\nu}(w-z)
\Tr [ (D^{(0)}_{f})^{-1}(w|z)  \tfrac{\delta D_f^{(1)}}{\delta A_{\nu}}(z)
(D^{(0)}_{f})^{-1}(z|w) \tfrac{\delta D_f^{(1)}}{\delta A_{\mu}}(w) ]
\ , \\
\begin{tikzpicture}[baseline=-.1cm,scale=0.6]\footnotesize
    \coordinate (z1) at (-.4,0);
    \coordinate (z2) at (.4,0);
    \draw[quark] (z1) node[left] {} arc[start angle=0, delta angle=360, radius=.5] node[left,pos=0.5] {$f$};
    \draw[quark] (z2) node[right] {} arc[start angle=180, delta angle=360, radius=.5] node[right,pos=0.5] {$g$};
    \draw[photon] (z1) -- (z2);
    \drawPhotonInsertion{z1};
    \drawPhotonInsertion{z2};
\end{tikzpicture}
\ = &
\frac{1}{8} \sum_{zw\mu\nu} \Lambda_{\mu\nu}(w-z)
\Tr [ (D^{(0)}_{f})^{-1}(z|z) \tfrac{\delta D_f^{(1)}}{\delta A_{\nu}}(z) ]
\nonumber \\ &  \hspace{74pt} \times
\Tr [ (D_{g}^{(0)})^{-1}(w|w) \tfrac{\delta D_f^{(1)}}{\delta A_{\mu}}(w)) ]
\ .
\end{align}
These diagrams are independent on $\mathcal{O}_0$, and therefore, they can be computed and recycled for different observables.

}

\section{Meson mass derivatives}
\label{sec:appendix-mass}
In this appendix, we show the results of the fits to the meson mass derivatives.
\begin{figure}[t] 
\begin{subfigure}{.475\linewidth}
  \includegraphics[width=\linewidth]{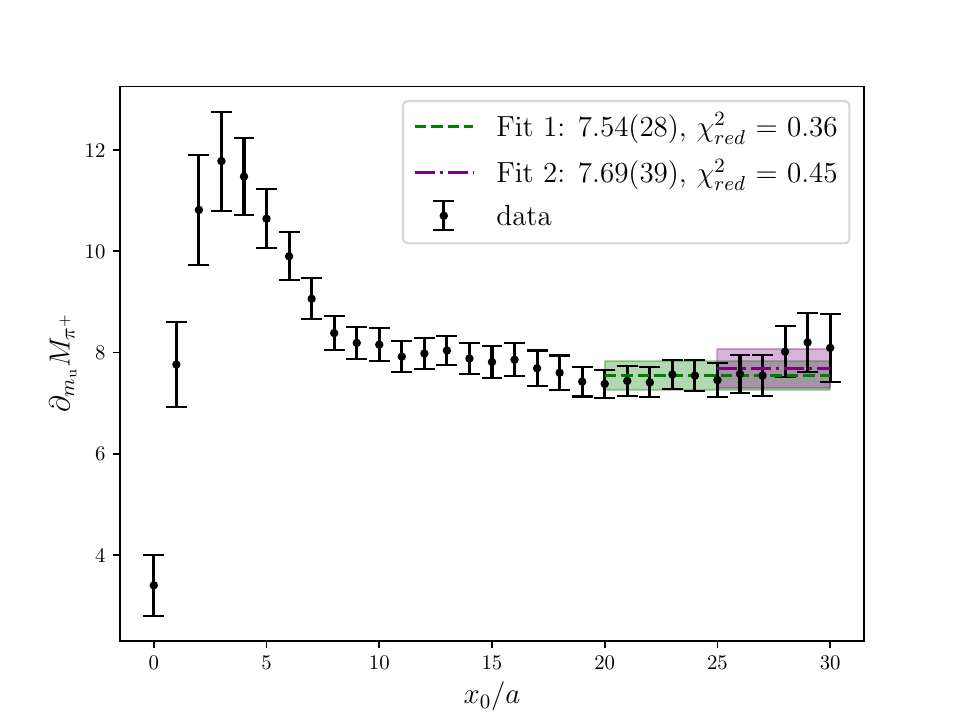}
  \caption{}
  \label{fig: pion mass der. a}
\end{subfigure}\hfill 
\begin{subfigure}{.475\linewidth}
  \includegraphics[width=\linewidth]{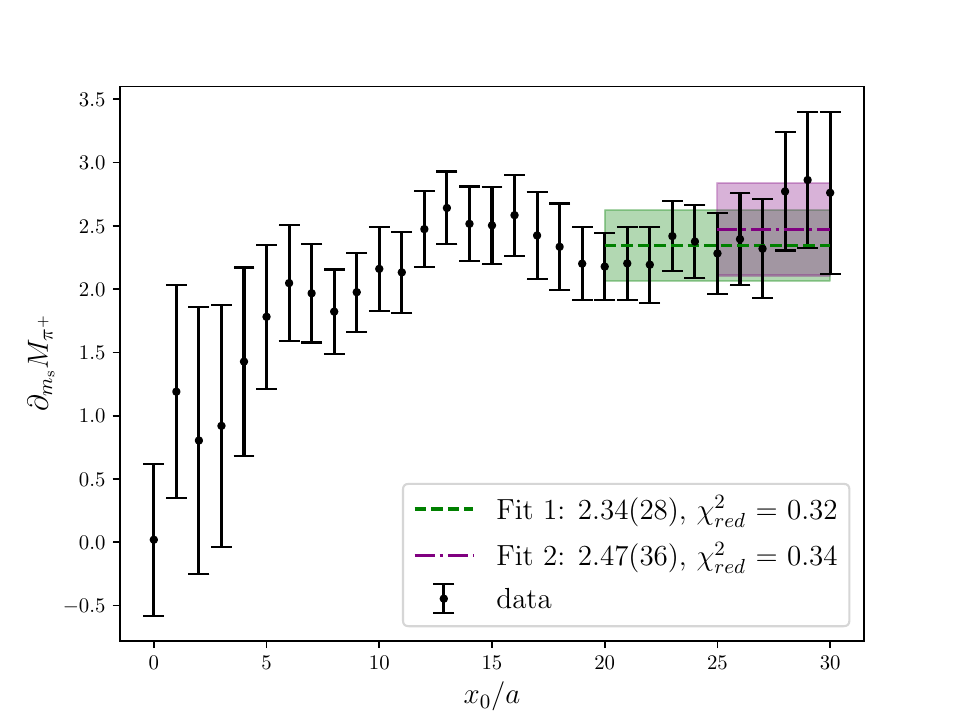}
  \caption{}
 \label{fig: pion mass der. b}
\end{subfigure}
\medskip 
\begin{subfigure}{.475\linewidth}
  \includegraphics[width=\linewidth]{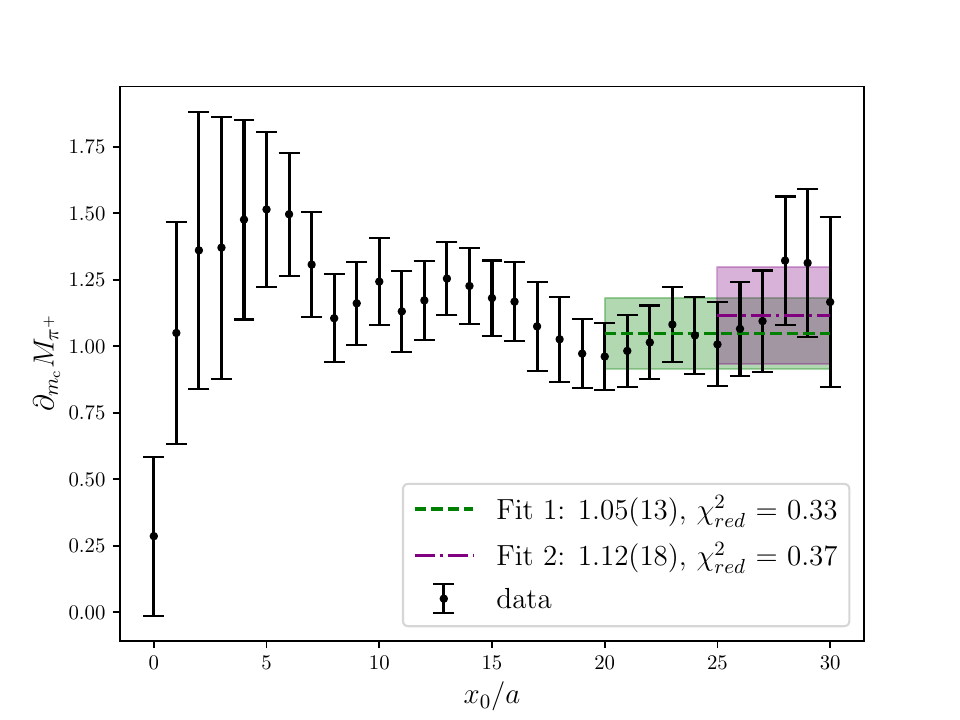}
  \caption{}
   \label{fig: pion mass der. c}
\end{subfigure}\hfill 
\begin{subfigure}{.475\linewidth}
  \includegraphics[width=\linewidth]{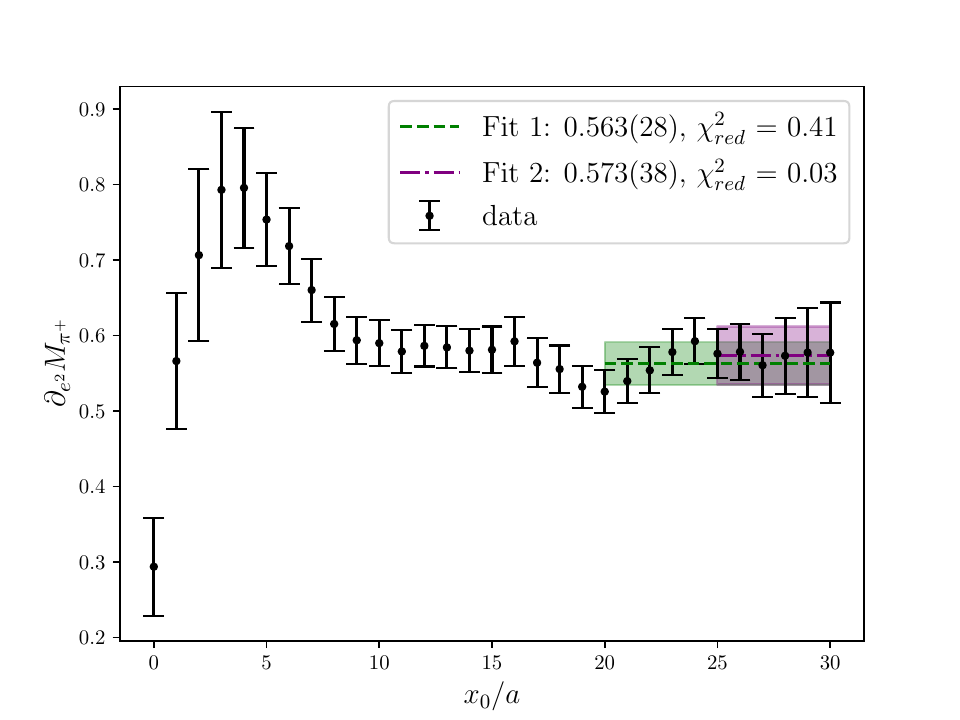}
  \caption{}
   \label{fig: pion mass der. d}
\end{subfigure}
\caption{Fits of the charged-pion mass derivative with respect to the up/down (a), strange (b), charm (c) quark mass, and to $e^2$ (d), computed on \texttt{A400a00}.}
\label{fig:Pimassder}
\end{figure}
\begin{figure}[t] 
\begin{subfigure}{.475\linewidth}
  \includegraphics[width=\linewidth]{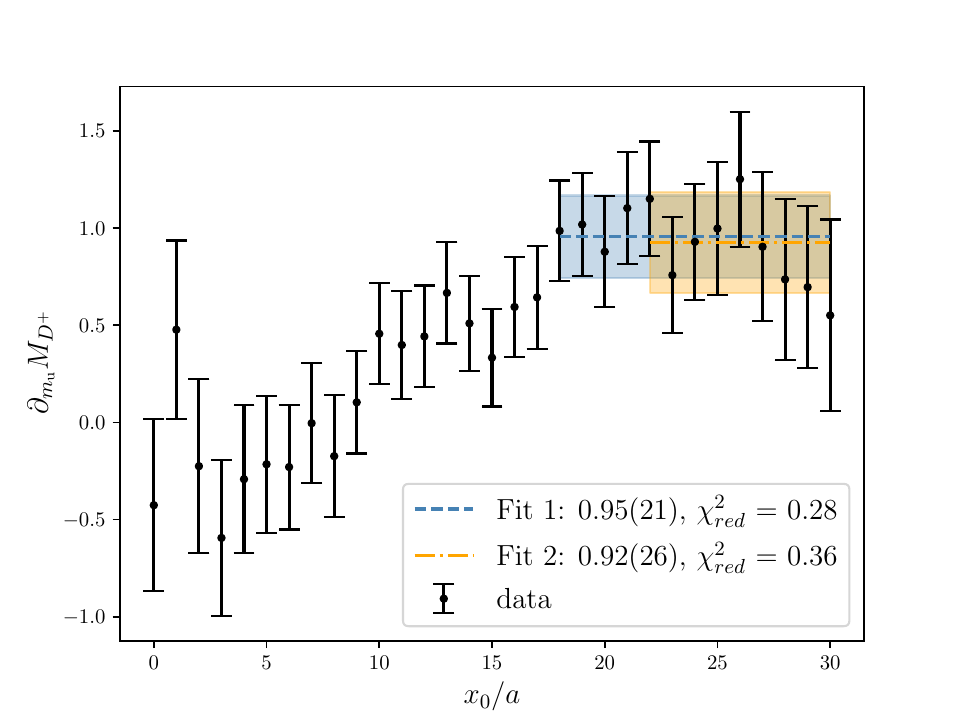}
  \caption{}
  \label{fig: D mass der. a}
\end{subfigure}\hfill 
\begin{subfigure}{.475\linewidth}
  \includegraphics[width=\linewidth]{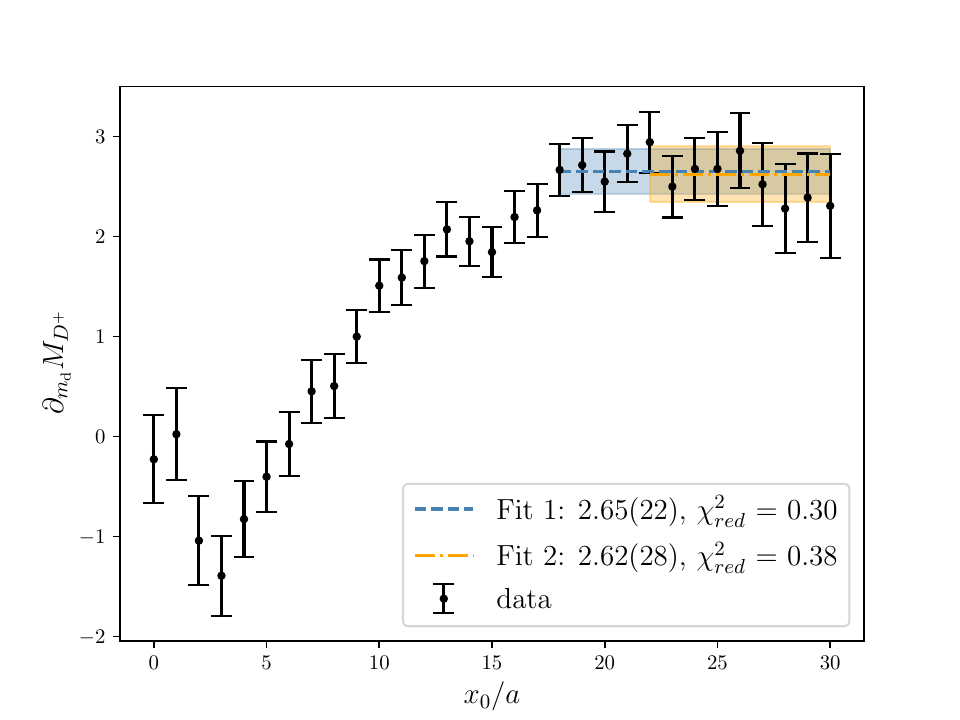}
  \caption{}
 \label{fig: D mass der. b}
\end{subfigure}
\medskip 
\begin{subfigure}{.475\linewidth}
  \includegraphics[width=\linewidth]{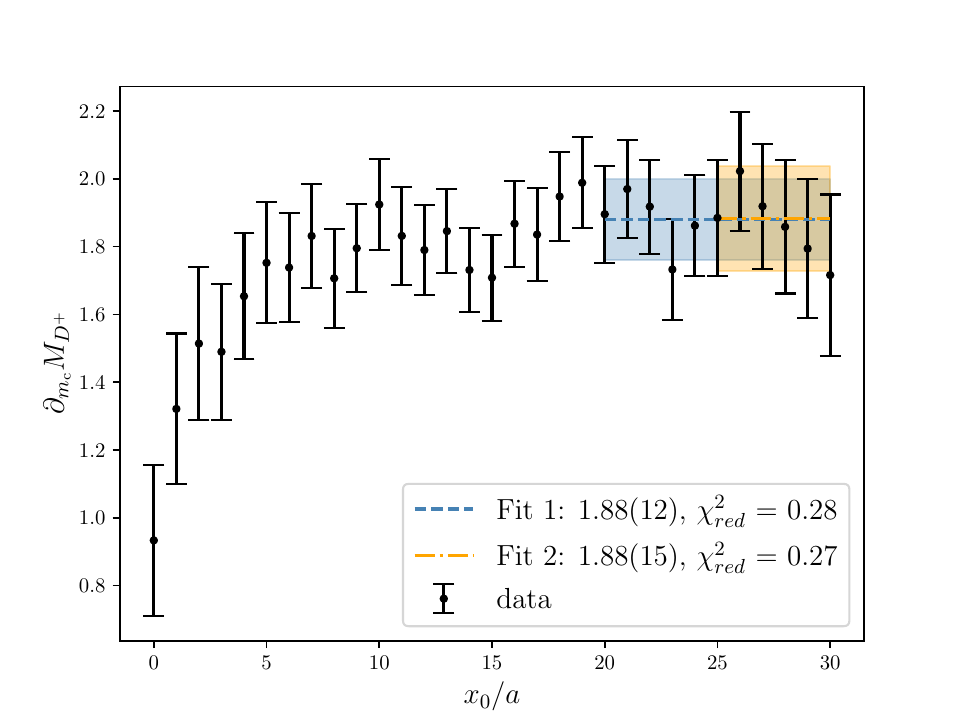}
  \caption{}
   \label{fig: D mass der. c}
\end{subfigure}\hfill 
\begin{subfigure}{.475\linewidth}
  \includegraphics[width=\linewidth]{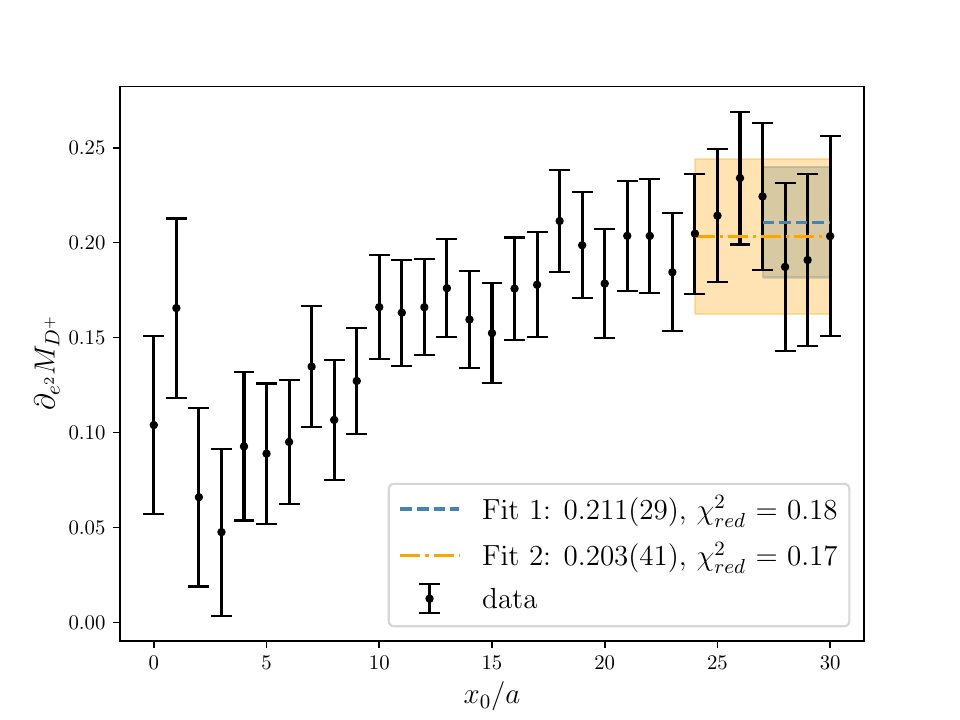}
  \caption{}
   \label{fig: D mass der. d}
\end{subfigure}
\caption{Fits of the charged $D$ mass derivative with respect to the up/strange (a), down (b), charm (c) quark mass, and to $e^2$ (d), computed on \texttt{A400a00}.}
\label{fig:Dmassder}
\end{figure}

In figure~\ref{fig:Pimassder}, we show example fits for the charged-pion mass derivatives.
The plots in the upper panel show the quantities $\partial_{m_{u/d }}
M_{\pi^+}$  and $\partial_{m_{s}} M_{\pi^+}$,  the plots in the lower panel
$\partial_{m_c} M_{\pi^+}$  and $\partial_{e^2} aM_{\pi^+}$.
In each plot, the lattice data and the results of the fit obtained for two fit
ranges are shown.
The fitting procedure minimizes the uncorrelated $\chi^2$ statistic.
Notice that the derivative with respect to $m_{u/d}$ shown in figure~\ref{fig: pion mass der. a} receives valence-valence and sea-sea contributions, while the derivatives with respect to $m_{s}$ and $m_c$, in figures~\ref{fig: pion mass der. b} and~\ref{fig: pion mass der. c} respectively, are due only to sea-sea effects.
The derivative with respect to $e^2$ shown in figure~\ref{fig: pion mass der.
d} receives all contributions. 
Due to the $\mathrm{SU(3)}$ symmetry of the ensemble \texttt{A400a00}, the sea-sea
effects from $m_u,m_d,m_s$ are equal.
Thus, $\partial_{m_{u/d }} M_{\pi^+}$ is the sum of two pieces, a quark-line connected graph correpsonding to the valence-valence effects and a sum of quark-line disconencted graphs for the sea-effects, effectively equal to $\partial_{m_{s}} M_{\pi^+}$.
Moreover, as a consequence of the $\mathrm{SU(3)}$ symmetry and that
$\sum_{f=\mathrm{u,d,s}}q_f=0$, all sea-valence effects to $\partial_{e^2} M_{\pi^+}$ due to light sea quarks
cancel out, as well as some of the sea-sea diagrams.
The fact that the absolute errors of $\partial_{m_{u/d }} M_{\pi^+}$ and
$\partial_{m_{s}} M_{\pi^+}$ are of comparable size indicates that the
uncertainty is dominated by the sea-sea contributions. 
Similar plots for the charged $D$ meson are shown in figure~\ref{fig:Dmassder}.

In tables ~\ref{tab:lmesonsderiv} and ~\ref{tab:cmesonsderiv}, we show the leading-order value and the derivatives of all meson masses appearing in the renormalization system~\eqref{eq:a380rcs}. Some of the derivatives are equal due to the unphysical SU(3) symmetry of the ensemble. The result for each quantity in the table is obtained by considering several fit ranges and combining the fit results based on the associated AIC weights~\cite{Jay_2021}. 
\begin{table}[t]
        \centering
        \begin{tabular}{ccccccc}
            \toprule
             $aM^{(0)}_{\pi} = aM^{(0)}_{K} $                      &
             $\frac{\partial M_{\pi^+,K^{+}}}{\partial m_u} $      &
             $\frac{\partial M_{\pi^+}}{\partial m_s} $            &
             $\frac{\partial M_{\pi^+,K^+,K^{0}}}{\partial m_c} $  &
             $\frac{a\partial M_{\pi^+,K^+}}{\partial e^2}$        &
             $\frac{a\partial M_{K^0}}{\partial e^2}$             \\
            \midrule
              0.1092(6) & 7.55(29) & 2.35(27) & 1.05(14)  & 0.564(29)  & 0.355(28) \\
            \bottomrule
        \end{tabular}
    \caption{Leading-order masses and derivatives of the light mesons computed
    on \texttt{A400a00}. The derivatives that are not listed can be obtained by using $\mathrm{SU}(3)$ flavour symmetry.}
    \label{tab:lmesonsderiv}
        \centering
        \begin{tabular}{cccccc}
            \toprule
          $aM^{(0)}_{D} = aM^{(0)}_{D_s} $                                &
           $ \frac{\partial M_{D_0,D^{+},D^{+}_s}}{\partial m_{u,d,s}} $  &
           $\frac{\partial M_{D^0}}{\partial m_s} $                 &
             $ \frac{\partial M_{D^0,D^+,D^{+}_s}}{\partial m_c} $        &
           $ \frac{a\partial M_{D^0}}{\partial e^2} $                     &
           $ \frac{a\partial M_{D^+}}{\partial e^2} $            \\
           \midrule
            0.5240(8) & 2.65(22)  & 0.95(21)  &  1.88(12) & 0.270(31) & 0.210(30)  \\
        \bottomrule
   \end{tabular}
    \caption{Leading-order masses and derivatives of the charmed mesons
    computed on \texttt{A400a00}. The derivatives that are not listed can
be obtained by using $\mathrm{SU}(3)$ flavour symmetry.}
    \label{tab:cmesonsderiv}
\end{table}

\clearpage
\newpage
\bibliographystyle{JHEP}
\bibliography{biblio.bib}


\end{document}